\documentclass[]{aa}  

\usepackage{natbib}
\bibpunct{(}{)}{;}{a}{}{,}
\usepackage{graphicx}
\usepackage{revsymb}	
\usepackage{amsmath}	
\usepackage[usenames]{color}	
\usepackage{txfonts}
\usepackage{amssymb}
\usepackage{color}
\usepackage{geometry} 
\usepackage[parfill]{parskip} 
\usepackage{amssymb}
\usepackage{slantsc}
\usepackage{array}
\usepackage{url}
\usepackage{amsfonts}
\usepackage[colorlinks=true,linkcolor=black, urlcolor=black, citecolor=black]{hyperref}
\usepackage{hyperref} 
\usepackage{sidecap}
\usepackage{graphicx}
\usepackage{xcolor}
\usepackage{nicefrac}
\usepackage{subfigure}
\usepackage{epsfig}
\colorlet{rouge}{red!70!darkgray}
\newcommand{\who}{WhoSGlAd\xspace}
\begin{document}
\title{Thorough characterisation of the 16 Cygni system.}
\subtitle{Part II. Seismic inversions of the internal structure.}
\author{G. Buldgen\inst{1} \and M. Farnir\inst{2,3} \and P. Eggenberger \inst{1} \and J. Bétrisey \inst{1} \and C. Pezzotti\inst{1} \and C. Pinçon\inst{4,3} \and M. Deal\inst{5} \and S. J. A. J. Salmon\inst{1}}
\institute{Observatoire de Genève, Université de Genève, 51 Ch. Pegasi, CH$-$1290 Sauverny, Suisse \and Department of Physics, University of Warwick, Coventry, CV4 7AL, UK \and STAR Institute, Université de Liège, Allée du Six Août 19C, B$-$4000 Liège, Belgium \and  LERMA, Observatoire de Paris, Sorbonne Université, Université PSL, CNRS, 75014 Paris, France \and Instituto de Astrof\'isica e Ci\^encias do Espa\c{c}o, Universidade do Porto, CAUP, Rua das Estrelas, PT4150-762 Porto, Portugal}
\date{June 2020}
\abstract{The advent of space-based photometry observations provided high-quality asteroseismic data for a large number of stars. These observations enabled the adaptation of advanced analyses techniques, until then restricted to the field of helioseismology, to study the best asteroseismic targets. Amongst these, the $16$Cyg binary system holds a special place, as they are the brightest solar twins observed by the \textit{Kepler} mission. For this specific system, modellers have access to high-quality asteroseismic, spectroscopic and interferometric data, making it the perfect testbed for the limitations of stellar models.}
{We aim to further constrain the internal structure and fundamental parameters of $16$CygA$\&$B using linear seismic inversion techniques of both global indicators and localised corrections of the hydrostatic structure.}{We start from the models defined by detailed asteroseismic modelling in our previous paper and extend our analysis by applying variational inversions to our evolutionary models. We carried out inversions of so-called seismic indicators and attempted to provide local corrections of the internal structure of the two stars.}{Our results indicate that linear seismic inversions alone are not able to discriminate between standard and non-standard models for 16CygA$\&$B. We confirm the results of our previous studies that used linear inversion techniques, but consider that the observed differences could be linked to small fundamental parameters variations rather than to a missing process in the models.}{We confirm the robustness and reliability of the results of the modelling we performed in our previous paper. We conclude that non-linear inversions are likely required to further investigate the properties of 16CygA$\&$B from a seismic point of view, but that these inversions have to be coupled to analyses of the depletion of light elements such as lithium and beryllium to constrain the macroscopic transport of chemicals in these stars and also to constrain potential non-standard evolutionary paths.}
\maketitle
\section{Introduction}

The advent of space-based photometry missions such as CoRoT \citep{Baglin}, \textit{Kepler} \citep{Borucki, ChaplinK2} and TESS \citep{Ricker2015} have made asteroseismic constraints standard tools for studying the internal structure and rotation of distant stars. Asteroseismology is also the golden path to determining precise and accurate stellar parameters, which is now an important goal and a necessary condition for the characterisation of exoplanetary systems. It even constitutes key requirements for the preparation of the PLATO mission \citep{Rauer2014}. Another major consequence of the availability of high-quality data was the generalisation of seismic analyses techniques until then only used in helioseismology, where they were very successful tremendous successes \citep[see e.g.][and references therein for reviews]{JCD2002,Basu08,Kosovichev2011,BuldgenReview,JCD2021}. In this context, the $16$Cygni binary system constitutes a prime target for their application because very high-quality seismic, spectroscopic, interferometric and astrometric data are available. This system has been extensively studied in the past years \citep[see e.g.][]{Mathur2012,Gruberbauer2013,Verma2014,Metcalfe2015,Bazot2020}, and is now considered a benchmark system for stellar modellers. 

The generalisation of linear inversion techniques has already been foreseen and tested on synthetic data \citep[see e.g.][]{Gough1993,Roxburgh98,Thompson2002,Takata2002,Basu2003}. An application to real data can be found a few years later in \citet{DiMauro2004} on Procyon A and more recently in \citet{Kosovichev2020}. Despite the successes of the space missions, the data quality in most cases remained far below that of helioseismic observations and the absence of high-degree modes, as a result of geometric cancellation, limited the potential of linear seismic inversions. However, applications of non-linear techniques can be found in \citet{Appourchaux2015}, who attempted to carry out a full inversion of the internal structure of HIP93511.

As a full scan of the internal structure of a distant star would not be possible with linear variational relations, \citet{Reese2012} focused on developing inversions for global quantities, starting with the mean density. The main goal was to exploit at best low-degree modes, for which some degree of localisation can only be achieved in the core, while they sometimes still show strong dependences on envelope properties. Thus, \citet{Reese2012} and subsequent works on indicator inversions studied target functions that could more easily be fitted and still provided relevant physical constraints on the stellar structure. \citet{Buldgen2015tau,Buldgen2015,Buldgen2018} defined additional quantities, denoted indicators, that could be determined for main-sequence stars observed by \textit{Kepler} \citep{Buldgen2017Legacy}. They applied their technique to a few targets, including the 16Cyg binary system \citep{Buldgen2016,Buldgen2016b}. In their latest study of 16Cyg, they concluded that a full re-modelling was required, especially in light of the publication of new opacity tables \citep{Mondet, Colgan, LePennec}. While both 16Cyg A$\&$B constitute excellent targets for testing non-standard physical processes such as turbulence at the base of the convective envelope and accretion of planetary material \citep{Deal2015}, these processes should be studied in a systematic way on a large set of models. In this approach, seismic inversions may be very attractive, as they may indicate the limitations of stellar evolution models. They are therefore highly valuable to improving the theory of stellar structure and evolution, but also to improve the determination of stellar fundamental parameters such as mass, radius, and age derived in a model-dependent way from stellar evolution computations. In turn, these models lead to more accurate planetary parameters that are crucial when exoplanetary systems are characterised in detail.

A full re-analysis with various evolutionary models was provided by the extensive modelling work of \citet{Farnir2020}, hereafter Paper I. The authors carried out a detailed study of $16$Cyg A$\&$B using the \who oscillation spectrum modelling technique. They defined relevant structural indicators \citep{Farnir2019}, and varied physical ingredients, hypotheses of the modelling (stars seen as independent or joint), and observational constraints. This work provides a suitable set of evolutionary models that can be further tested using seismic inversions of indicators and local corrections of the structure. We start in Sect. \ref{sec:Formalism} by recalling the principles of linear seismic inversions of stellar structure using the variational equations. We then present the evolutionary models we used in our study in Sect. \ref{sec:Models}. Section \ref{sec:Indic} presents our inversion results for the mean density, two core-condition indicators, and an envelope indicator for the 16Cyg binary system, and Sect. \ref{Sec:Prof} shows results of local corrections of the squared isothermal sound speed, denoted $u$ and defined as $u=\frac{P}{\rho}$, with $P$ the local pressure and $\rho$ the local density. Finally, in Sect. \ref{Sec:Disc}, we discuss our results of the inversions of indicators and local corrections and compare them to results reported in the literature \citep{Buldgen2016, Buldgen2016b, Bellinger2017}.

\section{Formalism of variational structural seismic inversions}\label{sec:Formalism}

The concept of helioseismic or asteroseismic inversions is quite broad. It encompasses all approaches of inferences from seismic data of the internal structure of the Sun or of a distant star. For example, the determination of a set of optimal parameters for an evolutionary stellar model from the adjustment of seismic data already constitutes an inference of what its internal structure is, given a set of hypotheses regarding its evolution and the physical processes that govern it. In addition to such inferences based on evolutionary models, other approaches, such as the computation of static models, have been used for SdB stars or white dwarfs \citep[see e.g.][]{Charpinet, Giammichele2018} and also constitute examples of seismic inversions. 

In the context of solar-like oscillators, structural inversions have taken various forms. Iterative inversions based on the reconnection of partial wave solutions have been developed \citep[e.g.][]{RoxTools2002,Roxburgh2010} and applied to a few \textit{Kepler} targets \citep{Appourchaux2015, RoxEps, RoxburgPerky} as well as to synthetic data \citep{Rox2002a, Rox2002b}. Applications of the variational inversions were also investigated early on by \citet{Gough932}, \citet{Gough1993}, \citet{Roxburgh98}, \citet{Thompson2002}, \citet{Basu2002}, and \citet{Basu2003}. In this context, the words "inversion techniques" most often referred to the application of the variational formulae to determine corrections to the internal structure of a star, thus only to a subcategory of seismic inversion techniques. 

The basic equations for the linear inversion problem derive from the variational principle of adiabatic stellar oscillations \citep{Chandrasekhar1964,Clement1964,LyndenBell1967}. In the case of a non-rotating, non-magnetic, isolated star, the problem can be written as a linear integral relation derived by \citet{Dziemboswki90},
\begin{align}
\frac{\delta \nu^{n,\ell}}{ \nu^{n,\ell}} = \int_{0}^{R} K^{n,\ell}_{s_{1},s_{2}}\frac{\delta s_{1}}{s_{1}}dr + \int_{0}^{R}K^{n,\ell}_{s_{2},s_{1}}\frac{\delta s_{2}}{s_{2}}dr, \label{eq:Variational}
\end{align}
where $\nu^{n,\ell}$ is the oscillation frequency of radial order $n$ and degree $\ell$, $s_{1}$ and $s_{2}$ are variables of the adiabatic oscillation equations such as the density, $\rho$, squared adiabatic sound speed, $c^{2}$, pressure, $P$, the first adiabatic exponent $\Gamma_{1}=\frac{\partial \ln \rho}{\partial \ln P}\vert_{S}$, etc. The functions $K^{n,\ell}_{s_{i},s_{j}}$ are the so-called kernel functions, which are related to the variable $s_{i}$ in the structural pair $(s_{i},s_{j})$. These kernel functions depend on the eigenfunctions of the oscillation modes and on the structure of the reference model.The expression $\delta$ denotes a linear perturbation of a given quantity such as the frequency or a structural variable, defined as
\begin{align}
\frac{\delta y}{y}=\frac{y_{\rm{Obs}}-y_{\rm{Ref}}}{y_{\rm{Ref}}},
\end{align}
where $y$ is a given quantity. The suffixes $\rm{Obs}$ and $\rm{Ref}$ refer to the observed quantity and that of the reference model respectively. The reference model should be already close enough of the actual target so that the linear approximation defined in Eq. \ref{eq:Variational} is valid. 

These relations can take various forms. The original expressions presented in \citet{Dziemboswki90} were derived for the density $\rho$ and the squared adiabatic sound speed, $c^{2}=\frac{\Gamma_{1}P}{\rho}$, but variable changes can be applied \citep[see amongst others][]{Elliott1996,Kosovichev,BuldgenKer} to derive expressions for any quantity of the adiabatic oscillation equations. Assuming also a linear development of the equation of state, we may define kernels related to chemical composition or temperature \citep[see e.g.][]{Gough1988}, which were used for example to derive kernels for the helium abundance, denoted $Y$. While this will lead to non-negligible biases in helioseismic inversions \citep[see][for a discussion]{Basu97EOS, BasuSun}, the larger uncertainties in the context of asteroseismology allow us to consider that these biases are negligible. \cite{Basu2009} found variations of about $1\%$ for density inversions and \cite{Basu97EOS} found variations of less than $0.1\%$ for $u$ inversions.  These equations can also be written for frequency combinations or frequency ratios, as in \citet{Oti2005}, and we can also define kernel functions related to Lagrangian perturbations instead of Eulerian ones, as in \citet{JCD1997}, or even perturbations at fixed acoustic radius, as noted in \citet{Pijpers}. 

Equation (\ref{eq:Variational}) suffers from a few caveats. First, it assumes a direct linear relation between the frequency differences and the structural corrections, which has only a limited validity that depends on the nature of the oscillation modes, as well as on the quality of the reference model, which will play a crucial role in asteroseismic inversions \citep{Thompson2002}. This impact justifies the construction of suitable sets of models to which the linear inversion is applied. In practice, this implies that the inference problem is separated into at least two steps, of which the linear inversion is the last. Second, a few hypotheses and simplifications regarding surface regions are made when deriving Eq. \ref{eq:Variational}. The hypothesis of adiabaticity of the oscillations, the use of simplified boundary conditions for the adiabatic oscillation problem (neutrality condition on the Lagrangian pressure perturbation), the poor modelling of the upper convective layers by the mixing-length theory, as well as a few surface terms from integrations by parts are neglected in the derivations. Thus, the formalism does not apply in the upper layers of stars, and ad-hoc corrections for the so-called surface effects have to be included. These take the forms of polynomial corrections, for example following \citet{RabelloParam} or more recent empirical correction by \citet{Ball2014} or \citet{Sonoi2015}. In practice, this implies that a third term needs to be added to the variational formula, written as  
\begin{align}
\mathcal{F}_{\rm{Surf}}=\sum_{i}a_{i}\nu^{i},
\end{align} 
which can then be limited to take the form of the two terms of the \citet{Ball2014} correction, a linearised formulation of the Lorentzian correction of \citet{Sonoi2015} or the polynomial formulation applied in helioseismology, and where the coefficients $a_{i}$ are related to the scaled inertias of the modes. 

In addition to these shortcomings, Eq. \ref{eq:Variational} also suffers from the implicit definition of the boundary of the integral relation. In the Eulerian expression of the perturbations, the radius is assumed to be the same for the observed target and the reference model. If this is not the case, some caution has to be taken regarding the interpretation of the inferred values from the variational expressions. This issue was discussed in \citet{Buldgen2015}, who reviewed the validity of the variational formulations for kernels related to the squared isothermal sound speed $u=\frac{P}{\rho}$, as well as in \citet{BuldgenKer}.

Additional regularisation terms can be applied to the inversion to take care of the radii differences, as in \citet{Takata2001} for the helioseismic case. In asteroseismology, the best scenario occurs when an independent radius determination is available from interferometry or from a combination of Gaia parallaxes and high-quality spectrocopy, although in the latter case, uncertainties regarding bolometric corrections and extinction have to be considered. A way to circumvent the issue is to add a term to the cost function of the inversion to scale the frequencies with respect to $M/R^{3}$ (i.e. the dynamical time) and make them adimensional. This solution was suggested by \citet{Gough932} and also applied in \citet{Roxburgh98}. 

In the case of $16$Cyg, the availibility of high-quality spectroscopic, seismic, astrometric, and interferometric data makes the two stars prime targets for variational seismic inversions. This context indeed motivated previous analyses of their internal structure with variational techniques, such as those of \citet{Buldgen2016}, \citet{Buldgen2016b}, and \citet{Bellinger2017}. 

To summarise, the goal of the seismic inversion techniques is to determine meaningful corrections to stellar evolutionary models by solving Eq. \ref{eq:Variational} using dedicated numerical techniques. Hereafter, we use the substractive optimally localised averages technique \citep[SOLA,][]{Pijpers}, implemented in the InversionKit software \citep{InversionKit}. 

\section{Reference evolutionary models}\label{sec:Models}

\begin{figure}
\begin{flushleft}
  	\includegraphics[trim=5 5 5 5, clip, width=0.95\linewidth]{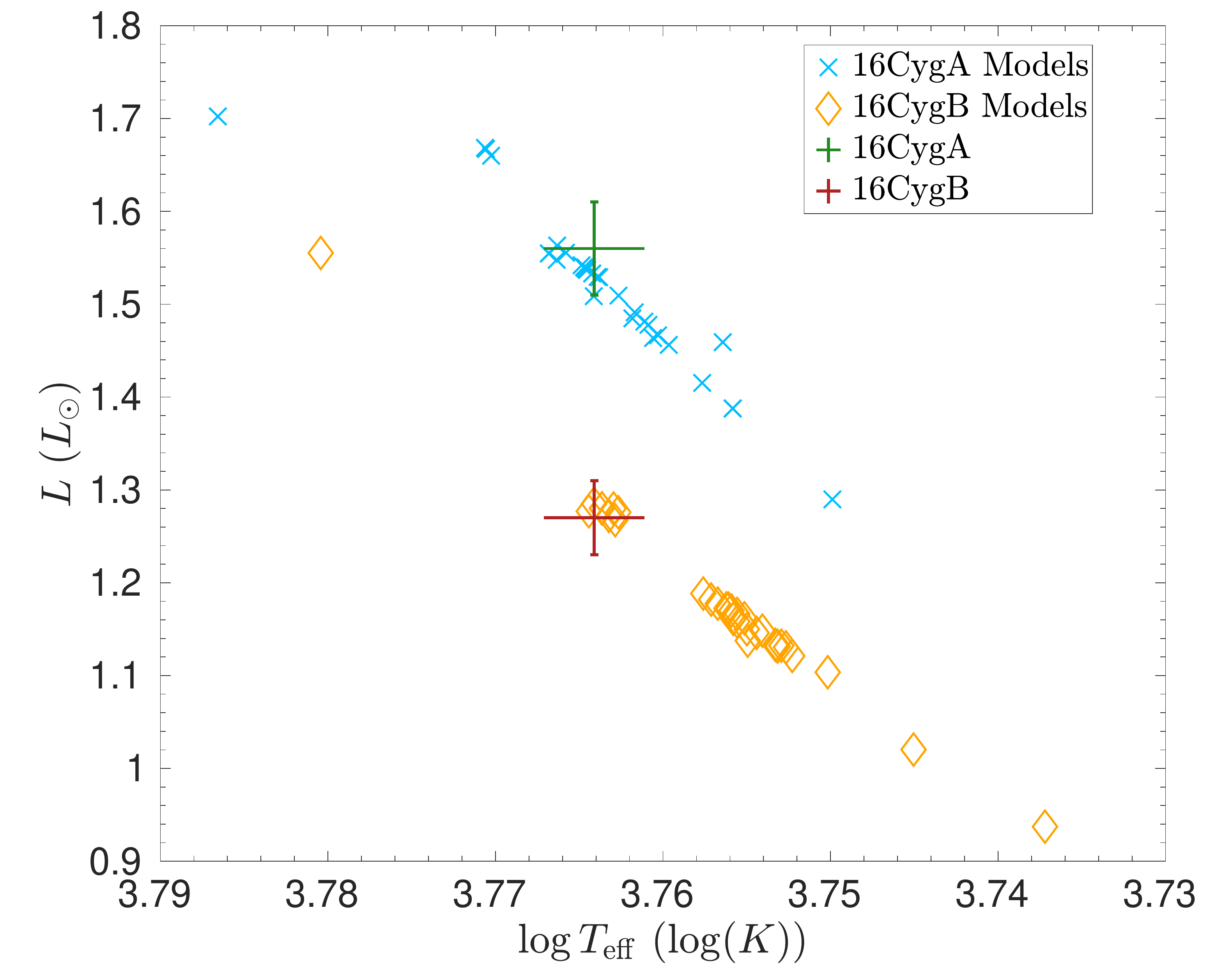}
 \end{flushleft}
	\caption{HR diagram showing the position of the reference models of Paper I with respect to both $16$CygA$\&$B. The green and red dots indicate the positions of 16CygA and 16CygB with their respective $1\sigma$ error bars.}
		\label{FigHR16Cyg}
\end{figure}

\begin{figure}
\begin{flushleft}
  	\includegraphics[trim=5 5 5 5, clip, width=0.95\linewidth]{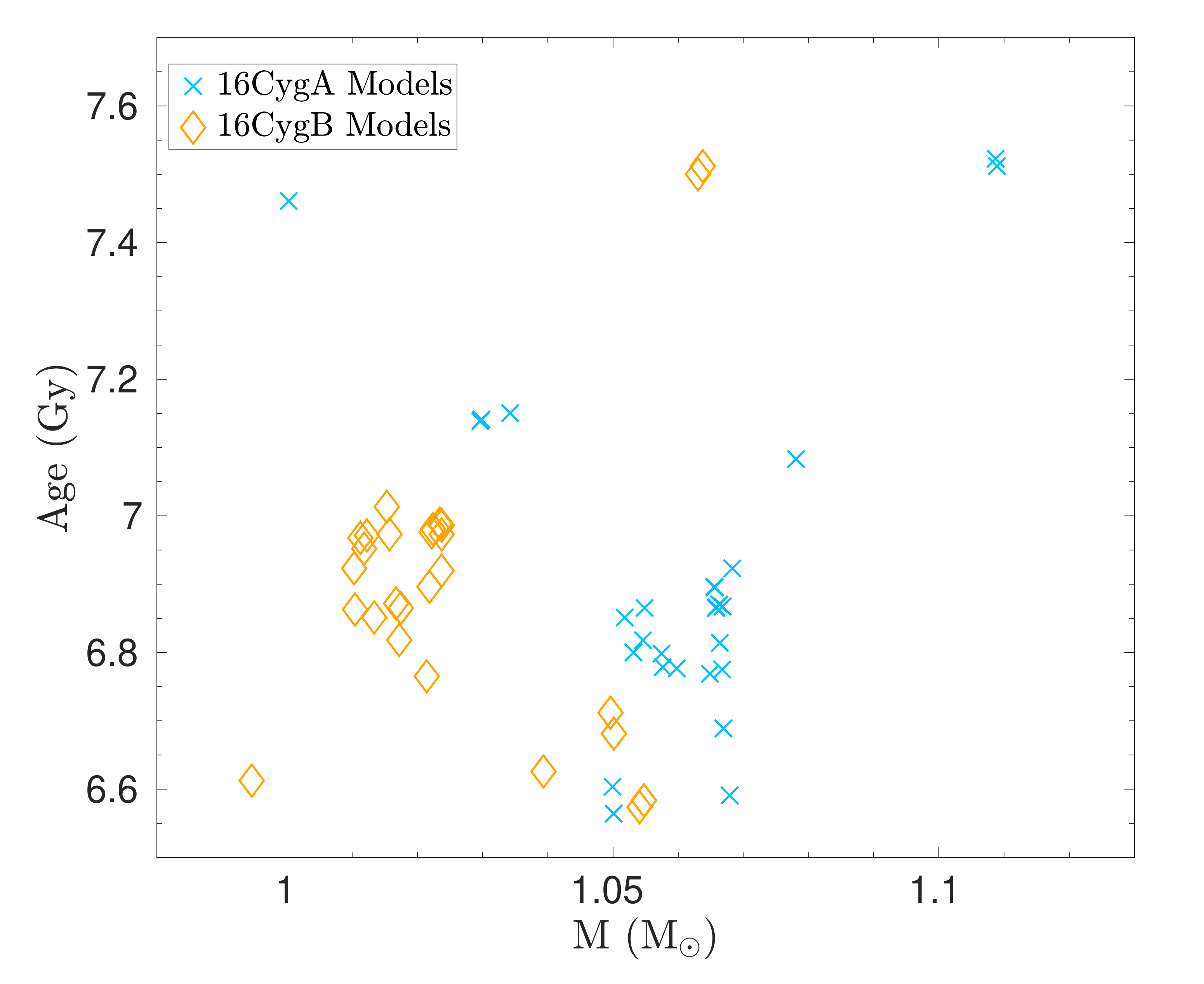}
 \end{flushleft}
	\caption{Age values of the various models determined from seismic evolutionary modelling in Paper I as a function of their mass.}
		\label{FigMAge}
\end{figure}

We used the full set of models of Paper I, computed with the Liège stellar evolution code \citep[CLES;][]{ScuflaireCles}. The adiabatic oscillations were computed using the Liège oscillation code \citep[LOSC;][]{ScuflaireOsc}. We show in Fig. \ref{FigHR16Cyg} the different positions in the Hertzsprung-Russel (HR) diagram of the set of reference models, and in Fig. \ref{FigMAge} we show their positions in a mass$-$age diagram. 

All models were computed using the method of \citet{Farnir2019}. We varied their physical ingredients, including or excluding classical constraints in the modelling and considered independent or joint modelling, where in the latter case, the same age and chemical composition is imposed to both components of the system. We count about $30$ models for each star. We refer to Paper I for the details of the evolutionary modelling procedure and the detailed results and discussion associated with the considered variations in physical ingredients and constraints. 

We briefly recall the variety of physical ingredients that we used to compute these models in Table \ref{tabModelProperties}. Multiple modifications were sometimes considered simultaneously, as mentioned in Paper I. Whenever turbulent diffusion was included, the following parametric mixing from \citet{Proffitt1991} was used
\begin{align}
D_{\mathrm{turb}} = D_{T}\left(\frac{\rho_{\mathrm{bcz}}}{\rho(r)}\right)^n,
\end{align}
with values of $D_{T}=1000,2500,10000$, $n$ being fixed to $3$, $\rho_{\rm{bcz}}$ denoting the density at the base of the convective envelope of the model, and $\rho(r)$ being the local value of the density at a radial coordinate $r$ in the model. For some models, microscopic and turbulent diffusion were entirely neglected. Envelope and core overshooting, when included, were treated as leading to an instantaneous mixing, with the radiative and adiabatic temperature gradient enforced, respectively, in the overshooting regions, following recommendations of \citet{Viallet2015}. The value of the efficiency was fixed to $0.1\rm{H_{P}}$ in both cases, with $\rm{H_{P}}=\frac{-dr}{d\ln P}$ the local pressure scale height. 

This set of stellar evolutionary models includes various physical ingredients and provides an extensive exploration of the possible solutions of detailed seismic modelling for 16Cyg A$\&$B. Moreover, thanks to the optimisation procedure, we have an ensemble of models that is suitable for linear seismic inversions, built with a technique that takes the smooth component of the spectrum and the helium glitch into account simultaneously. As stated in Paper I, the convective envelope glitch could not be constrained in this case as its signal was too small with respect to the uncertainties. The main goal is to determine whether a subset of models stands out from the analysis of linear variational inversions that break free from the evolutionary framework, thus providing constraints on the physical processes acting inside the two stars. 

\begin{table*}[t]
\begin{center}
\caption{Considered variations of physical ingredients in the evolutionary models}
\label{tabModelProperties}
  \centering
\begin{tabular}{r | c }
\hline \hline
\textbf{Physical ingredient}&\textbf{Considered variations}\\ \hline
Chemical Mixture&AGSS09$^{1}$, GN93$^{2}$\\
Opacities &OPAL$^{3}$, OP$^{4}$, OPLIB$^{5}$\\ 
Equation of state& CEFF$^{6}$, FreeEOS$^{7}$, OPAL$^{8}$\\
Atmosphere& Eddington$^{9}$, Model-C$^{10}$\\
Element diffusion& Miscrocopic$^{11,12}$ and Parametric Turbulence$^{13}$\\
Envelope Overshooting& Instantaneous - $\nabla_{Rad}$\\
Core Overshooting& Instantaneous - $\nabla_{Ad}$\\
\hline
\end{tabular}
\\
\small{\textit{References:} $^{1}$\citet{AGSS09}, $^{2}$\citet{GrevNoels},$^{3}$\citet{OPAL}, $^{4}$\citet{Badnell},$^{5}$\citet{Colgan},$^{6}$\citet{CEFF}, $^{7}$\citet{Irwin}, $^{8}$\citet{Opal2002}, $^{9}$\citet{Eddington}, $^{10}$\citet{Vernazza}, $^{11}$\citet{Thoul}, $^{12}$\citet{Paquette}, $^{13}$\citet{Proffitt1991}}
\end{center}
\end{table*}

The dataset we used to carry out the linear inversions is the same as we used for the evolutionary modelling in Paper I. It is the oscillation spectrum determined from the full duration of the \textit{Kepler} mission for these stars presented by \citet{Davies2015}. Non-seismic data were taken as in Paper I from \citet{Ramirez2009} for $\left[\rm{ Fe/H} \right]$ and \citet{White2013} for $T_{eff}$ and $R$.

\section{Inversions of seismic indicators}\label{sec:Indic}

The motivation for the inversion of seismic indicators has been briefly described above. It mainly results from the fact that the observations of only low-degree modes will limit the physical information that can be inferred from the frequencies. In other words, a full tomography of the internal structure of distant stars cannot be achieved, but a few meaningful physical quantities can be defined. 

The SOLA inversion technique is well suited for this goal. \citet{Pijpers} already mentioned that it can be used to determine rotation gradients, for example, and that the target function can be adapted to more general forms than the classical Gaussian function used in helioseismic inversions. The original paper of \citet{Pijpers} started by giving a general form for the target function, the Gaussian target being a specific case aimed at improving results with respect to the original MOLA techniques regarding oscillatory wings of the averaging kernels.. The cost function of the SOLA method for indicator inversions is defined as
\begin{align}
\mathcal{J}(c_{i})=&\int_{0}^{R}\left[ K_{\rm{Avg}}(r)-\mathcal{T}(r)\right] dr + \beta \int_{0}^{R}K^{2}_{\rm{Cross}}(r)dr \nonumber \\ & +\lambda \left[ n - \sum_{i}c_{i}\right]+\tan \theta \frac{\sum_{i}(c_{i}\sigma_{i})^{2}}{<\sigma^{2}>}+\sum_{k}a_{k}\sum_{i}c_{i}\psi_{k}(\nu_{i}), \label{eq:CostFunction}
\end{align}
with $\mathcal{T}$ the target function of the inversion, $\lambda$ a Lagrange multiplier, $c_{i}$ the inversion coefficients, $n$ an integer linked to the indicator definition, $\theta$ and $\beta$ the trade-off parameters, $\sigma_{i}$ the uncertainties of the individual frequencies, $<\sigma^{2}>=\frac{1}{N}\sum_{i=1}^{N}\sigma^{2}_{i}$, and $\sum_{k}a_{k}\psi_{k}(\nu_{i})$ is the polynomial expression of the surface correction \citep[e.g. the correction of][]{Ball2014}. In addition to these quantities, we define in Eq. \ref{eq:CostFunction} two terms, $K_{\rm{Avg}}$ and $K_{\rm{Cross}}$, the averaging and cross-term kernels. They are defined from the recombination of the structural kernels of Eq. \ref{eq:Variational} using the inversion coefficients.

In the most general case, the target function $\mathcal{T}$ of an indicator $A$, related to the structural variable $s_{1}$, is defined using an integral relation of the form
\begin{align}
\frac{\delta A}{A}=\int_{0}^{R}\mathcal{T}(r)\frac{\delta s_{1}}{s_{1}}dr. \label{eq:IndicDef}
\end{align}
For this indicator, the averaging and cross-term kernels are defined from the inversion coefficients $c_{i}$ as
\begin{align}
K_{\rm{Avg}}=\sum_{i}c_{i}K^{i}_{s_{1},s_{2}}, \\
K_{\rm{Cross}}=\sum_{i}c_{i}K^{i}_{s_{2},s_{1}}.
\end{align}
The series of coefficients, $c_{i}$, that are determined from the minimisation of Eq. \ref{eq:CostFunction}, allow us to estimate an inverted value for the indicator defined by Eq. \ref{eq:IndicDef} from the frequencies following
\begin{align}
\left(\frac{\delta A}{A}\right)_{\rm{Inv}}=\sum_{i}c_{i}\frac{\delta \nu_{i}}{\nu_{i}}.
\end{align}
The accuracy of the inversion depends on the quality of the fit of the averaging kernel to the target function, on the damping of the contribution from the cross-term kernel, and on the accurate reproduction of the surface effects. Meanwhile, its precision is determined by the fourth term in Eq. \ref{eq:CostFunction}, related to the propagation of the observational uncertainties of the individual frequencies. The determination of the trade-off parameters is made using a so-called L-curve analysis \citep[see e.g.][]{Backus1970,Pijpers,RabelloParam}.

The integer $n$ is a form of normalisation, introduced for mean density inversions by \citet{Reese2012} and generalised in later works. It is an adaptation of the unimodularity constraint that is commonly used in helioseismic inversions, related to the link between the indicator $A$ and the inverse of the dynamical time, going as $\bar{\rho}^{1/2}$, with $\bar{\rho}$ the mean density. In other words, $n$ is defined from the relation $A \propto \bar{\rho}^{n/2}$ and can be used as an additional regularisation constraint on the inversion coefficients because it can be shown that their sum should be close to $n$ \citep{Reese2012}. 

In practice, each indicator will not be proportional to the dynamical time itself. This requires further discussion regarding the implicit scaling applied by the inversion technique, which has a direct impact on the verification of the integral relations between frequency and structure. The issue is for example seen when the inverted $\frac{\delta u}{u}$ values from synthetic data are compared to the actual differences between two models. If the radii of the two models are different, the relative differences of any quantity that do not scale with the mean density are automatically rescaled by the implicit hypothesis made on the radius when the variational equations are used. It can be shown that they behave as if the ``observed'' model had a mass defined as $\tilde{M}_{\rm{Obs}}=\frac{4 \pi \bar{\rho}_{\rm{Obs}}R^{3}_{\rm{ref}}}{3}$, where the suffix ``Obs'' denotes the target model and ``Ref'' denotes the reference model. 

The issue was already noted in early works, where it was argued that the seismic mean density and the spectroscopic surface gravity values would provide the ratio of $M/R$, which would solve the problem. Other works \citep{Gough932,Roxburgh98} also included a term in the SOLA cost function that is directly related to the minimisation of the mean density. While this certainly is an important effect, it does not disqualify the use of asteroseismic inversions, but shows that comparisons of inversion results from models with different radii should take this scaling into account. 

As the inversion problem is defined as a trade-off between precision and accuracy, the evaluation of the optimal set of trade-off parameters from the sole use of the L-curve can sometimes be difficult. This is especially true for asteroseismic inversions, where the quality of the averaging kernels is not necessarily always good and the applicability of the variational equations is uncertain. In some cases, high accuracy and precision can be accidentally achieved through compensations, meaning that the inversion appears artificially robust. Consequently, when asteroseismic inversions are tested on artificial data, it is useful to compute the actual differences between the known solution and the inverted one using the following expressions:
\begin{align}
\epsilon_{\rm{Avg}}&=\int_{0}^{R}\left( K_{\rm{Avg}} - \mathcal{T}_{A} \right) \frac{\delta s_{1}}{s_{1}} dr,\label{eq:AvgError}\\
\epsilon_{\rm{Cross}}&=\int_{0}^{R}\left( K_{\rm{Cross}} \right) \frac{\delta s_{2}}{s_{2}}dr,\label{eq:CrossError}\\
\epsilon_{\rm{Res}}&= \frac{A_{\rm{Inv}}-A_{\rm{Obs}}}{A_{\rm{Ref}}}-\epsilon_{\rm{Avg}}- \epsilon_{\rm{Cross}},\label{eq:ResError}
\end{align}
with $A_{\rm{Inv}}$ the inverted value of the indicator, $A_{\rm{Obs}}$ the actual value of the indicator for the artificial target, and $A_{\rm{Ref}}$ the value of the indicator for the reference model of the inversion. For an artificial target, these quantities are trivial to compute because $s_{1}$ and $s_{2}$ are known. These expressions can be used to detect potential non-linearities, issues with surface effect corrections, or a compensation of errors within the computation of the integrals that would lead to an artificially high accuracy. 

\subsection{Mean density}\label{Sec:MeanDens}
Mean density inversions were defined in \citet{Reese2012} and then further studied by \citet{Buldgen2015tau} and \cite{Buldgen2019} and were applied to various observed targets. They are applicable to a large number of stars due to their easily fitted target function. 

The definition of the target function is given by
\begin{align}
\mathcal{T}_{\bar{\rho}}=\frac{4\pi\rho x^{2}}{\rho_{R}}, \label{eq:Targetrho}
\end{align}
with $x=\frac{r}{R}$, $\rho_{R}=\frac{M}{R^{3}}$, and $n$ is equal to $2$ in Eq. \ref{eq:CostFunction}. 

\begin{figure*}
\begin{flushleft}
	\begin{minipage}{\textwidth}
  	\includegraphics[trim=5 5 5 5, clip, width=0.47\linewidth]{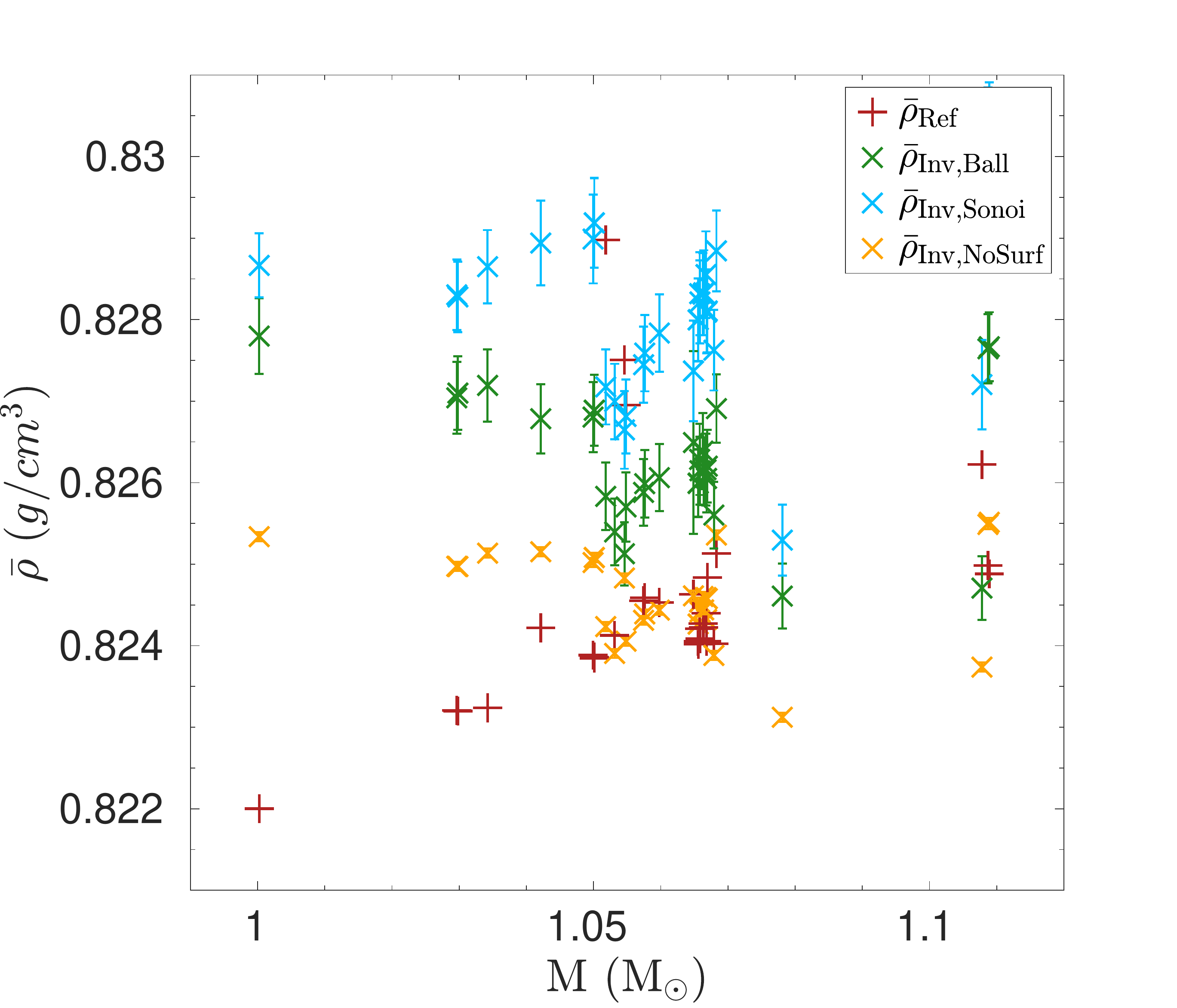}
 	\includegraphics[trim=5 5 5 5, clip, width=0.47\linewidth]{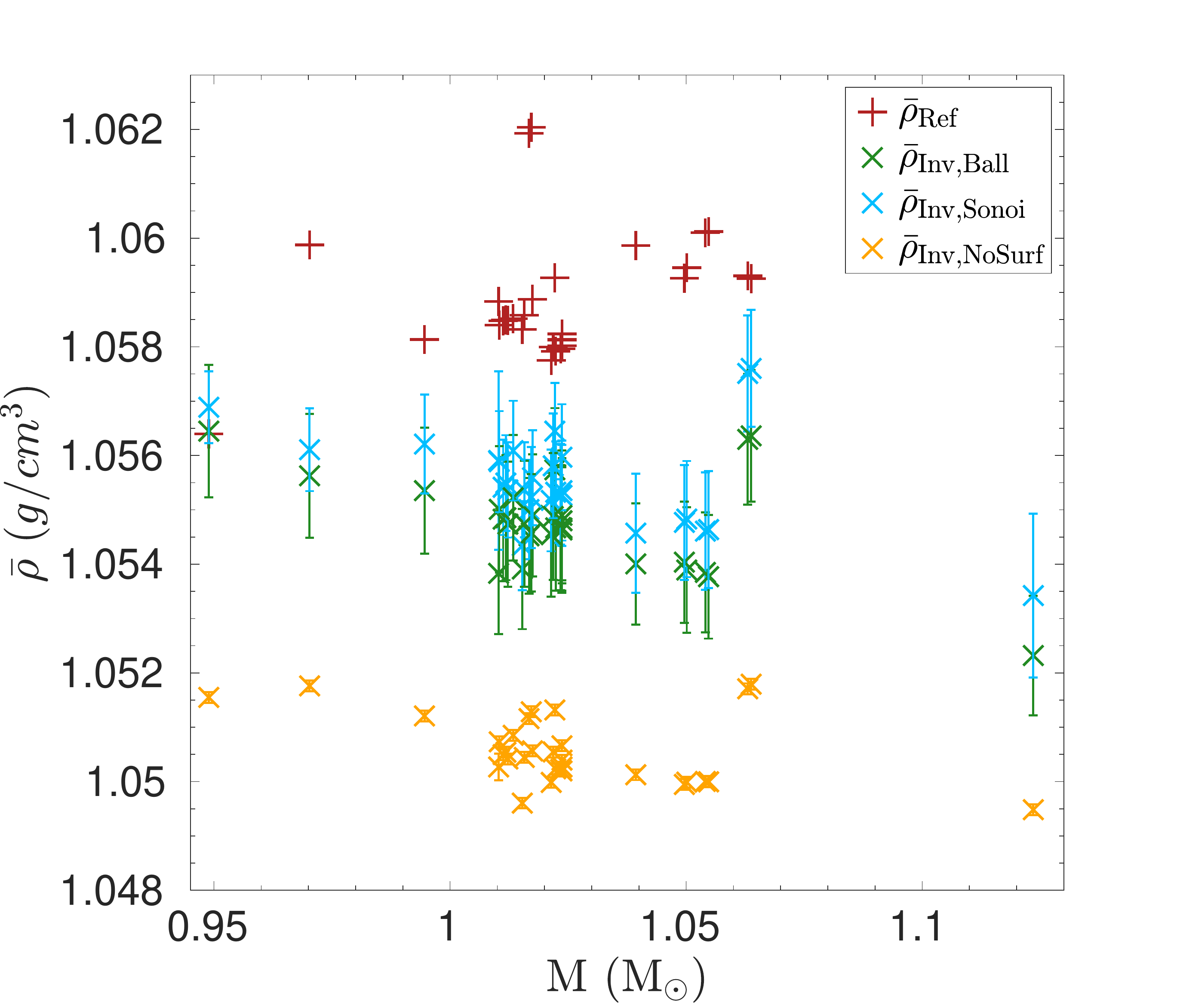}
 \end{minipage}	
 \end{flushleft}
	\caption{Inverted mean density as a function of mass for the set of reference models for 16Cyg A (left panel) and 16Cyg B (right panel). The red crosses are the reference values of the mean density, the orange, green, and blue crosses are the inversion results without any surface corrections, with the surface corrections of \citet{Ball2014} and \citet{Sonoi2015} respectively.}
		\label{FigRhoInvCygAB}
\end{figure*}

The inversion results for both 16Cyg A$\&$B are shown in Fig. \ref{FigRhoInvCygAB} as a function of the mass of the reference models. The verifications of robustness we carried out are presented in Appendix \ref{Sec:RhoCheckups}. We used the models of Paper I and computed the inversions using either the two-terms of \citet{Ball2014}, the \citet{Sonoi2015} surface correction prescriptions, or no correction at all. The reference models are already in excellent agreement with the inversion, especially for 16CygA, where some of them are within the range provided by the SOLA inversion. In the case of 16CygB,  a trend is observed where the reference models always show a slightly higher mean density than the inversion. The disagreement is, significantly reduced, however, if surface corrections are included. Because these corrections themselves have their own uncertainties and the variations between the reference values and the inverted ones are about $0.2\%$, we can consider that most of the evolutionary models already agree very well with the inversions.

Overall, the precision of the inversion is very high. It reaches $0.3\%$ even taking the dispersion due to model-dependency and surface effects into account. We can also note that the values remain consistent even though they span a relatively wide mass range between $0.1$ and $0.2\rm{M}_{\odot}$ for the two stars. The final values for the inverted mean densities of 16CygA$\&$B determined from our study are $\bar{\rho}_{Inv,A}=0.827\pm0.003$ $g/cm^{3}$ and $\bar{\rho}_{Inv,B}=1.055\pm0.003$ $g/cm^{3}$. The already good agreement of the reference models with these inverted values gives confidence in the \who oscillation spectrum modelling technique that we used in Paper I to determine reliable fundamental stellar parameters. 
 
\subsection{Core Condition Indicators}\label{Sec:Core}

Two core condition indicators were developed in \citet{Buldgen2015} and \citet{Buldgen2018}, denoted $t_{u}$ and $S_{\rm{Core}}$. These inversions are more demanding in terms of the quality of the seismic data and are therefore only applicable to a more limited number of targets. They are also more prone to exhibit non-linear behaviours. The confirmations for these inversions are shown in Appendix \ref{Sec:CoreCheckups}.

The target function for the $t_{u}$ indicator is given by
\begin{align}
\mathcal{T}_{t_{u}}=-\frac{2u}{t_{u}}\frac{d}{dr}\left(f(r)\frac{du}{dr} \right), \label{eq:Targettu}
\end{align}
with $f(r)$ defined by
\begin{align}
f(r)=r (r-R) \exp \left(-7\frac{r^{2}}{R} \right), 
\end{align}
with $r$ the radial coordinate, $R$ the stellar radius and $u=\frac{P}{\rho}$, the squared isothermal sound speed. In addition, $t_{u}$ is defined by the integral
\begin{align}
t_{u}=\int_{0}^{R}f(r)\left(\frac{du}{dr}\right)^{2}dr.
\end{align}
From these definitions, \citet{Buldgen2015} showed that $n$ is equal to $4$ in Eq. \ref{eq:CostFunction}.

The target function for the $S_{\rm{Core}}$ indicator is given by
\begin{align}
\mathcal{T}_{S_{\rm{Core}}}=\frac{-g(r)}{S_{\rm{Core}}S_{5/3}},
\end{align}
with $S_{5/3}=\frac{P}{\rho^{5/3}}$ a proxy for the entropy of the stellar plasma. The weight function $g(r)$ is defined by
\begin{align}
g(r)=&r\left(\alpha_{1}\exp\left(-\alpha_{2}(\frac{r}{R}-\alpha_{3})^{2} \right) + \alpha_{4} \exp\left(-\alpha_{5} (\frac{r}{R}-\alpha_{6}) \right)\right)\nonumber \\
 &\tanh\left(\alpha_{7}(1-\frac{r}{R})^{4}\right),
\end{align}
with $\alpha_{1}=16$, $\alpha_{2}=26$, $\alpha_{3}=0.06$, $\alpha_{4}=\alpha_{5}=6.0$, $\alpha_{6}=0.07$, and $\alpha_{7}=50$. These parameters may be varied depending on the star that is studied or the seismic dataset. The integral defition of the $S_{\rm{Core}}$ indicator is given by
\begin{align}
S_{\rm{Core}}=\int_{0}^{R}\frac{g(r)}{S_{5/3}}dr \label{eq:DefSCore}
\end{align}
In this case, \citet{Buldgen2018} showed that $n$ is equal to $\frac{-2}{3}$ in Eq. \ref{eq:CostFunction}.

Both indicators are defined from the expected physical behaviours of core conditions and the rather intricate definitions of the target functions stem from the difficulties of accomodating a limited number of observed frequencies (compared to helioseismology) and the degeneracies linked to the availability of only low $\ell$ modes, often with larger uncertainties. Consequently, not all target functions are equivalent for various sets of observed modes and stars. The differences in mass and evolutionary stages directly affect the behaviour of the structural kernels and the target they can fit. 

\begin{figure*}
\begin{flushleft}
	\begin{minipage}{\textwidth}
  	\includegraphics[trim=5 5 5 5, clip, width=0.47\linewidth]{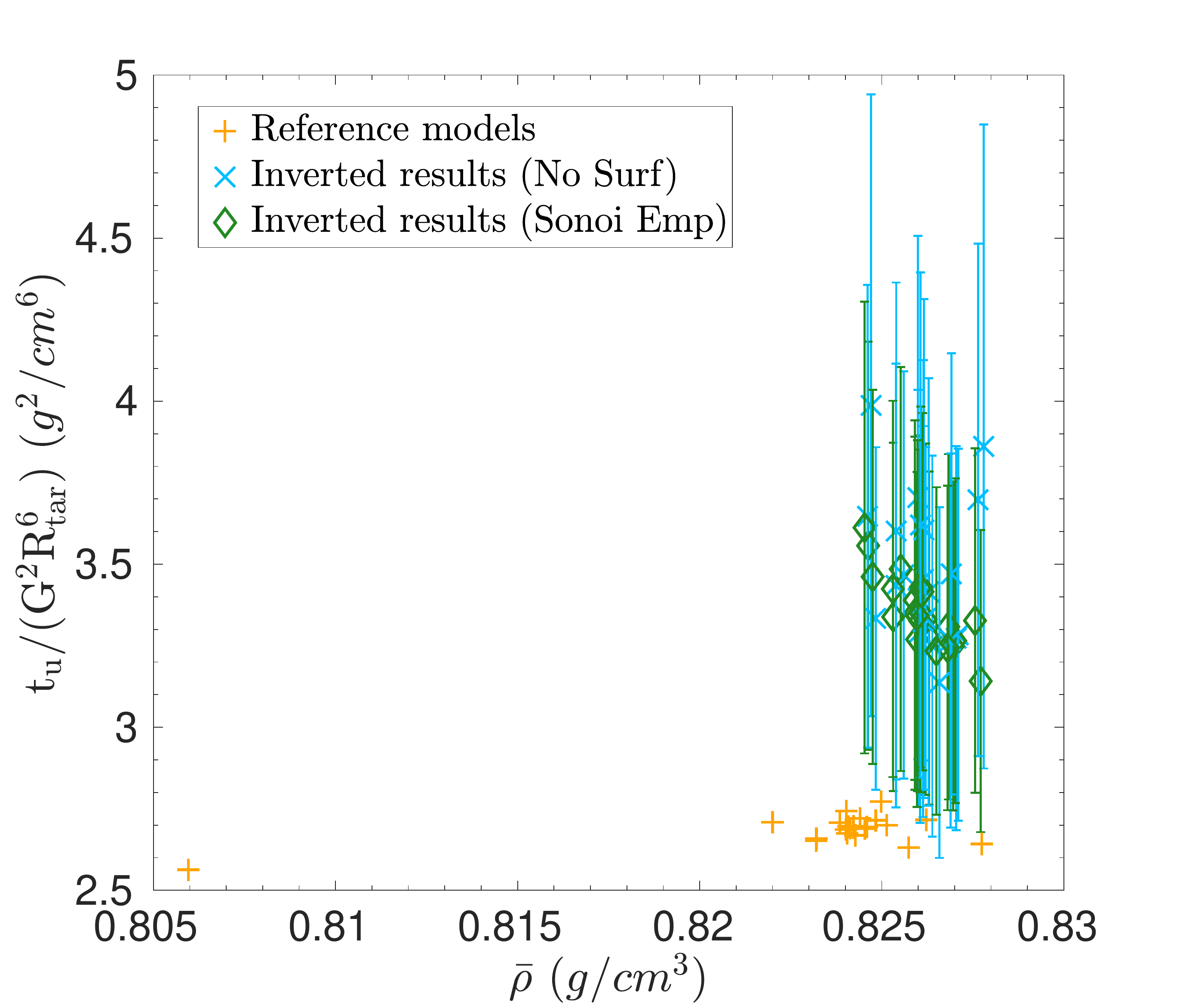}
 	\includegraphics[trim=5 5 5 5, clip, width=0.47\linewidth]{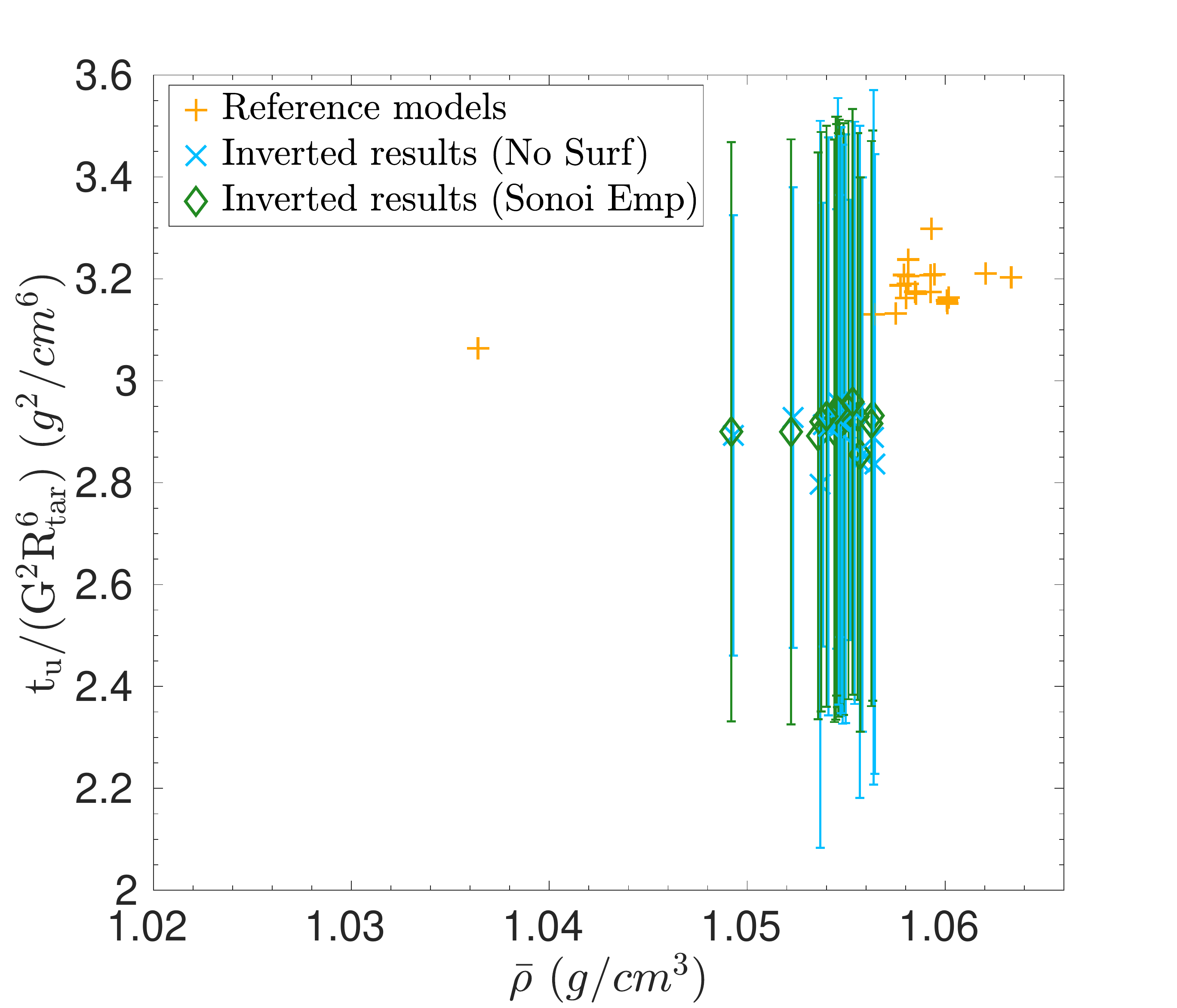}
 \end{minipage}	
 \end{flushleft}
	\caption{Inverted core condition indicator $t_{u}$ as a function of the inverted mean density for the set of reference models for 16Cyg A (left panel) and 16Cyg B (right panel).}
		\label{FigTuInvCygAB}
\end{figure*}

In Fig. \ref{FigTuInvCygAB}, we present the results for the $t_{u}$ indicator for 16CygA and 16CygB as a function of the mean density. In each case, we carried out the inversion without and with surface corrections, using the \citet{Sonoi2015} empirical formula in $\log g$ and $T_{eff}$. This approach allowed us to measure the impact of surface effects while keeping the same quality of fit of the target function by the averaging kernel, as adding two additional parameters to the cost function of the inversion reduces its accuracy, without affecting the conclusions of our study. We started with the case of 16CygB, for which the analysis is straightforward. All models constructed by the \who method agree with the inversion results, with or without surface corrections. A narrow domain of mean density is defined by the mean density inversion, as already noted in Fig. \ref{FigRhoInvCygAB}, but no significant correction in terms of $t_{u}$ is observed. This indicates that the models built from the \who technique agree much better than the models of \citet{Buldgen2016}, which were built by fitting the individual small frequency separations and the inverted mean density or the inverted acoustic radius. 

For 16CygA, the results are in line with those of \citet{Buldgen2016} regarding the $t_{u}$ values, with the inversion favouring a higher range of inverted indicator values. However, as already noted by these authors, the uncertainties of the inversion imply that the indicator cannot be directly used in a modelling procedure. The disagreement appears less important than the one found in \citet{Buldgen2015}. We discuss its potential origin below, but it none of the models of Paper I appears to be in the higher range of values of the inversion results, despite the wide variety of tested physical ingredients. We also note that slightly increasing the trade-off parameter $\theta$ to reduce the error amplification leads to a smaller correction than is illustrated in Fig. \ref{FigTuInvCygAB}, but still shows an increase in the $t_{u}$ value for $16$CygA. Similarly, including a surface correction also leads to slightly lower $t_{u}$ values.

\begin{figure*}
\begin{flushleft}
	\begin{minipage}{\textwidth}
  	\includegraphics[trim=5 5 5 5, clip, width=0.47\linewidth]{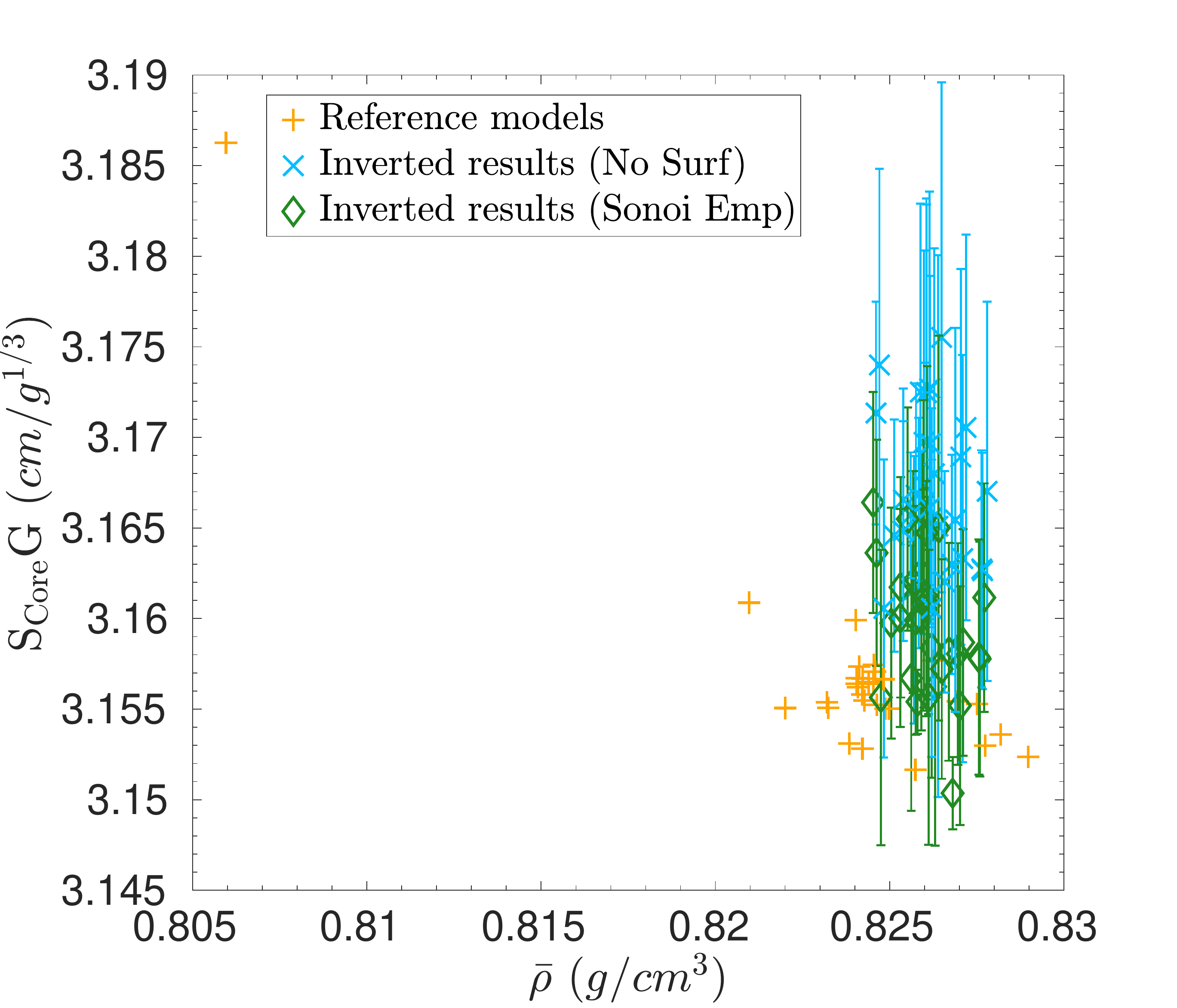}
 	\includegraphics[trim=5 5 5 5, clip, width=0.47\linewidth]{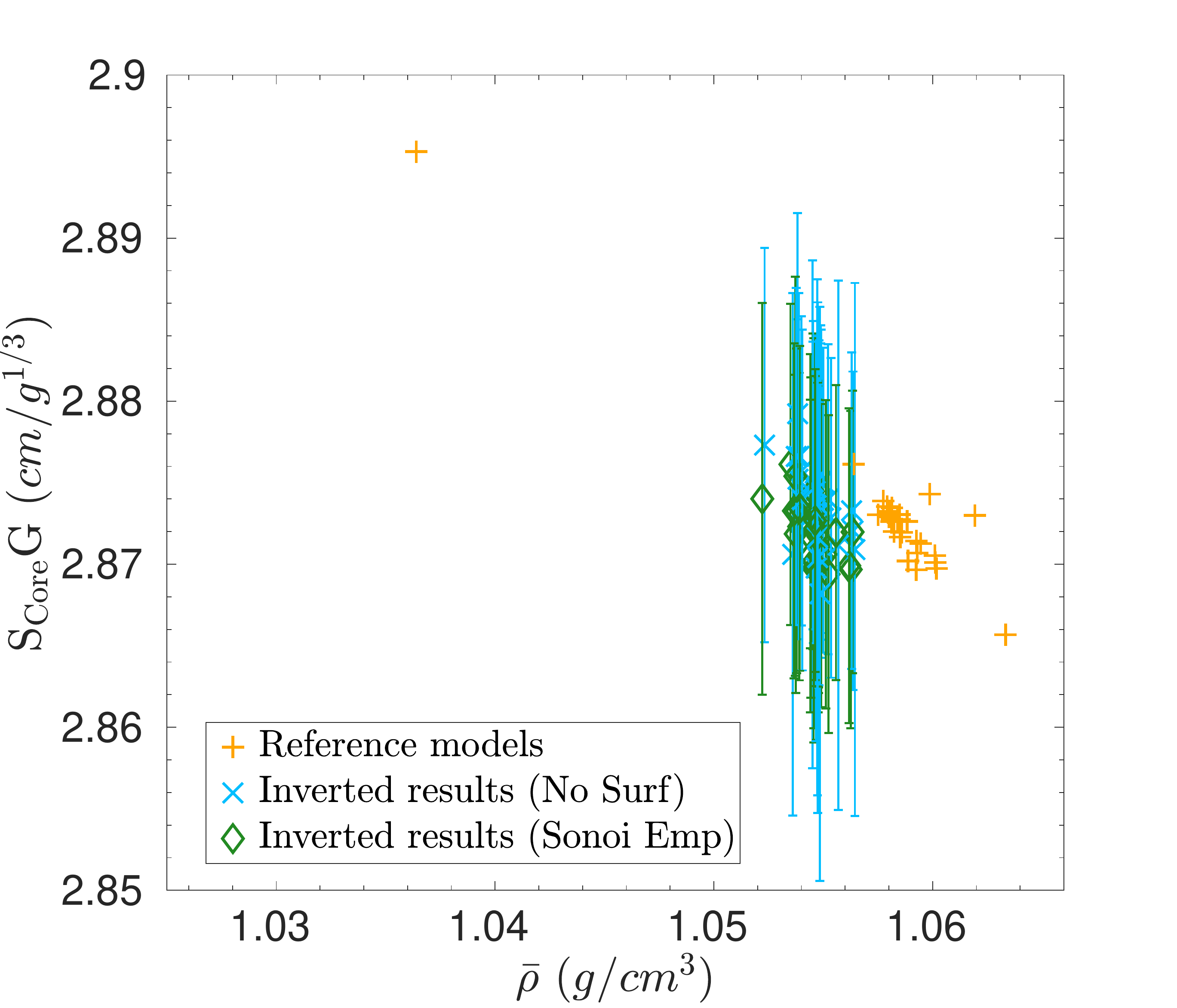}
 \end{minipage}	
 \end{flushleft}
	\caption{Inverted core condition indicator $S_{\rm{Core}}$ as a function of the inverted mean density for the reference models for 16Cyg A (left panel) and 16Cyg B (right panel).}
		\label{FigSCoreInvCygAB}
\end{figure*} 

In Fig. \ref{FigSCoreInvCygAB}, we illustrate the results for the two stars determined using the $S_{Core}$ indicator. The conclusions are quite similar to those found for the $t_{u}$ inversion. For $16$CygA, a slight disagreement is found, with the inversion favouring models with a slightly lower mass or a larger radius, whereas for $16$CygB, the models built from the evolutionary modelling procedure agree excellently with the inverted results. The slight disagreement for $16$CygA almost completely disappears when a surface correction is applied to the frequencies. This implies that the corrections we have seen for the $t_{u}$ inversion might not be the result of a disagreement in the internal structure of the models of $16$CygA.

These results emphasize the importance of the quality of the evolutionary modelling procedure to carry out the inversion. For $16$CygB, all models are validated by the inversion, but not all models agree perfectly with the classical constraints or the interferometric radii. This contrasts with the results of \citet{Buldgen2016}, where by construction, the weighting between the classical and asteroseismic constraints in the cost function was made so that the seismic constraints did not dominate completely. In this case, the \who method allows us to reproduce very well the seismic for 16CygB very well, but as already noted in Paper I, simultaneously reproducing all classical constraints would require additional free parameters. For $16$CygA, the disagreement might be due to surface effects. Nevertheless, it seems that it remains overall less important than the disagreement found in \citet{Buldgen2016}, most likely as a result of the more efficient evolutionary modelling technique. We further discuss in Sect. \ref{Sec:Disc} what the implication of this results in terms of modifications of the physical inputs of the models.

\subsection{Envelope indicator}\label{Sec:Env}

The last seismic indicator we used is the one defined as a convective envelope indicator in \citet{Buldgen2018}, denoted $S_{\rm{Env}}$. This indicator aims at estimating the height of the entropy proxy plateau in the convective envelope, which indicates potential limitations of the temperature and mean molecular weight gradients in the upper radiative layers of a solar-type star. The target function of this indicator is defined as
\begin{align}
\mathcal{T_{\rm{Env}}}=\frac{h(r)S_{5/3}}{S_{\rm{Env}}},
\end{align}
with the weight function $h(r)$ defined as
\begin{align}
h(r)=\left[\alpha_{1} \exp\left(-\alpha_{2}(\frac{r}{R}-\alpha_{3})^{2}\right)+\alpha_{4} \exp\left(-\alpha_{5}(\frac{r}{R}-\alpha_{6})^{2}\right)\nonumber \right. \\ \left.+\frac{0.78}{1+\left(\exp\left((\frac{R}{r}-\frac{1}{\alpha_{7}})/\alpha_{8}\right)\right)}\right] r^{\alpha_{9}}\tanh \left(\alpha_{10}\left(1-\left(\frac{r}{R}\right)^{4}\right)\right),
\end{align}
with $\alpha_{1}=30$, $\alpha_{2}=120$, $\alpha_{3}=0.31$, $\alpha_{4}=7.3$, $\alpha_{5}=26$, $\alpha_{6}=0.33$, $\alpha_{7}=1.7$ $\alpha_{8}=1.2$, $\alpha_{9}=1.5$, and $\alpha_{10}=50$. The $S_{\rm{Env}}$ indicator is defined as 
\begin{align}
S_{Env}=\int_{0}^{R}h(r) S_{5/3}dr. \label{eq:DefSEnv}
\end{align}
For this indicator, \citet{Buldgen2018} showed that $n$ is equal to $\frac{2}{3}$ in Eq. \ref{eq:CostFunction}.  As for the $S_{\rm{Core}}$ indicator, the $\alpha_{i}$ parameters will vary depending on the target and the available oscillation modes. In practice, the $\alpha_{i}$ parameters will vary with the observed star to determine the best target function. These will be mostly $\alpha_{1}$, $\alpha_{2}$, and $\alpha_{3}$ as they define the amplitude, position, and localisation of the highest peak of the function (see e.g. the left panel of Fig. \ref{FigKerSEnvCheckA}) that leads to the largest variations of the indicator.

\begin{figure*}
\begin{flushleft}
	\begin{minipage}{\textwidth}
  	\includegraphics[trim=5 5 5 5, clip, width=0.47\linewidth]{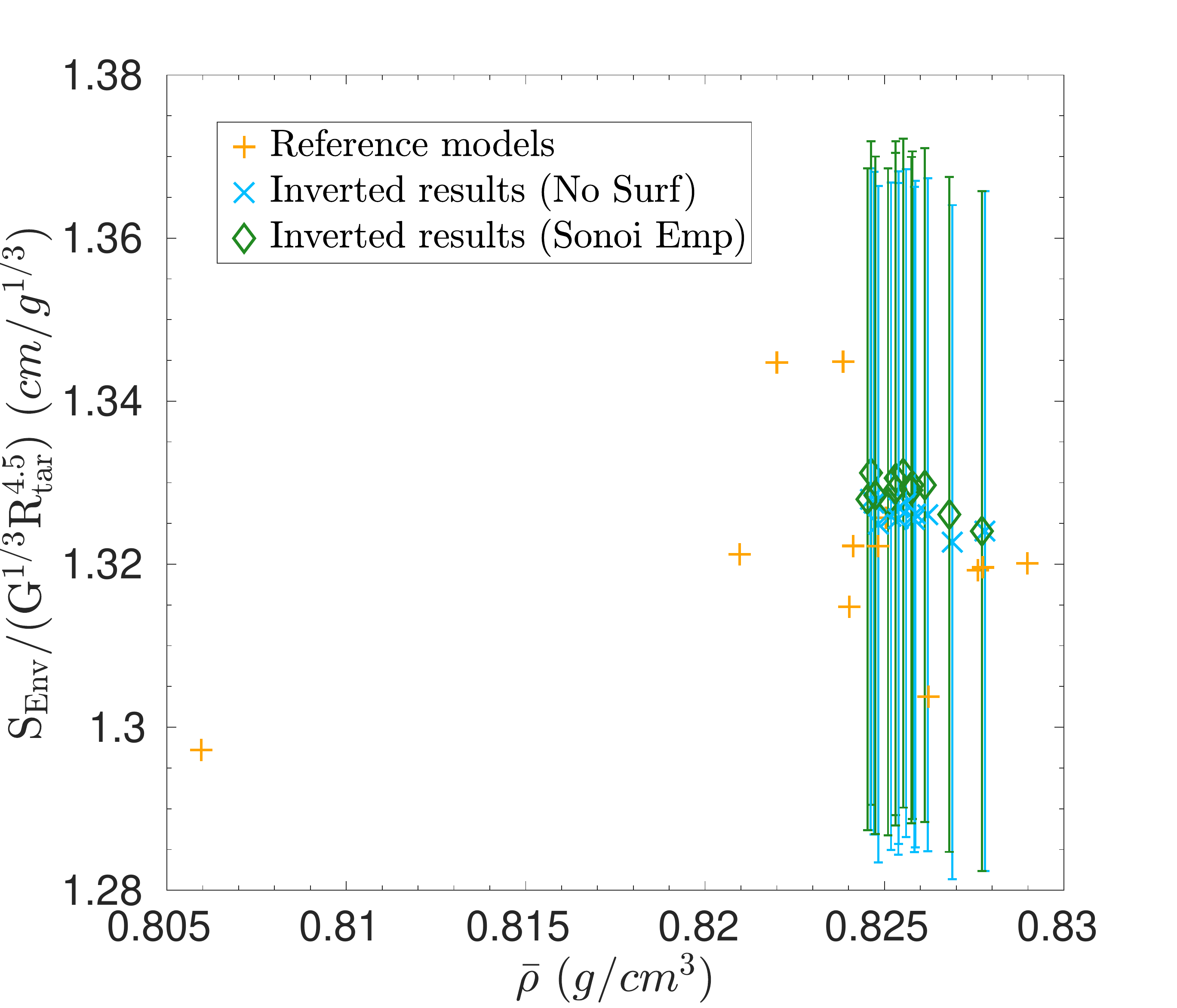}
 	\includegraphics[trim=5 5 5 5, clip, width=0.47\linewidth]{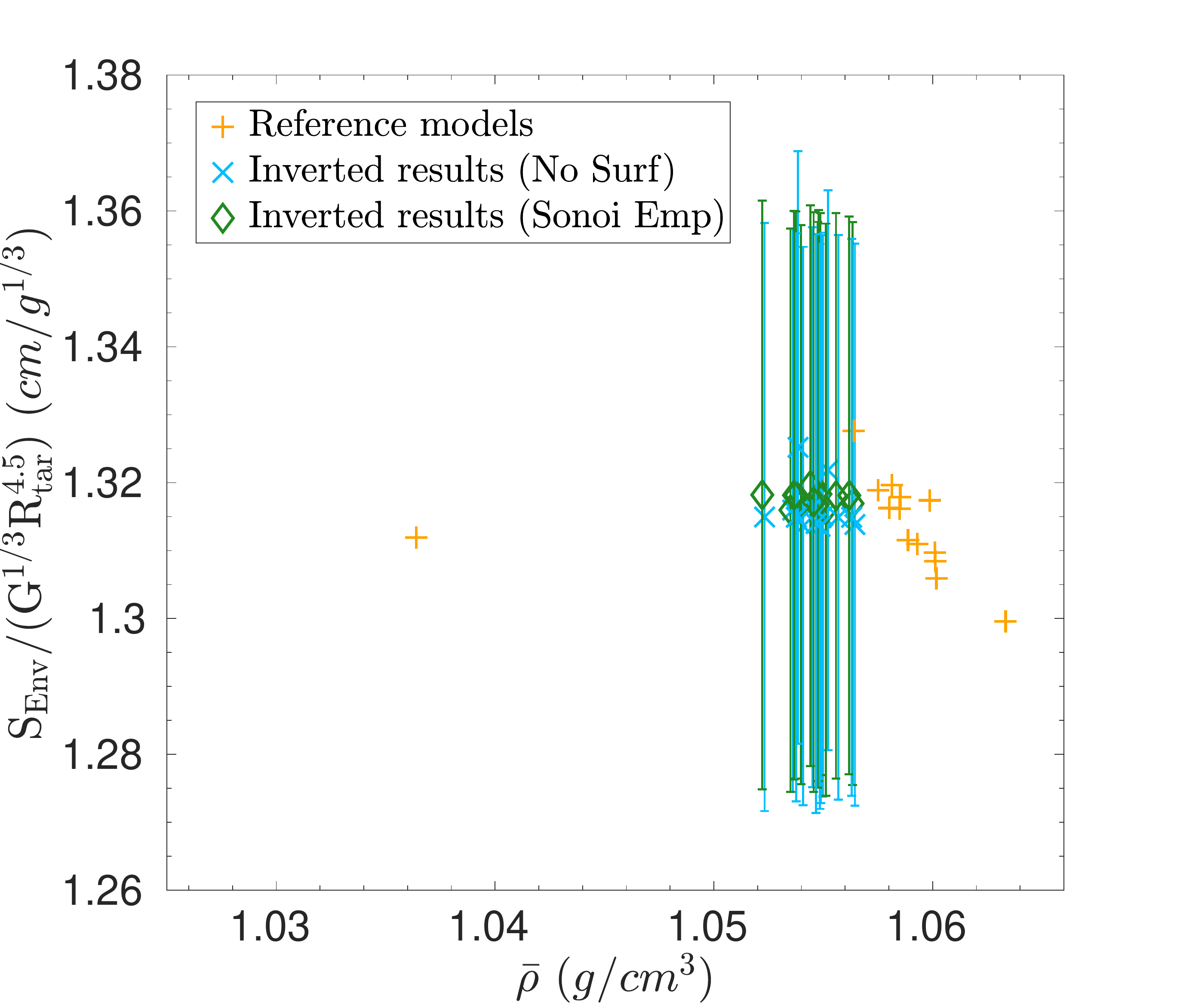}
 \end{minipage}	
 \end{flushleft}
	\caption{Inverted envelope condition indicator $S_{\rm{Env}}$ as a function of the inverted mean density for a subset of reference models for 16Cyg A (left panel) and 16Cyg B (right panel).}
		\label{FigSEnvInvCygAB}
\end{figure*}

In Fig. \ref{FigSEnvInvCygAB} we illustrate the inversion results for both $16$CygA$\&$B for the $S_{Env}$ indicator, and the tests of robustness are shown in Appendix \ref{Sec:EnvCheckups}. We carried out the inversion with and without surface corrections and applied the same approach as in Sect. \ref{Sec:Core}. The results show no significant corrections for the evolutionary models. In hindsight, this is expected because the method of \citet{Farnir2019} focused on reproducing a series of seismic indexes and the helium ionisation glitch signature simultaneously. The variations in the properties of the convective envelope are actually small amongst the models and do not exceed $2\%$ for the two stars. This is beyond the resolution of the inversion technique.

\section{Localised inversions using a Gaussian target}\label{Sec:Prof}

In addition to studying global indicators, we attempted to carry out localised inversions of the profile of the squared isothermal sound speed, $u$. The most suitable structural kernels for this inversion are those of the $\left( u,Y \right)$ pair \citep{BasuJCD2002,Thompson2002,Basu2003}. With these kernels, the cross-term of the inversion will be naturally reduced by the low amplitude of the $\Gamma_{1}$ derivatives with respect to $Y$ and contributes to almost nothing to the cost function, almost transforming Eq \ref{eq:Variational} into a single-integral relation as for rotation inversions. In addition, we can argue that the fitting technique used in Paper I has also intrinsically reduced the cross-term contribution.

We followed the definition of \citet{RabelloParam}, used for the solar case
\begin{align}
\mathcal{T_{\rm{Gauss}}}=Ar \exp\left( \left( \frac{r-r_{0}}{R\Delta} + \frac{R\Delta}{2r_{0}} \right)^{2} \right),
\end{align}
with $A$ a normalisation constant, $r_{0}$ the position at which one wishes to localise the averaging kernel and $\Delta=\frac{\Delta_{A}c(r_{0})}{c_{A}}$ relates to the width of the Gaussian, with $\Delta_{A}$ a free parameter of the SOLA inversion and $c(r_{0})$ and $c_{A}$ the adiabatic sound speed at the inverted point and at a reference radius of $0.2R$, respectively.

The unimodularity term in this case is the one found in usual helioseismic inversions
\begin{align}
\int_{0}^{R}K_{\rm{Avg}}dr=1.
\end{align}

As for the previous cases, we started by carrying out inversions in specific test cases using various evolutionary models that are illustrated in Appendix. \ref{Sec:LocCheckups}. 

For the inversion of the observed data, we chose five models for 16CygA$\&$B in our set of reference models. We note that results remained very similar for other cases, and we only present five results to avoid redundancy. The inversions were carried out using the two-term surface correction of \cite{Ball2014}. Because the stellar radius and mass are unknown, the frequencies need to be adimensionalised \citep{Gough932}. In other words, this implies that the target and reference models have mean density values as close to each other as possible so that we can consider that relative differences in dimensional and adimensional frequencies are equal, such that
\begin{align}
\frac{\nu^{\rm{Obs}}_{n,\ell}-\nu^{\rm{Ref}}_{n,\ell}}{\nu^{\rm{Ref}}_{n,\ell}} \simeq \frac{\tilde{\nu}^{\rm{Obs}}_{n,\ell}-\tilde{\nu}^{\rm{Ref}}_{n,\ell}}{\tilde{\nu}^{\rm{Ref}}_{n,\ell}},
\end{align}
with $\tilde{\nu}=(\frac{GM}{R^{3}})^{-1/2}\nu$ the adimensional frequency. We ensured that the reference model had the same mean density as the target well within $1\%$ by using mean density inversions (see Sect. \ref{Sec:MeanDens}). 

\begin{figure*}
\begin{flushleft}
	\begin{minipage}{\textwidth}
  	\includegraphics[trim=5 5 5 5, clip, width=0.47\linewidth]{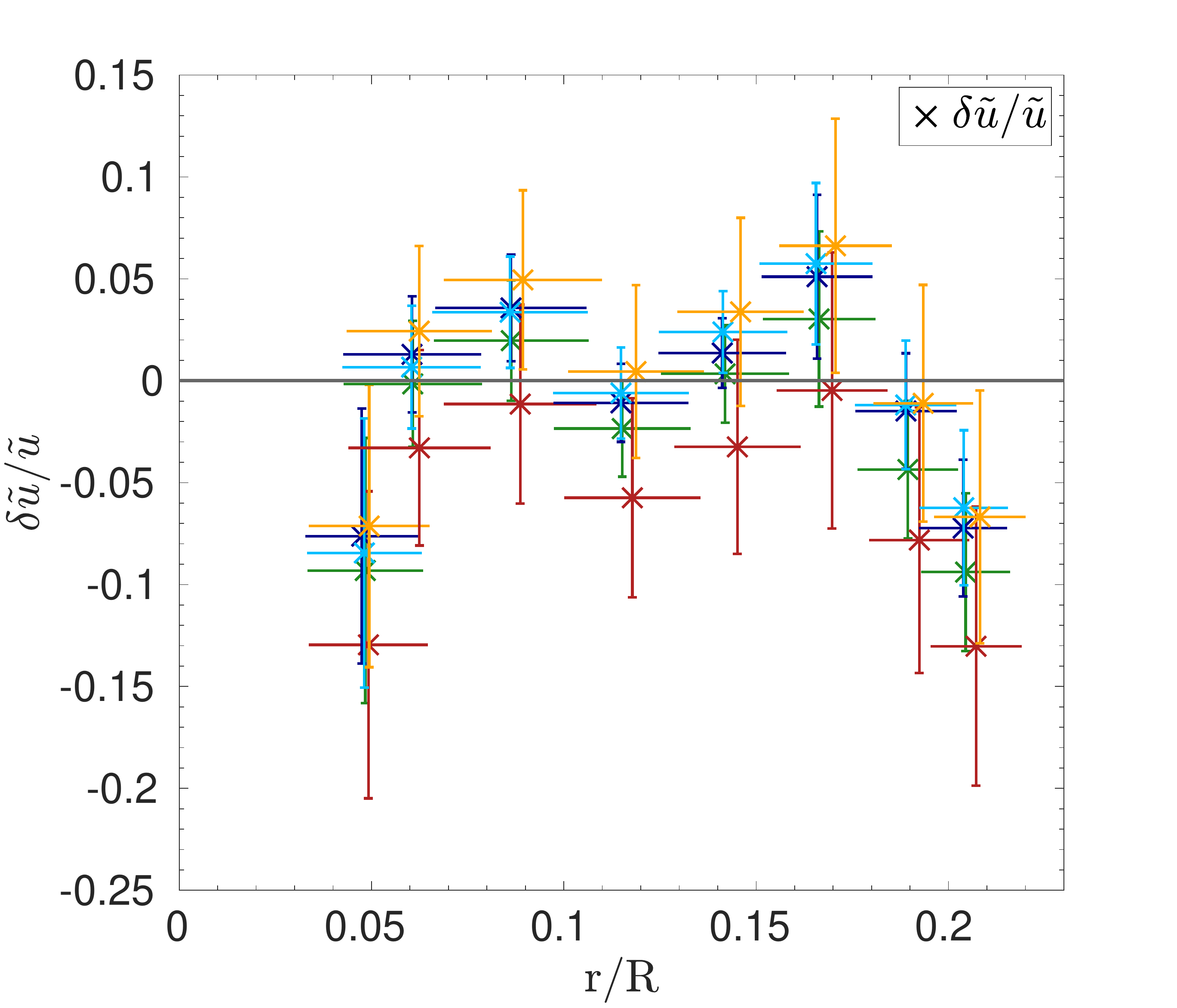}
 	\includegraphics[trim=5 5 5 5, clip, width=0.47\linewidth]{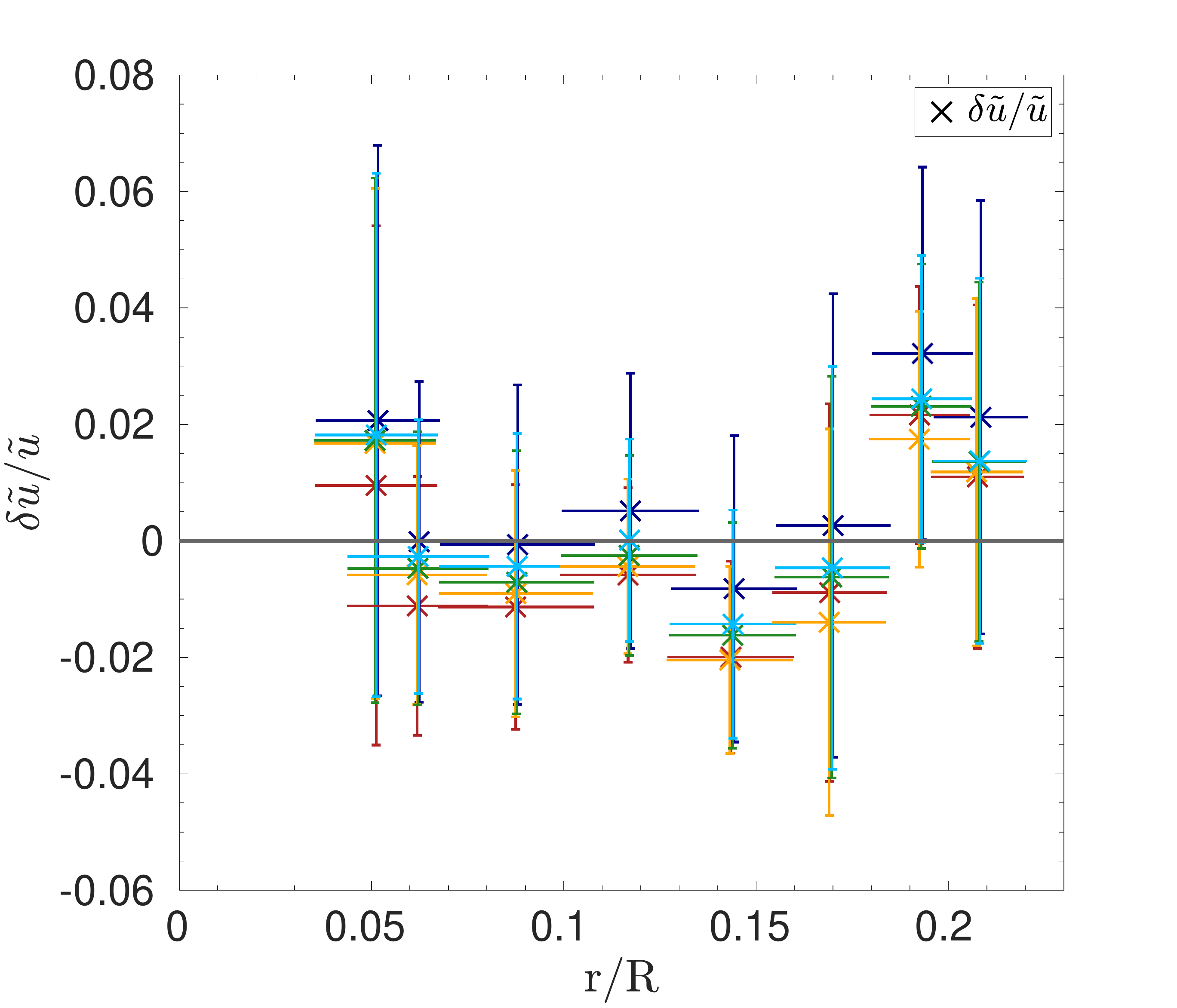}
 \end{minipage}	
 \end{flushleft}
	\caption{Inversion of the adimensional squared isothermal sound speed $(\tilde{u})$ at various points of the internal structure as a function of the normalised radius for 16Cyg A (left panel) and 16Cyg B (right panel) for five models, including the \citet{Ball2014} correction in the cost function. The models and their associated colour code are provided in Tables \ref{Tab:ModelLocalA} and \ref{Tab:ModelLocalB}.}
		\label{FigStrucInvAB}
\end{figure*}

In Fig. \ref{FigStrucInvAB}, we illustrate our inversion results for 16CygA$\&$B, using the surface correction \citet{Ball2014} in the cost function of the inversion technique as for the test cases. We rejected the models without an appropriate mean density value for both cases because we saw in the test cases that they could lead to spurious inversion results.

The properties of the models summarised in Tables \ref{Tab:ModelLocalA} and \ref{Tab:ModelLocalB} for 16CygA and 16CygB respectively. They were chosen to present a variety of physical ingredients in their chemical mixture, opacities, and transport of chemicals they included but also in the constraints and free parameters used in their modelling in Paper I. The constraint ``Same age'' implies that the modelling has been carried out simultaneously for 16CygA$\&$B imposing that the two stars have the same age. 

\begin{table*}[t]
\begin{center}
\caption{Properties of the subset of models we used for localised inversions of $16$CygA}
\label{Tab:ModelLocalA}
  \centering
\begin{tabular}{r | c | c | c | c | c }
\hline \hline
&\textbf{Model 1}&\textbf{Model 2}&\textbf{Model 3}&\textbf{Model 4}&\textbf{Model 5}\\ \hline
Chemical Mixture&AGSS09&AGSS09&AGSS09&AGSS09&AGSS09\\
Opacities &OPAL&OPAL&OPAL&OPLIB&OPAL\\ 
Equation of state&FreeEOS&FreeEOS&FreeEOS&FreeEOS&FreeEOS\\
Element diffusion& Micro&Micro + Turbulence&None&Micro&Micro\\
Constraints& Same age, $X_{0}$, $Z_{0}$ - $\nu_{i}$-$T_{eff}$& Same age - $\nu_{i}$& Same age -$\nu_{i}$&$\nu_{i}$ - $\left[ Fe/H\right]$& $\nu_{i}$ - $T_{eff}$ - $R$\\
Free parameters& M, age, X, Z,$\alpha_{\rm{MLT}}$ & M, age, X, Z & M, age, X, Z & M, age, X, Z & M, age, X, Z,$\alpha_{\rm{MLT}}$ \\
Color in plots& Green & Red & Blue& Cyan & Orange\\
\hline
\end{tabular}
\end{center}
\end{table*}

\begin{table*}[t]
\begin{center}
\caption{Properties of the subset of models we used for localised inversions of $16$CygB}
\label{Tab:ModelLocalB}
  \centering
\begin{tabular}{r | c | c | c | c | c }
\hline \hline
&\textbf{Model 1}&\textbf{Model 2}&\textbf{Model 3}&\textbf{Model 4}&\textbf{Model 5}\\ \hline
Chemical Mixture&AGSS09&AGSS09&GN93&AGSS09&AGSS09\\
Opacities &OPAL&OPAL&OPAL&OPAL&OPAL\\ 
Equation of state&FreeEOS&FreeEOS&FreeEOS&FreeEOS&FreeEOS\\
Element diffusion& Micro&Micro + Turbulence&Micro&Micro&None\\
Constraints& Same age, $X_{0}$, $Z_{0}$ - $\nu_{i}$-$T_{eff}$& Same age, $X_{0}$, $Z_{0}$ - $\nu_{i}$& Same age -$\nu_{i}$&L-$\nu_{i}$-$T_{eff}$& $\nu_{i}$ \\
Free parameters& M, age, X, Z,$\alpha_{\rm{MLT}}$ & M, age, X, Z & M, age, X, Z & M, age, X, Z & M, age, X, Z\\
Color in plots& Green & Red & Blue& Cyan & Orange\\
\hline
\end{tabular}
\end{center}
\end{table*}

Localised kernels could be obtained up to $~0.25$ stellar radii. Above this limit, we consider that the oscillatory wings (see Fig. \ref{FigKerLocalGauss}) of the kernels make the inversion unreliable for our set of trade-off parameters. \citet{Bellinger2017} showed kernels with a slightly less oscillatory behaviour for their kernels and trust their results up to $0.3$ stellar radii. We were able achieve this with our software using different trade-off parameters, namely a larger width of the target function that allowed us to have kernels with an almost Gaussian form at a few additional locations. We also note that they do not show the amplitude of their kernels above the photosphere, where these have high amplitudes that change the results (see e.g. Fig $2$ of \citet{Basu2003}) and also strongly influence the choice of the trade-off parameters. Nevertheless, in our case, adding a few points between $0.25$ and $0.3$ stellar radii to the inversion does not affect our conclusions. 

Figure \ref{FigStrucInvAB} shows that the results for 16CygB indicate that there is no additional information to be extracted from the inversion. Variations of the trade-off parameters lead to similar conclusions, as the inverted values are essentially fully consistent with $0$. 

The case of $16$CygA is slightly more interesting. The inversion seems to pinpoint slight differences at $0.05$ and $0.23$ stellar radii. It might even be argued that the points above $0.15$ stellar radii seem to indicate a different slope in the $u$ profile. Nevertheless, the differences are basically in agreement with $0$ at $1.5\sigma$ for all models at all inverted points. They are also entirely suppressed for some values of the trade-off parameters. 

The model that pecifically fits the interferometric radius and the effective temperature shows the best overall agreement with the inversion. In particular, it excellently agrees in effective temperature and luminosity, but has a higher metallicity by $2\sigma$ than the model that was determined from spectroscopy by \citet{Ramirez2009} and \citet{Tucci2014}, but it agrees slightly better with the values determined by \citet{Morel2021}. From the inversion alone, it seems difficult to advocate that the models need strong corrections of their internal structure.A slight variation in radius of $0.006R_{\odot}$ between the light blue model results and those of the orange model changes from being consistent with $0$ at $1\sigma$ regarding the inversion results. It thus appears that only slight modifications would likely bring most of the models in full agreement with the inversion, as we already devised from the indicator inversions. From a closer analysis, it also appears that non-standard models including additional transport at the base of the convective zone do not fare much better regarding the core inversions of isothermal sound speed. The orange model is indeed essentially a standard model while the red model includes an additional transport of chemicals in the form of turbulent diffusion. There is no clear argument to be made in favour of one or the other, implying that in this specific case, the SOLA inversion of the localised relative squared isothermal sound speed difference has reached its limits. 

\section{Discussion}\label{Sec:Disc}

The combination of various inversions, clearly shows that the models derived for the two stars do not show significant discrepancies in their inversion results. This can be expected for most models based on the dispersion of the results in terms of mass and age in Fig \ref{FigMAge} and on the results of \citet{Bazot2020}. Most of the models lie in a relatively narrow range, and as the tests of robustness of the inversion technique show that their internal structure in terms of indicators and adimensional isothermal sound speed are very similar; they differ by about $1\%$ or less at most. This is well below the resolution limit of the inversion techniques and implies that if the models disagree with an inversion result for $16$CygA$\&$B, they should all do so in similar fashion. 

To some extent, this is shown in the left panels of Figs. \ref{FigTuInvCygAB}, \ref{FigSCoreInvCygAB} and \ref{FigStrucInvAB}. However, we should keep in mind that slight mismatches in the averaging kernels can influence the results and that the accuracy of the inversion itself is not perfect. Moreover, adding surface corrections causes the discrepancies to disappear completely for the $S_{Core}$ indicator and reduces them slightly for the $t_{u}$ indicator, implying that the cause of the differences might well not be the internal structure of the models but, the treatment of the surface effects in the cost function of the SOLA method. Furthermore, potential correlations between the determined individual frequencies(recently reported by O. Benomar for the specific case of 16CygA), could also lead to slightly larger uncertainties on the inverted results, bringing them in agreement with the evolutionary models used here. Slight departures from the solar-scaled chemical mixture could also have an impact on the agreement of the models with the inversions, without the need for invoking non-standard processes. 

\begin{figure}
\begin{flushleft}
  	\includegraphics[trim=5 5 5 5, clip, width=0.98\linewidth]{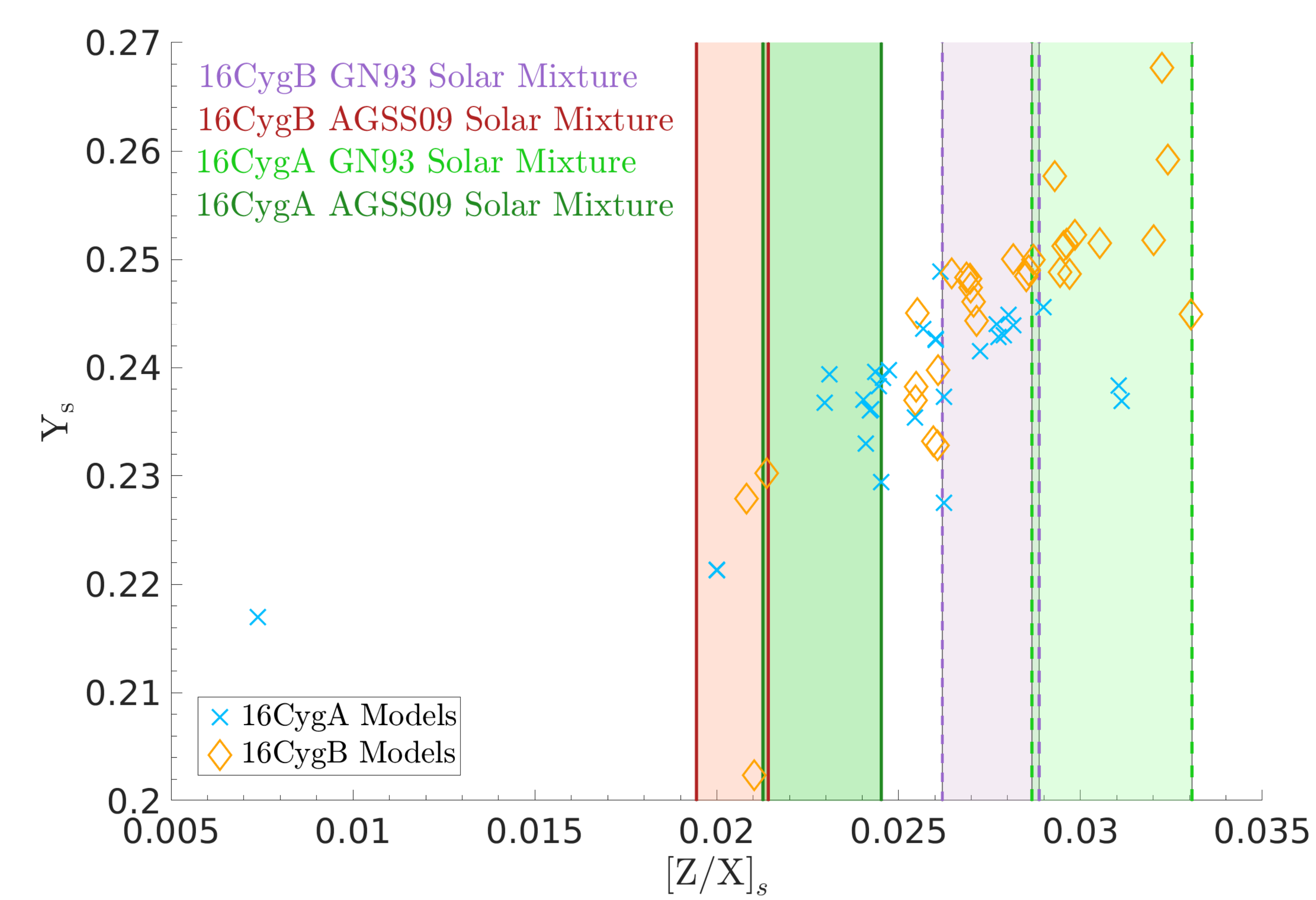}
 \end{flushleft}
	\caption{Surface helium abundance as a function of the surface metallicity for our reference models. The vertical lines indicate the limits derived from spectroscopic estimates by \citet{Ramirez2009}. The helium abundance range is that given from the fit using the \who method.}
		\label{FigZXY16Cyg}
\end{figure}

The fact that both standard and non-standard models fare equally well with respect to the inversions implies that to some extent, the internal structure of the star and its mean density are reproduced within a similar degree of agreement with various physical ingredients, as expected, and that the effects of the change in physics are compensated for by a change in chemical composition, mass and age. This result means that a certain degree of degeneracy exists in asteroseismic constraints. In other words, additional observational constraints on chemical abundances, interferometric radii, parallaxes, and so on will help lift some degeneracies and help provide more reliable models. For example, the disagreement of some of our models with spectroscopic constraints, already seen in Paper I, demonstrates some degree of difference between the various observations, as illustrated in Fig. \ref{FigZXY16Cyg}. Only a few models agree with the constraints for the surface chemical composition, especially for the case of 16CygB. This was already found in paper I and may be our best indicator, alongside lithium and beryllium depletion, that macroscopic mixing is acting inside these stars or that opacity modifications could be required. In this sense, both stars, being solar twins, are directly influenced by the revision of key ingredients of solar models.

Therefore, high-quality spectroscopic data cannot be neglected and the results of seismic inversions do not allow us to select a subset of models that agree significantly better with these non-seismic constraints. Most likely, the most efficient solution is to include them in the evolutionary modelling procedure, as was done for some models of Paper I.

Our models do not confirm the conclusions of \citet{Bellinger2017}, who claimed that missing physical processes, improper applications of known processes or improper inputs in the model computations could be at the origin of their observed differences. We note that in another paper, \citet{Bellinger2019} reported that the uncertainties in their studies of 16Cygni are too large to conclude on differences with the predictions of stellar evolutionary models, as we find here. However, \citet{Bellinger2021} mentioned that the internal sound speed in the core of 16Cygni must be corrected and compared it to the case of HR7322, for which they invoked radiative accelerations. However, \citet{Buldgen2016} were able to recover models with high inverted values of their $t_{u}$ indicator by recomputing models that included more efficient microscopic diffusion, leading to significantly lower masses and younger ages for both stars. The results of \citet{Buldgen2016} are not confirmed here either, as using a larger set of models built with a more robust seismic modelling technique leads to a much less obvious need for corrections in $t_{u}$ and $S_{Core}$. While the trend remains, its origin is not entirely clear and might even potentially be spurious. In this respect we consider the results of the evolutionary modelling of \citet{Farnir2020} to be far more robust than the claimed accepted parameters of \citet{Buldgen2016} for the 16Cyg system.

Regarding the comparison made in \cite{Bellinger2021} between HR7322 and 16Cyg, we can safely say that radiative accelerations can be directly ruled out, as they are not expected to drastically change the internal structure of a solar twin such as $16$CygA \citep{Deal2015}. The inclusion of macroscopic mixing at the base of the convective zone is a promising candidate, however, not to explain the inversion results, as seen here, but rather to reproduce the lithium and beryllium depletions in both stars \citep{King, Deliyannis}. In this respect, the availability of rotation inversions from \citet{Bazot2019} allows us to test prescriptions of transport related to the effects of shear-induced turbulence and magnetic instabilities, as well as to place both stars in the context of the lithium and beryllium destruction in the Sun. The analysis of the impact of this transport on the intriguing discrepancies in light element abundances between 16CygA and 16CygB as well as the potential impact of accretion of planetary matter needs to be placed in perspective with formation, migration and engulfment scenarios for these stars. This requires a multidisciplinary approach as well as a robust seismic modelling procedure, here provided by the \who technique. 

\section{Conclusion}\label{Sec:Conc}

We studied the 16Cyg binary system using variational inversion techniques and the stellar evolutionary models from \citet{Farnir2020}, to determine whether non-standard evolutionary models including macroscopic transport, or models including more recent opacity tables, can be differentiated from standard models by means of linear seismic inversions alone.
 
Our findings are similar to the results of \citet{Buldgen2016} for $16$CygA. Inversions of the core condition indicator, $t_{u}$, seem to indicate the need for a slightly lower mass and larger radius for $16$CygA. The need for corrections is not as clear however, especially for inversions of the $S_{\rm{Core}}$ indicator and internal profile inversions, wich appear to agree with the internal structure of the models we used. Consequently, it seems difficult to advocate for a particular non-standard process that should be acting to reconcile the $16$CygA models. Based on a larger set of models, we see that inversion results often agree with the reference models. When showing corrections, these are barely beyond the $1\sigma$ uncertainties. This implies that model dependency, uncertainty on the surface correction or on the individual frequencies could explain such small differences.

In the case of $16$CygB, the models built using the \who technique agree excellently with all inversions, for standard and non-standard models including additional mixing. 

Our results do not find the trends observed in \citet{Bellinger2017} and show the importance of a detailed and robust seismic modelling procedure before carrying out linear inversions of the stellar structure, coupled to models built including variations of their physical ingredients. In this particular case, the linear inversions validate the \who technique and show its efficiency and robustness for solar twins.

Nevertheless, the presence of non-standard mixing of chemical elements during the evolution of $16$CygA$\&$B can still be investigated using the lithium and beryllium abundances of the two stars, as in \citet{Deal2015}. The hypothesis of the accretion of planetary matter needs to be placed in context with a detailed study of the two stars that takes the rotation inversion results of \citet{Bazot2019} and their implication for the depletion of light elements from macroscopic transport of chemicals and angular momentum  into account. This will be the subject of a forthcoming paper in this series.

\section*{Acknowledgements}
G.B and J.B. acknowledge fundings from the SNF AMBIZIONE grant No 185805 (Seismic inversions and modelling of transport processes in stars). M. F. is supported by the FRIA (Fond pour la Recherche en Industrie et Agriculture) - FNRS PhD Grant and acknowledges the support STFC consolidated grant ST/T000252/1. C.P. acknowledges fundings from the Swiss National Science Foundation (project Interacting Stars, number 200020-172505). P.E. and S.J.A.J.S. have received funding from the European Research Council (ERC) under the European Union's Horizon 2020 research and innovation programme (grant agreement No 833925, project STAREX). C.P. acknowledges support from F. R. S.-FNRS (Belgium) as a Chargé de Recherche. M.D. acknowledge the support of FCT/MCTES through the research grants UIDB/04434/2020, UIDP/04434/2020 and PTDC/FIS-AST/30389/2017, and by FEDER - Fundo Europeu de Desenvolvimento Regional through COMPETE2020 - Programa Operacional Competitividade e Internacionalização (grant: POCI-01-0145-FEDER-030389). M.D. is supported by national funds through FCT in the form of a work contract.

\bibliography{biblioarticle16Cyg}

\begin{thebibliography}{94}
\expandafter\ifx\csname natexlab\endcsname\relax\def\natexlab#1{#1}\fi

\bibitem[{{Appourchaux} {et~al.}(2015){Appourchaux}, {Antia}, {Ball},
  {Creevey}, {Lebreton}, {Verma}, {Vorontsov}, {Campante}, {Davies}, {Gaulme},
  {R{\'e}gulo}, {Horch}, {Howell}, {Everett}, {Ciardi}, {Fossati}, {Miglio},
  {Montalb{\'a}n}, {Chaplin}, {Garc{\'\i}a}, \& {Gizon}}]{Appourchaux2015}
{Appourchaux}, T., {Antia}, H.~M., {Ball}, W., {et~al.} 2015, \aap, 582, A25

\bibitem[{{Asplund} {et~al.}(2009){Asplund}, {Grevesse}, {Sauval}, \&
  {Scott}}]{AGSS09}
{Asplund}, M., {Grevesse}, N., {Sauval}, A.~J., \& {Scott}, P. 2009, ARA\&A,
  47, 481

\bibitem[{{Backus} \& {Gilbert}(1970)}]{Backus1970}
{Backus}, G. \& {Gilbert}, F. 1970, Philosophical Transactions of the Royal
  Society of London Series A, 266, 123

\bibitem[{{Badnell} {et~al.}(2005){Badnell}, {Bautista}, {Butler}, {Delahaye},
  {Mendoza}, {Palmeri}, {Zeippen}, \& {Seaton}}]{Badnell}
{Badnell}, N.~R., {Bautista}, M.~A., {Butler}, K., {et~al.} 2005, \mnras, 360,
  458

\bibitem[{{Baglin} {et~al.}(2009){Baglin}, {Auvergne}, {Barge}, {Deleuil},
  {Michel}, \& {CoRoT Exoplanet Science Team}}]{Baglin}
{Baglin}, A., {Auvergne}, M., {Barge}, P., {et~al.} 2009, in IAU Symposium,
  Vol. 253, IAU Symposium, ed. F.~{Pont}, D.~{Sasselov}, \& M.~J. {Holman},
  71--81

\bibitem[{{Ball} \& {Gizon}(2014)}]{Ball2014}
{Ball}, W.~H. \& {Gizon}, L. 2014, \aap, 568, A123

\bibitem[{{Basu}(2003)}]{Basu2003}
{Basu}, S. 2003, \apss, 284, 153

\bibitem[{{Basu} \& {Antia}(2008)}]{Basu08}
{Basu}, S. \& {Antia}, H.~M. 2008, \physrep, 457, 217

\bibitem[{{Basu} {et~al.}(2009{\natexlab{a}}){Basu}, {Chaplin}, {Elsworth},
  {New}, \& {Serenelli}}]{BasuSun}
{Basu}, S., {Chaplin}, W.~J., {Elsworth}, Y., {New}, R., \& {Serenelli}, A.~M.
  2009{\natexlab{a}}, ApJ, 699, 1403

\bibitem[{{Basu} {et~al.}(2009{\natexlab{b}}){Basu}, {Chaplin}, {Elsworth},
  {New}, \& {Serenelli}}]{Basu2009}
{Basu}, S., {Chaplin}, W.~J., {Elsworth}, Y., {New}, R., \& {Serenelli}, A.~M.
  2009{\natexlab{b}}, \apj, 699, 1403

\bibitem[{{Basu} \& {Christensen-Dalsgaard}(1997)}]{Basu97EOS}
{Basu}, S. \& {Christensen-Dalsgaard}, J. 1997, \aap, 322, L5

\bibitem[{{Basu} {et~al.}(2002{\natexlab{a}}){Basu}, {Christensen-Dalsgaard},
  \& {Thompson}}]{Basu2002}
{Basu}, S., {Christensen-Dalsgaard}, J., \& {Thompson}, M.~J.
  2002{\natexlab{a}}, in ESA Special Publication, Vol. 485, Stellar Structure
  and Habitable Planet Finding, ed. B.~{Battrick}, F.~{Favata}, I.~W.
  {Roxburgh}, \& D.~{Galadi}, 249--252

\bibitem[{{Basu} {et~al.}(2002{\natexlab{b}}){Basu}, {Christensen-Dalsgaard},
  \& {Thompson}}]{BasuJCD2002}
{Basu}, S., {Christensen-Dalsgaard}, J., \& {Thompson}, M.~J.
  2002{\natexlab{b}}, in ESA Special Publication, Vol. 485, Stellar Structure
  and Habitable Planet Finding, ed. B.~{Battrick}, F.~{Favata}, I.~W.
  {Roxburgh}, \& D.~{Galadi}, 249--252

\bibitem[{{Bazot}(2020)}]{Bazot2020}
{Bazot}, M. 2020, \aap, 635, A26

\bibitem[{{Bazot} {et~al.}(2019){Bazot}, {Benomar}, {Christensen-Dalsgaard},
  {Gizon}, {Hanasoge}, {Nielsen}, {Petit}, \& {Sreenivasan}}]{Bazot2019}
{Bazot}, M., {Benomar}, O., {Christensen-Dalsgaard}, J., {et~al.} 2019, \aap,
  623, A125

\bibitem[{{Bellinger} {et~al.}(2017){Bellinger}, {Basu}, {Hekker}, \&
  {Ball}}]{Bellinger2017}
{Bellinger}, E.~P., {Basu}, S., {Hekker}, S., \& {Ball}, W.~H. 2017, \apj, 851,
  80

\bibitem[{{Bellinger} {et~al.}(2021){Bellinger}, {Basu}, {Hekker},
  {Chrisensen-Dalsgaard}, \& {Ball}}]{Bellinger2021}
{Bellinger}, E.~P., {Basu}, S., {Hekker}, S., {Chrisensen-Dalsgaard}, J., \&
  {Ball}, W.~H. 2021, arXiv e-prints, arXiv:2105.04564

\bibitem[{{Bellinger} {et~al.}(2019){Bellinger}, {Basu}, {Hekker}, \&
  {Christensen-Dalsgaard}}]{Bellinger2019}
{Bellinger}, E.~P., {Basu}, S., {Hekker}, S., \& {Christensen-Dalsgaard}, J.
  2019, \apj, 885, 143

\bibitem[{{Borucki} {et~al.}(2010){Borucki}, {Koch}, {Basri}, {Batalha},
  {Brown}, {Caldwell}, {Caldwell}, {Christensen-Dalsgaard}, {Cochran},
  {DeVore}, {Dunham}, {Dupree}, {Gautier}, {Geary}, {Gilliland}, {Gould},
  {Howell}, {Jenkins}, {Kondo}, {Latham}, {Marcy}, {Meibom}, {Kjeldsen},
  {Lissauer}, {Monet}, {Morrison}, {Sasselov}, {Tarter}, {Boss}, {Brownlee},
  {Owen}, {Buzasi}, {Charbonneau}, {Doyle}, {Fortney}, {Ford}, {Holman},
  {Seager}, {Steffen}, {Welsh}, {Rowe}, {Anderson}, {Buchhave}, {Ciardi},
  {Walkowicz}, {Sherry}, {Horch}, {Isaacson}, {Everett}, {Fischer}, {Torres},
  {Johnson}, {Endl}, {MacQueen}, {Bryson}, {Dotson}, {Haas}, {Kolodziejczak},
  {Van Cleve}, {Chandrasekaran}, {Twicken}, {Quintana}, {Clarke}, {Allen},
  {Li}, {Wu}, {Tenenbaum}, {Verner}, {Bruhweiler}, {Barnes}, \&
  {Prsa}}]{Borucki}
{Borucki}, W.~J., {Koch}, D., {Basri}, G., {et~al.} 2010, Science, 327, 977

\bibitem[{{Buldgen} {et~al.}(2017{\natexlab{a}}){Buldgen}, {Reese}, \&
  {Dupret}}]{Buldgen2017Legacy}
{Buldgen}, G., {Reese}, D., \& {Dupret}, M.-A. 2017{\natexlab{a}}, in European
  Physical Journal Web of Conferences, Vol. 160, European Physical Journal Web
  of Conferences, 03005

\bibitem[{{Buldgen} {et~al.}(2015{\natexlab{a}}){Buldgen}, {Reese}, \&
  {Dupret}}]{Buldgen2015}
{Buldgen}, G., {Reese}, D.~R., \& {Dupret}, M.~A. 2015{\natexlab{a}}, \aap,
  583, A62

\bibitem[{{Buldgen} {et~al.}(2016{\natexlab{a}}){Buldgen}, {Reese}, \&
  {Dupret}}]{Buldgen2016}
{Buldgen}, G., {Reese}, D.~R., \& {Dupret}, M.~A. 2016{\natexlab{a}}, \aap,
  585, A109

\bibitem[{{Buldgen} {et~al.}(2017{\natexlab{b}}){Buldgen}, {Reese}, \&
  {Dupret}}]{BuldgenKer}
{Buldgen}, G., {Reese}, D.~R., \& {Dupret}, M.~A. 2017{\natexlab{b}}, \aap,
  598, A21

\bibitem[{{Buldgen} {et~al.}(2018){Buldgen}, {Reese}, \&
  {Dupret}}]{Buldgen2018}
{Buldgen}, G., {Reese}, D.~R., \& {Dupret}, M.~A. 2018, \aap, 609, A95

\bibitem[{{Buldgen} {et~al.}(2015{\natexlab{b}}){Buldgen}, {Reese}, {Dupret},
  \& {Samadi}}]{Buldgen2015tau}
{Buldgen}, G., {Reese}, D.~R., {Dupret}, M.~A., \& {Samadi}, R.
  2015{\natexlab{b}}, \aap, 574, A42

\bibitem[{{Buldgen} {et~al.}(2019{\natexlab{a}}){Buldgen}, {Rendle}, {Sonoi},
  {Davies}, {Miglio}, {Salmon}, {Reese}, {Bossini}, {Eggenberger}, {Noels}, \&
  {Scuflaire}}]{Buldgen2019}
{Buldgen}, G., {Rendle}, B., {Sonoi}, T., {et~al.} 2019{\natexlab{a}}, \mnras,
  482, 2305

\bibitem[{{Buldgen} {et~al.}(2019{\natexlab{b}}){Buldgen}, {Salmon}, \&
  {Noels}}]{BuldgenReview}
{Buldgen}, G., {Salmon}, S., \& {Noels}, A. 2019{\natexlab{b}}, Frontiers in
  Astronomy and Space Sciences, 6, 42

\bibitem[{{Buldgen} {et~al.}(2016{\natexlab{b}}){Buldgen}, {Salmon}, {Reese},
  \& {Dupret}}]{Buldgen2016b}
{Buldgen}, G., {Salmon}, S.~J.~A.~J., {Reese}, D.~R., \& {Dupret}, M.~A.
  2016{\natexlab{b}}, \aap, 596, A73

\bibitem[{{Chandrasekhar}(1964)}]{Chandrasekhar1964}
{Chandrasekhar}, S. 1964, \apj, 139, 664

\bibitem[{{Chaplin} {et~al.}(2015){Chaplin}, {Lund}, {Handberg}, {Basu},
  {Buchhave}, {Campante}, {Davies}, {Huber}, {Latham}, {Latham}, {Serenelli},
  {Antia}, {Appourchaux}, {Ball}, {Benomar}, {Casagrande},
  {Christensen-Dalsgaard}, {Coelho}, {Creevey}, {Elsworth}, {Garc{\'i}a},
  {Gaulme}, {Hekker}, {Kallinger}, {Karoff}, {Kawaler}, {Kjeldsen},
  {Lundkvist}, {Marcadon}, {Mathur}, {Miglio}, {Mosser}, {R{\'e}gulo},
  {Roxburgh}, {Silva Aguirre}, {Stello}, {Verma}, {White}, {Bedding},
  {Barclay}, {Buzasi}, {Dehuevels}, {Gizon}, {Houdek}, {Howell}, {Salabert}, \&
  {Soderblom}}]{ChaplinK2}
{Chaplin}, W.~J., {Lund}, M.~N., {Handberg}, R., {et~al.} 2015, Publications of
  the Astronomical Society of Pacific, 127, 1038

\bibitem[{{Charpinet} {et~al.}(2008){Charpinet}, {Van Grootel}, {Reese},
  {Fontaine}, {Green}, {Brassard}, \& {Chayer}}]{Charpinet}
{Charpinet}, S., {Van Grootel}, V., {Reese}, D., {et~al.} 2008, A\&Ap, 489, 377

\bibitem[{{Christensen-Dalsgaard}(2002)}]{JCD2002}
{Christensen-Dalsgaard}, J. 2002, Reviews of Modern Physics, 74, 1073

\bibitem[{{Christensen-Dalsgaard}(2021)}]{JCD2021}
{Christensen-Dalsgaard}, J. 2021, Living Reviews in Solar Physics, 18, 2

\bibitem[{{Christensen-Dalsgaard} \& {Daeppen}(1992)}]{CEFF}
{Christensen-Dalsgaard}, J. \& {Daeppen}, W. 1992, \aapr, 4, 267

\bibitem[{{Christensen-Dalsgaard} \& {Thompson}(1997)}]{JCD1997}
{Christensen-Dalsgaard}, J. \& {Thompson}, M.~J. 1997, \mnras, 284, 527

\bibitem[{{Clement}(1964)}]{Clement1964}
{Clement}, M.~J. 1964, \apj, 140, 1045

\bibitem[{{Colgan} {et~al.}(2016){Colgan}, {Kilcrease}, {Magee}, {Sherrill},
  {Abdallah}, {Hakel}, {Fontes}, {Guzik}, \& {Mussack}}]{Colgan}
{Colgan}, J., {Kilcrease}, D.~P., {Magee}, N.~H., {et~al.} 2016, ApJ, 817, 116

\bibitem[{{Davies} {et~al.}(2015){Davies}, {Chaplin}, {Farr}, {Garc{\'\i}a},
  {Lund}, {Mathis}, {Metcalfe}, {Appourchaux}, {Basu}, {Benomar}, {Campante},
  {Ceillier}, {Elsworth}, {Handberg}, {Salabert}, \& {Stello}}]{Davies2015}
{Davies}, G.~R., {Chaplin}, W.~J., {Farr}, W.~M., {et~al.} 2015, \mnras, 446,
  2959

\bibitem[{{Deal} {et~al.}(2015){Deal}, {Richard}, \& {Vauclair}}]{Deal2015}
{Deal}, M., {Richard}, O., \& {Vauclair}, S. 2015, \aap, 584, A105

\bibitem[{{Deliyannis} {et~al.}(2000){Deliyannis}, {Cunha}, {King}, \&
  {Boesgaard}}]{Deliyannis}
{Deliyannis}, C.~P., {Cunha}, K., {King}, J.~R., \& {Boesgaard}, A.~M. 2000,
  Astr. J., 119, 2437

\bibitem[{{di Mauro}(2004)}]{DiMauro2004}
{di Mauro}, M.~P. 2004, in ESA Special Publication, Vol. 559, SOHO 14 Helio-
  and Asteroseismology: Towards a Golden Future, ed. D.~{Danesy}, 186

\bibitem[{{Dziembowski} {et~al.}(1990){Dziembowski}, {Pamyatnykh}, \&
  {Sienkiewicz}}]{Dziemboswki90}
{Dziembowski}, W.~A., {Pamyatnykh}, A.~A., \& {Sienkiewicz}, R. 1990, \mnras,
  244, 542

\bibitem[{{Eddington}(1959)}]{Eddington}
{Eddington}, A.~S. 1959, {The internal constitution of the stars}

\bibitem[{{Elliott}(1996)}]{Elliott1996}
{Elliott}, J.~R. 1996, \mnras, 280, 1244

\bibitem[{{Farnir} {et~al.}(2020){Farnir}, {Dupret}, {Buldgen}, {Salmon},
  {Noels}, {Pin{\c{c}}on}, {Pezzotti}, \& {Eggenberger}}]{Farnir2020}
{Farnir}, M., {Dupret}, M.~A., {Buldgen}, G., {et~al.} 2020, \aap, 644, A37

\bibitem[{{Farnir} {et~al.}(2019){Farnir}, {Dupret}, {Salmon}, {Noels}, \&
  {Buldgen}}]{Farnir2019}
{Farnir}, M., {Dupret}, M.~A., {Salmon}, S.~J.~A.~J., {Noels}, A., \&
  {Buldgen}, G. 2019, \aap, 622, A98

\bibitem[{{Giammichele} {et~al.}(2018){Giammichele}, {Charpinet}, {Fontaine},
  {Brassard}, {Green}, {Van Grootel}, {Bergeron}, {Zong}, \&
  {Dupret}}]{Giammichele2018}
{Giammichele}, N., {Charpinet}, S., {Fontaine}, G., {et~al.} 2018, \nat, 554,
  73

\bibitem[{{Gough} \& {Kosovichev}(1988)}]{Gough1988}
{Gough}, D.~O. \& {Kosovichev}, A.~G. 1988, in ESA Special Publication, Vol.
  286, Seismology of the Sun and Sun-Like Stars, ed. E.~J. {Rolfe}, 195--201

\bibitem[{{Gough} \& {Kosovichev}(1993{\natexlab{a}})}]{Gough932}
{Gough}, D.~O. \& {Kosovichev}, A.~G. 1993{\natexlab{a}}, in Astronomical
  Society of the Pacific Conference Series, Vol.~40, IAU Colloq. 137: Inside
  the Stars, ed. W.~W. {Weiss} \& A.~{Baglin}, 541

\bibitem[{{Gough} \& {Kosovichev}(1993{\natexlab{b}})}]{Gough1993}
{Gough}, D.~O. \& {Kosovichev}, A.~G. 1993{\natexlab{b}}, \mnras, 264, 522

\bibitem[{{Grevesse} \& {Noels}(1993)}]{GrevNoels}
{Grevesse}, N. \& {Noels}, A. 1993, in Origin and Evolution of the Elements,
  ed. N.~{Prantzos}, E.~{Vangioni-Flam}, \& M.~{Casse}, 15--25

\bibitem[{{Gruberbauer} {et~al.}(2013){Gruberbauer}, {Guenther}, {MacLeod}, \&
  {Kallinger}}]{Gruberbauer2013}
{Gruberbauer}, M., {Guenther}, D.~B., {MacLeod}, K., \& {Kallinger}, T. 2013,
  \mnras, 435, 242

\bibitem[{{Iglesias} \& {Rogers}(1996)}]{OPAL}
{Iglesias}, C.~A. \& {Rogers}, F.~J. 1996, ApJ, 464, 943

\bibitem[{{Irwin}(2012)}]{Irwin}
{Irwin}, A.~W. 2012, {FreeEOS: Equation of State for stellar interiors
  calculations}, Astrophysics Source Code Library

\bibitem[{{King} {et~al.}(1997){King}, {Deliyannis}, {Hiltgen}, {Stephens},
  {Cunha}, \& {Boesgaard}}]{King}
{King}, J.~R., {Deliyannis}, C.~P., {Hiltgen}, D.~D., {et~al.} 1997, Astr. J.,
  113, 1871

\bibitem[{{Kosovichev}(1999)}]{Kosovichev}
{Kosovichev}, A.~G. 1999, Journal of Computational and Applied Mathematics,
  109, 1

\bibitem[{{Kosovichev}(2011)}]{Kosovichev2011}
{Kosovichev}, A.~G. 2011, {Advances in Global and Local Helioseismology: An
  Introductory Review}, in Lecture Notes in Physics, Berlin Springer Verlag,
  ed. J.-P. {Rozelot} \& C.~{Neiner}, Vol. 832, 3

\bibitem[{{Kosovichev} \& {Kitiashvili}(2020)}]{Kosovichev2020}
{Kosovichev}, A.~G. \& {Kitiashvili}, I.~N. 2020, in Solar and Stellar Magnetic
  Fields: Origins and Manifestations, ed. A.~{Kosovichev}, S.~{Strassmeier}, \&
  M.~{Jardine}, Vol. 354, 107--115

\bibitem[{{Le Pennec} {et~al.}(2015){Le Pennec}, {Turck-Chi{\`e}ze}, {Salmon},
  {Blancard}, {Coss{\'e}}, {Faussurier}, \& {Mondet}}]{LePennec}
{Le Pennec}, M., {Turck-Chi{\`e}ze}, S., {Salmon}, S., {et~al.} 2015, \apjl,
  813, L42

\bibitem[{{Lynden-Bell} \& {Ostriker}(1967)}]{LyndenBell1967}
{Lynden-Bell}, D. \& {Ostriker}, J.~P. 1967, \mnras, 136, 293

\bibitem[{{Mathur} {et~al.}(2012){Mathur}, {Metcalfe}, {Woitaszek}, {Bruntt},
  {Verner}, {Christensen-Dalsgaard}, {Creevey}, {Do{\v{g}}an}, {Basu},
  {Karoff}, {Stello}, {Appourchaux}, {Campante}, {Chaplin}, {Garc{\'\i}a},
  {Bedding}, {Benomar}, {Bonanno}, {Deheuvels}, {Elsworth}, {Gaulme}, {Guzik},
  {Handberg}, {Hekker}, {Herzberg}, {Monteiro}, {Piau}, {Quirion},
  {R{\'e}gulo}, {Roth}, {Salabert}, {Serenelli}, {Thompson}, {Trampedach},
  {White}, {Ballot}, {Brand{\~a}o}, {Molenda-{\.Z}akowicz}, {Kjeldsen},
  {Twicken}, {Uddin}, \& {Wohler}}]{Mathur2012}
{Mathur}, S., {Metcalfe}, T.~S., {Woitaszek}, M., {et~al.} 2012, \apj, 749, 152

\bibitem[{{Metcalfe} {et~al.}(2015){Metcalfe}, {Creevey}, \&
  {Davies}}]{Metcalfe2015}
{Metcalfe}, T.~S., {Creevey}, O.~L., \& {Davies}, G.~R. 2015, \apjl, 811, L37

\bibitem[{{Mondet} {et~al.}(2015){Mondet}, {Blancard}, {Coss{\'e}}, \&
  {Faussurier}}]{Mondet}
{Mondet}, G., {Blancard}, C., {Coss{\'e}}, P., \& {Faussurier}, G. 2015, ApJs,
  220, 2

\bibitem[{{Morel} {et~al.}(2021){Morel}, {Creevey}, {Montalb{\'a}n}, {Miglio},
  \& {Willett}}]{Morel2021}
{Morel}, T., {Creevey}, O.~L., {Montalb{\'a}n}, J., {Miglio}, A., \& {Willett},
  E. 2021, \aap, 646, A78

\bibitem[{{Ot{\'\i} Floranes} {et~al.}(2005){Ot{\'\i} Floranes},
  {Christensen-Dalsgaard}, \& {Thompson}}]{Oti2005}
{Ot{\'\i} Floranes}, H., {Christensen-Dalsgaard}, J., \& {Thompson}, M.~J.
  2005, \mnras, 356, 671

\bibitem[{{Paquette} {et~al.}(1986){Paquette}, {Pelletier}, {Fontaine}, \&
  {Michaud}}]{Paquette}
{Paquette}, C., {Pelletier}, C., {Fontaine}, G., \& {Michaud}, G. 1986, \apjs,
  61, 177

\bibitem[{{Pijpers} \& {Thompson}(1994)}]{Pijpers}
{Pijpers}, F.~P. \& {Thompson}, M.~J. 1994, \aap, 281, 231

\bibitem[{{Proffitt} \& {Michaud}(1991)}]{Proffitt1991}
{Proffitt}, C.~R. \& {Michaud}, G. 1991, \apj, 380, 238

\bibitem[{{Rabello-Soares} {et~al.}(1999){Rabello-Soares}, {Basu}, \&
  {Christensen-Dalsgaard}}]{RabelloParam}
{Rabello-Soares}, M.~C., {Basu}, S., \& {Christensen-Dalsgaard}, J. 1999,
  MNRAS, 309, 35

\bibitem[{{Ram{\'\i}rez} {et~al.}(2009){Ram{\'\i}rez}, {Mel{\'e}ndez}, \&
  {Asplund}}]{Ramirez2009}
{Ram{\'\i}rez}, I., {Mel{\'e}ndez}, J., \& {Asplund}, M. 2009, \aap, 508, L17

\bibitem[{{Rauer} {et~al.}(2014){Rauer}, {Catala}, {Aerts}, {Appourchaux},
  {Benz}, {Brandeker}, {Christensen-Dalsgaard}, {Deleuil}, {Gizon}, {Goupil},
  {G{\"u}del}, {Janot-Pacheco}, {Mas-Hesse}, {Pagano}, {Piotto}, {Pollacco},
  {Santos}, {Smith}, {Su{\'a}rez}, {Szab{\'o}}, {Udry}, {Adibekyan}, {Alibert},
  {Almenara}, {Amaro-Seoane}, {Eiff}, {Asplund}, {Antonello}, {Barnes},
  {Baudin}, {Belkacem}, {Bergemann}, {Bihain}, {Birch}, {Bonfils}, {Boisse},
  {Bonomo}, {Borsa}, {Brand{\~a}o}, {Brocato}, {Brun}, {Burleigh}, {Burston},
  {Cabrera}, {Cassisi}, {Chaplin}, {Charpinet}, {Chiappini}, {Church},
  {Csizmadia}, {Cunha}, {Damasso}, {Davies}, {Deeg}, {D{\'\i}az}, {Dreizler},
  {Dreyer}, {Eggenberger}, {Ehrenreich}, {Eigm{\"u}ller}, {Erikson}, {Farmer},
  {Feltzing}, {de Oliveira Fialho}, {Figueira}, {Forveille}, {Fridlund},
  {Garc{\'\i}a}, {Giommi}, {Giuffrida}, {Godolt}, {Gomes da Silva}, {Granzer},
  {Grenfell}, {Grotsch-Noels}, {G{\"u}nther}, {Haswell}, {Hatzes},
  {H{\'e}brard}, {Hekker}, {Helled}, {Heng}, {Jenkins}, {Johansen},
  {Khodachenko}, {Kislyakova}, {Kley}, {Kolb}, {Krivova}, {Kupka}, {Lammer},
  {Lanza}, {Lebreton}, {Magrin}, {Marcos-Arenal}, {Marrese}, {Marques},
  {Martins}, {Mathis}, {Mathur}, {Messina}, {Miglio}, {Montalban}, {Montalto},
  {Monteiro}, {Moradi}, {Moravveji}, {Mordasini}, {Morel}, {Mortier},
  {Nascimbeni}, {Nelson}, {Nielsen}, {Noack}, {Norton}, {Ofir}, {Oshagh},
  {Ouazzani}, {P{\'a}pics}, {Parro}, {Petit}, {Plez}, {Poretti}, {Quirrenbach},
  {Ragazzoni}, {Raimondo}, {Rainer}, {Reese}, {Redmer}, {Reffert},
  {Rojas-Ayala}, {Roxburgh}, {Salmon}, {Santerne}, {Schneider}, {Schou},
  {Schuh}, {Schunker}, {Silva-Valio}, {Silvotti}, {Skillen}, {Snellen}, {Sohl},
  {Sousa}, {Sozzetti}, {Stello}, {Strassmeier}, {{\v{S}}vanda}, {Szab{\'o}},
  {Tkachenko}, {Valencia}, {Van Grootel}, {Vauclair}, {Ventura}, {Wagner},
  {Walton}, {Weingrill}, {Werner}, {Wheatley}, \& {Zwintz}}]{Rauer2014}
{Rauer}, H., {Catala}, C., {Aerts}, C., {et~al.} 2014, Experimental Astronomy,
  38, 249

\bibitem[{{Reese} \& {Zharkov}(2016)}]{InversionKit}
{Reese}, D. \& {Zharkov}, S. 2016, {InversionKit: Linear inversions from
  frequency data}

\bibitem[{{Reese} {et~al.}(2012){Reese}, {Marques}, {Goupil}, {Thompson}, \&
  {Deheuvels}}]{Reese2012}
{Reese}, D.~R., {Marques}, J.~P., {Goupil}, M.~J., {Thompson}, M.~J., \&
  {Deheuvels}, S. 2012, \aap, 539, A63

\bibitem[{{Ricker} {et~al.}(2015){Ricker}, {Winn}, {Vanderspek}, {Latham},
  {Bakos}, {Bean}, {Berta-Thompson}, {Brown}, {Buchhave}, {Butler}, {Butler},
  {Chaplin}, {Charbonneau}, {Christensen-Dalsgaard}, {Clampin}, {Deming},
  {Doty}, {De Lee}, {Dressing}, {Dunham}, {Endl}, {Fressin}, {Ge}, {Henning},
  {Holman}, {Howard}, {Ida}, {Jenkins}, {Jernigan}, {Johnson}, {Kaltenegger},
  {Kawai}, {Kjeldsen}, {Laughlin}, {Levine}, {Lin}, {Lissauer}, {MacQueen},
  {Marcy}, {McCullough}, {Morton}, {Narita}, {Paegert}, {Palle}, {Pepe},
  {Pepper}, {Quirrenbach}, {Rinehart}, {Sasselov}, {Sato}, {Seager},
  {Sozzetti}, {Stassun}, {Sullivan}, {Szentgyorgyi}, {Torres}, {Udry}, \&
  {Villasenor}}]{Ricker2015}
{Ricker}, G.~R., {Winn}, J.~N., {Vanderspek}, R., {et~al.} 2015, Journal of
  Astronomical Telescopes, Instruments, and Systems, 1, 014003

\bibitem[{{Rogers} \& {Nayfonov}(2002)}]{Opal2002}
{Rogers}, F.~J. \& {Nayfonov}, A. 2002, \apj, 576, 1064

\bibitem[{{Roxburgh}(2002)}]{RoxTools2002}
{Roxburgh}, I.~W. 2002, in ESA Special Publication, Vol. 485, Stellar Structure
  and Habitable Planet Finding, ed. B.~{Battrick}, F.~{Favata}, I.~W.
  {Roxburgh}, \& D.~{Galadi}, 75--85

\bibitem[{{Roxburgh}(2010)}]{Roxburgh2010}
{Roxburgh}, I.~W. 2010, \apss, 328, 3

\bibitem[{{Roxburgh}(2015)}]{RoxburgPerky}
{Roxburgh}, I.~W. 2015, \aap, 574, A45

\bibitem[{{Roxburgh}(2016)}]{RoxEps}
{Roxburgh}, I.~W. 2016, A\&Ap, 585, A63

\bibitem[{{Roxburgh} {et~al.}(1998){Roxburgh}, {Audard}, {Basu},
  {Christensen-Dalsgaard}, \& {Vorontsov}}]{Roxburgh98}
{Roxburgh}, I.~W., {Audard}, N., {Basu}, S., {Christensen-Dalsgaard}, J., \&
  {Vorontsov}, S.~V. 1998, in Proceedings of the IAU Symposium 181: Sounding
  Solar and Stellar Interiors, ed. J.~{Provost} \& F.~X. {Schmider}, Vol. 181,
  245--246

\bibitem[{{Roxburgh} \& {Vorontsov}(2002{\natexlab{a}})}]{Rox2002a}
{Roxburgh}, I.~W. \& {Vorontsov}, S.~V. 2002{\natexlab{a}}, in ESA Special
  Publication, Vol. 485, Stellar Structure and Habitable Planet Finding, ed.
  B.~{Battrick}, F.~{Favata}, I.~W. {Roxburgh}, \& D.~{Galadi}, 337--339

\bibitem[{{Roxburgh} \& {Vorontsov}(2002{\natexlab{b}})}]{Rox2002b}
{Roxburgh}, I.~W. \& {Vorontsov}, S.~V. 2002{\natexlab{b}}, in ESA Special
  Publication, Vol. 485, Stellar Structure and Habitable Planet Finding, ed.
  B.~{Battrick}, F.~{Favata}, I.~W. {Roxburgh}, \& D.~{Galadi}, 341--343

\bibitem[{{Scuflaire} {et~al.}(2008{\natexlab{a}}){Scuflaire}, {Montalb{\'a}n},
  {Th{\'e}ado}, {Bourge}, {Miglio}, {Godart}, {Thoul}, \&
  {Noels}}]{ScuflaireOsc}
{Scuflaire}, R., {Montalb{\'a}n}, J., {Th{\'e}ado}, S., {et~al.}
  2008{\natexlab{a}}, ApSS, 316, 149

\bibitem[{{Scuflaire} {et~al.}(2008{\natexlab{b}}){Scuflaire}, {Th{\'e}ado},
  {Montalb{\'a}n}, {Miglio}, {Bourge}, {Godart}, {Thoul}, \&
  {Noels}}]{ScuflaireCles}
{Scuflaire}, R., {Th{\'e}ado}, S., {Montalb{\'a}n}, J., {et~al.}
  2008{\natexlab{b}}, ApSS, 316, 83

\bibitem[{{Sonoi} {et~al.}(2015){Sonoi}, {Samadi}, {Belkacem}, {Ludwig},
  {Caffau}, \& {Mosser}}]{Sonoi2015}
{Sonoi}, T., {Samadi}, R., {Belkacem}, K., {et~al.} 2015, \aap, 583, A112

\bibitem[{{Takata} \& {Gough}(2001)}]{Takata2001}
{Takata}, M. \& {Gough}, D.~O. 2001, in ESA Special Publication, Vol. 464, SOHO
  10/GONG 2000 Workshop: Helio- and Asteroseismology at the Dawn of the
  Millennium, ed. A.~{Wilson} \& P.~L. {Pall{\'e}}, 543--546

\bibitem[{{Takata} \& {Montgomery}(2002)}]{Takata2002}
{Takata}, M. \& {Montgomery}, M.~H. 2002, in Astronomical Society of the
  Pacific Conference Series, Vol. 259, IAU Colloq. 185: Radial and Nonradial
  Pulsationsn as Probes of Stellar Physics, ed. C.~{Aerts}, T.~R. {Bedding}, \&
  J.~{Christensen-Dalsgaard}, 606

\bibitem[{{Thompson} \& {Christensen-Dalsgaard}(2002)}]{Thompson2002}
{Thompson}, M.~J. \& {Christensen-Dalsgaard}, J. 2002, in ESA Special
  Publication, Vol. 485, Stellar Structure and Habitable Planet Finding, ed.
  B.~{Battrick}, F.~{Favata}, I.~W. {Roxburgh}, \& D.~{Galadi}, 95--101

\bibitem[{{Thoul} {et~al.}(1994){Thoul}, {Bahcall}, \& {Loeb}}]{Thoul}
{Thoul}, A.~A., {Bahcall}, J.~N., \& {Loeb}, A. 1994, \apj, 421, 828

\bibitem[{{Tucci Maia} {et~al.}(2014){Tucci Maia}, {Mel{\'e}ndez}, \&
  {Ram{\'\i}rez}}]{Tucci2014}
{Tucci Maia}, M., {Mel{\'e}ndez}, J., \& {Ram{\'\i}rez}, I. 2014, \apjl, 790,
  L25

\bibitem[{{Verma} {et~al.}(2014){Verma}, {Faria}, {Antia}, {Basu}, {Mazumdar},
  {Monteiro}, {Appourchaux}, {Chaplin}, {Garc{\'\i}a}, \&
  {Metcalfe}}]{Verma2014}
{Verma}, K., {Faria}, J.~P., {Antia}, H.~M., {et~al.} 2014, \apj, 790, 138

\bibitem[{{Vernazza} {et~al.}(1981){Vernazza}, {Avrett}, \&
  {Loeser}}]{Vernazza}
{Vernazza}, J.~E., {Avrett}, E.~H., \& {Loeser}, R. 1981, \apjs, 45, 635

\bibitem[{{Viallet} {et~al.}(2015){Viallet}, {Meakin}, {Prat}, \&
  {Arnett}}]{Viallet2015}
{Viallet}, M., {Meakin}, C., {Prat}, V., \& {Arnett}, D. 2015, \aap, 580, A61

\bibitem[{{White} {et~al.}(2013){White}, {Huber}, {Maestro}, {Bedding},
  {Ireland}, {Baron}, {Boyajian}, {Che}, {Monnier}, {Pope}, {Roettenbacher},
  {Stello}, {Tuthill}, {Farrington}, {Goldfinger}, {McAlister}, {Schaefer},
  {Sturmann}, {Sturmann}, {ten Brummelaar}, \& {Turner}}]{White2013}
{White}, T.~R., {Huber}, D., {Maestro}, V., {et~al.} 2013, \mnras, 433, 1262

\end{thebibliography}

\appendix
\section{Inversion robustness tests}\label{Sec:Checkups}

Because the linear approximation used to derive Eq. \ref{eq:Variational} has intrinsic limitations, the robustness and stability of the inversion must be tested thoroughly before it is applied to the actual observed data. For this study, we chose to carry out verifications for each inversion and for each star. We detail our approach in this section. 

The reference and target models for the inversion were chosen from evolutionary models of Paper I. We considered that if the inversion technique is able to distinguish between various evolutionary models and the actual stars, it should be able to distinguish between evolutionary models built with various physical ingredients. Thus we chose the models by varying their reference solar abundances, opacity tables, including or excluding additional turbulence at the base of their convective envelope, as well as the inclusion or exclusion of non-seismic constraints in their modelling with the \who technique. The dataset for the inversion was chosen to be as similar as possible to that of the real targets, so that the trade-off problem for the verification is the same as for the inversion of the actual data. In other words, we used the exact same oscillation modes (in terms of $n$ and $\ell$) as were detected in 16CygA$\&$B and assigned to each mode their actual observed uncertainties. We also provide the actual error contributions in Appendix. \ref{SecSupplTables} for each indicator inversion verification presented below. 

\subsection{Mean density}\label{Sec:RhoCheckups}

\begin{figure*}
\begin{flushleft}
	\begin{minipage}{\textwidth}
  	\includegraphics[trim=5 5 5 5, clip, width=0.47\linewidth]{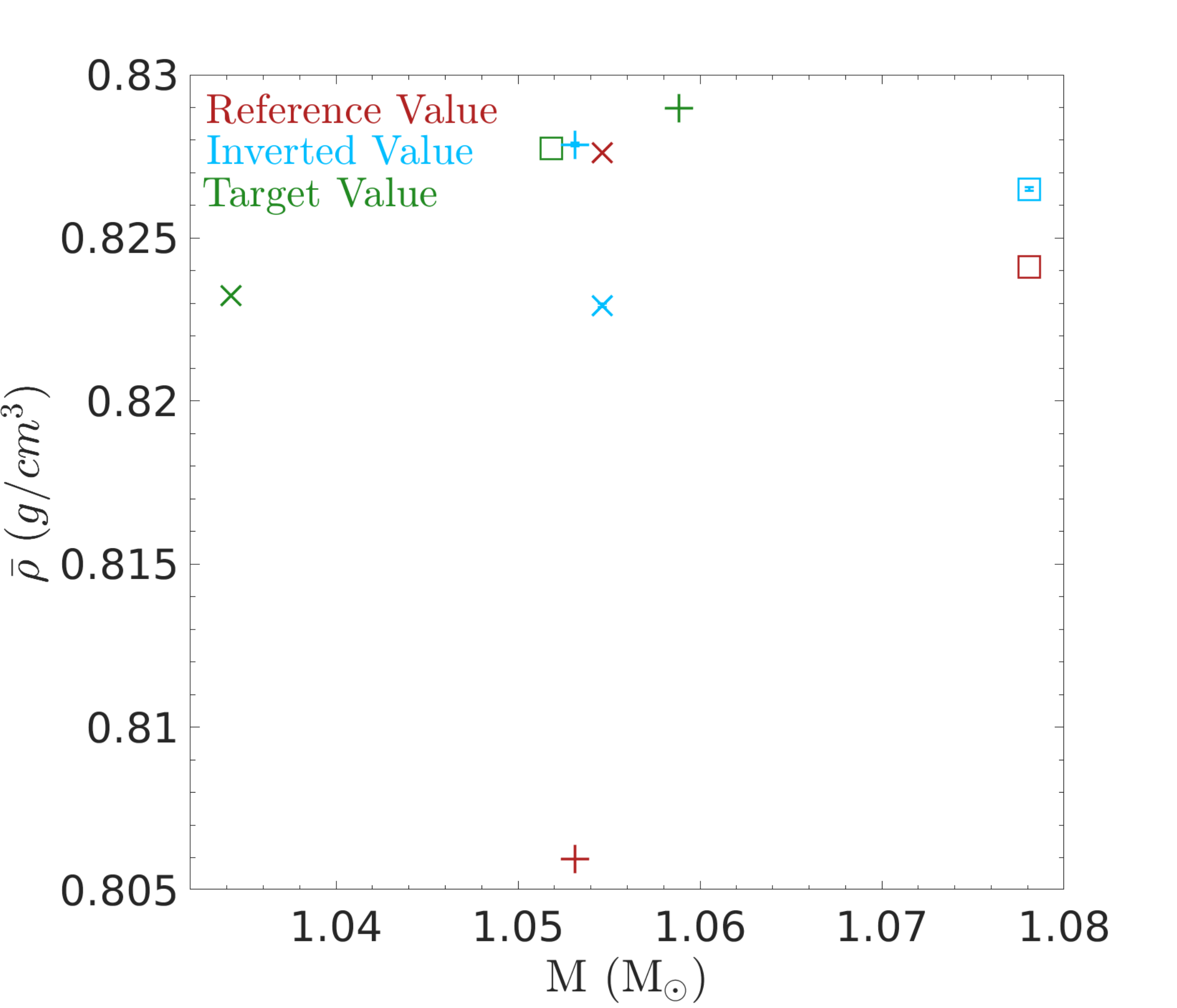}
 	\includegraphics[trim=5 5 5 5, clip, width=0.47\linewidth]{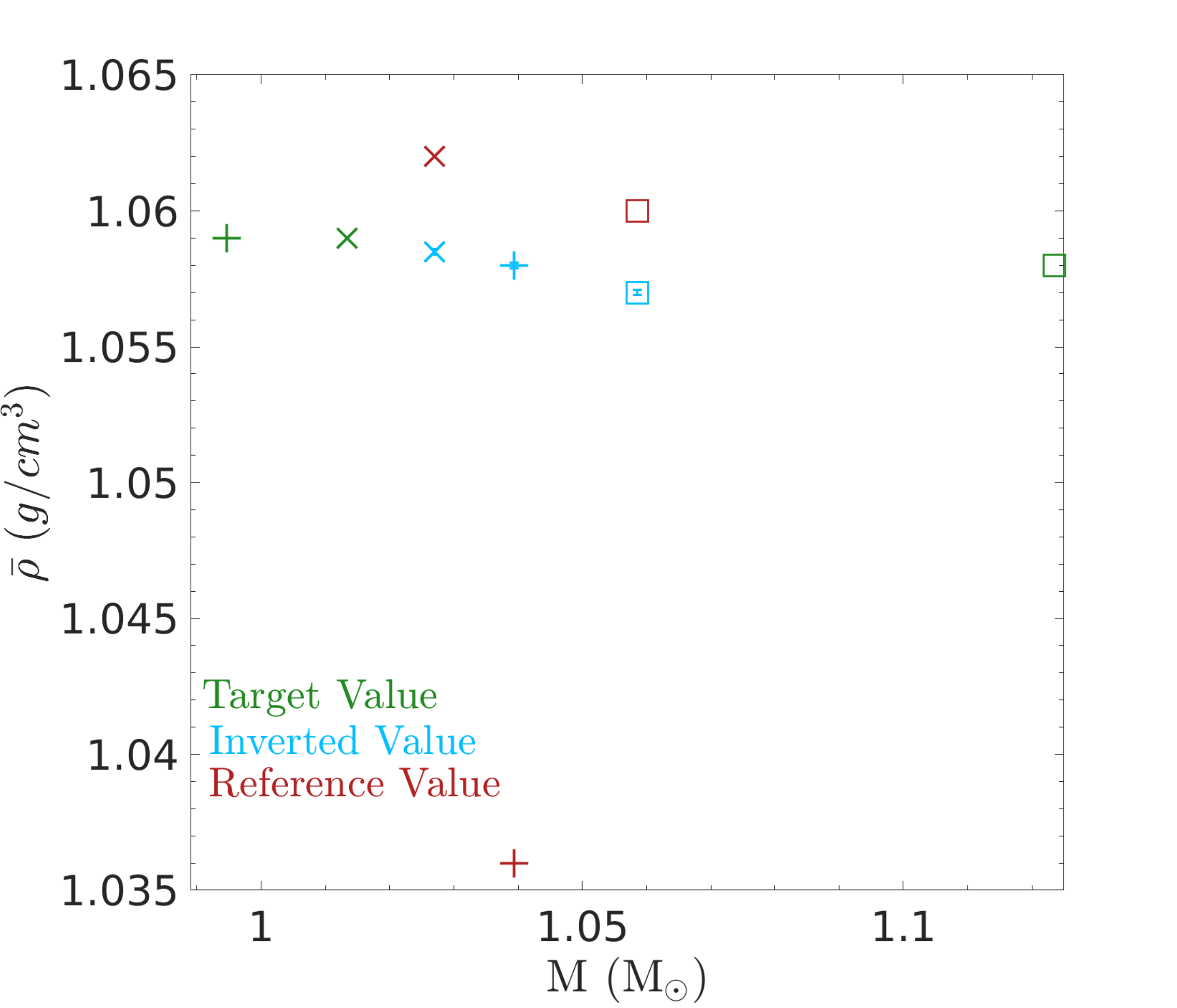}
 \end{minipage}	
 \end{flushleft}
	\caption{Inversion results for the mean density, as a function of mass, using various stellar evolutionary models of 16CygA (left panel) and 16CygB (right panel) as artificial targets and the exact same dataset as the observed dataset in both cases. The results include error bars but they are almost invisible, and lead to an overestimated precision, especially in the case without surface corrections that we used here for the verification step.}
		\label{FigRhoCheckCygAB}
\end{figure*}

\begin{figure*}
\begin{center}
  	\includegraphics[trim=5 5 5 5, clip, width=0.80\linewidth]{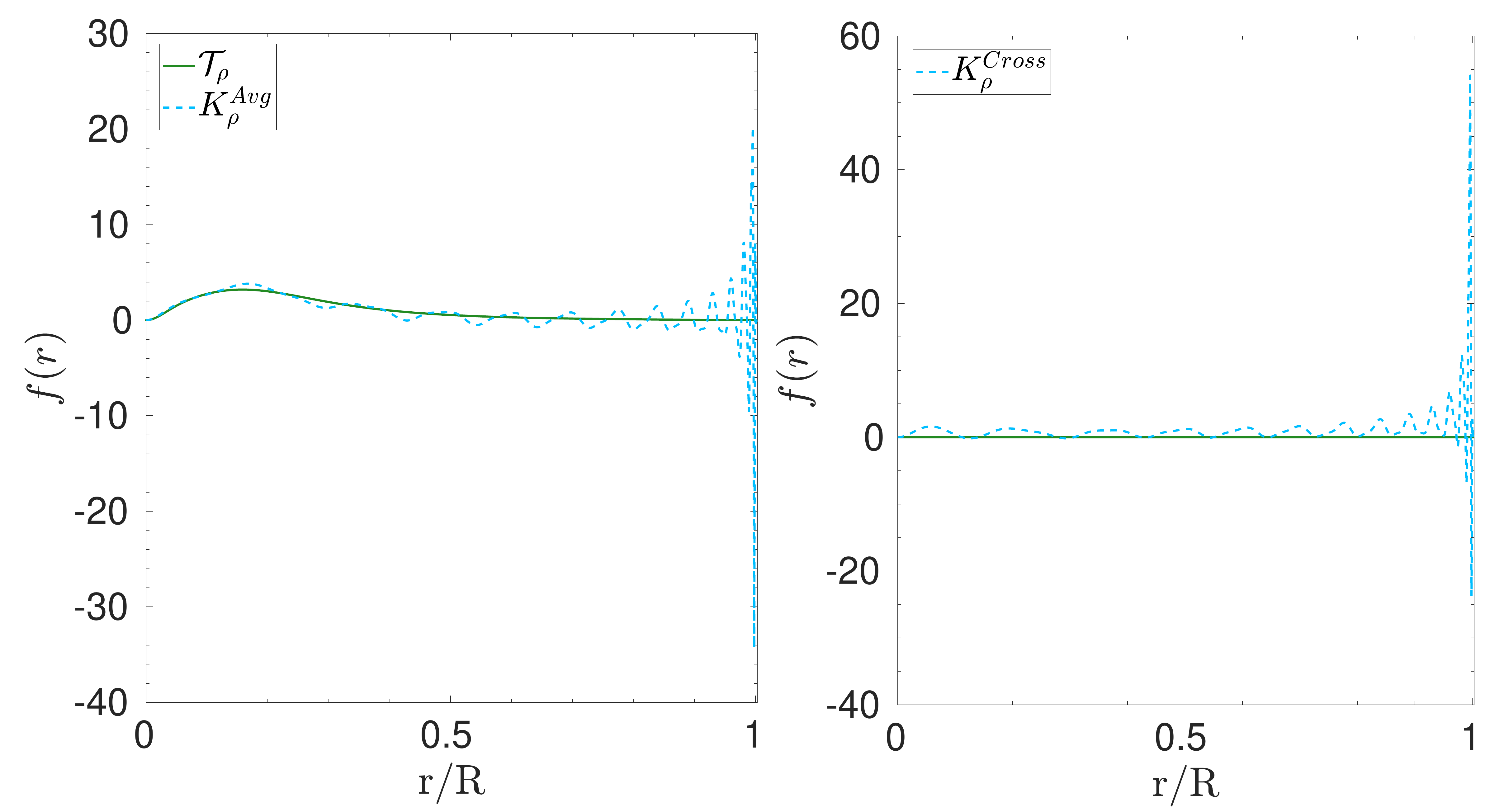}
 \end{center}
	\caption{Averaging kernel for the mean density inversion of 16CygA (left, blue) and its associated target function (green). Cross-term kernel for the mean density inversion of 16CygA (right, blue) and its associated target function (green).}
		\label{FigKerRhoCheckA}
\end{figure*}

\begin{figure*}
\begin{center}
  	\includegraphics[trim=5 5 5 5, clip, width=0.80\linewidth]{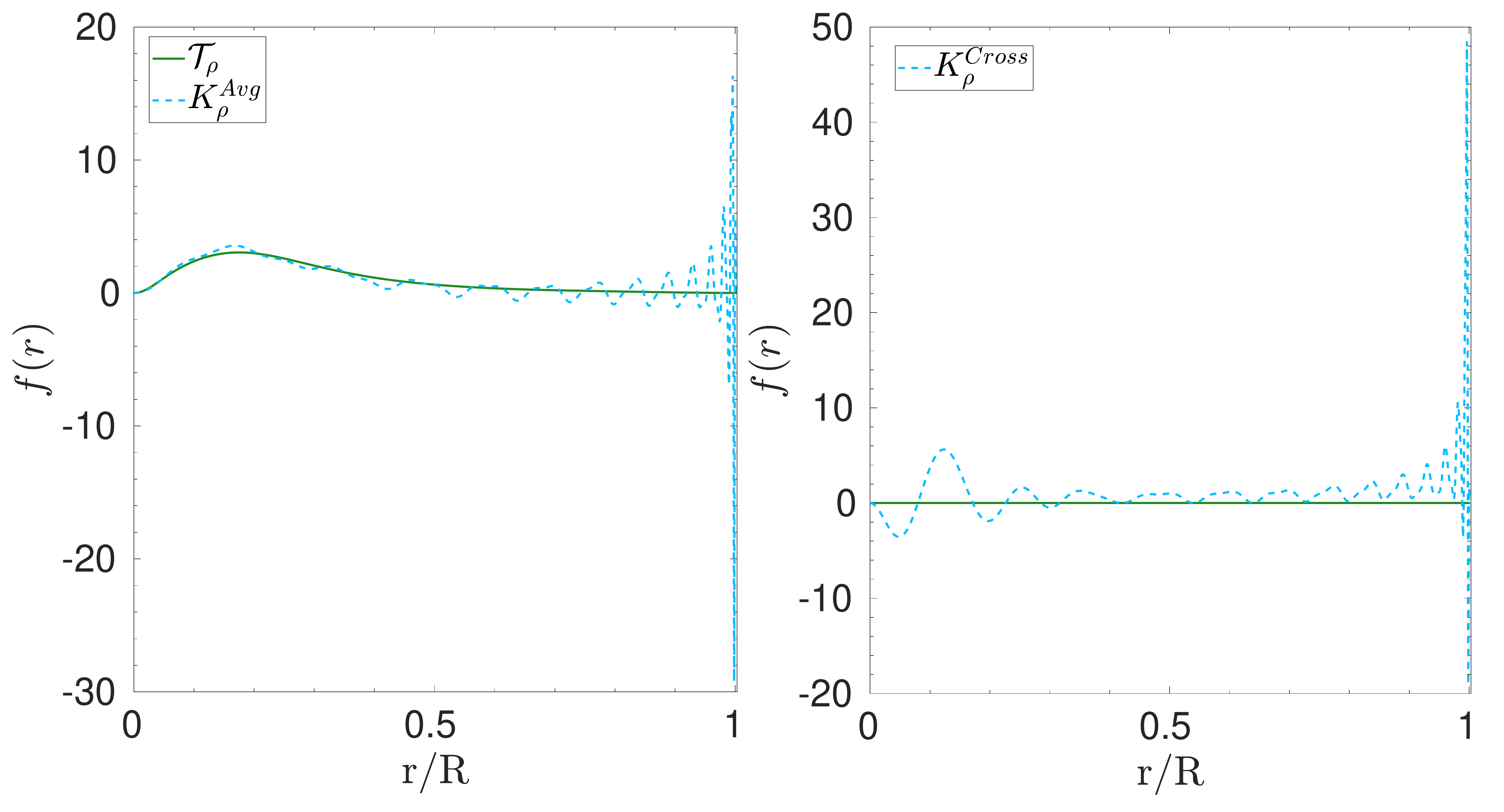}
 \end{center}
	\caption{Averaging kernel for the mean density inversion of 16CygB (left, blue) and its associated target function (green). Cross-term kernel for the mean density inversion of 16CygB (right, blue) and its associated target function (green).}
		\label{FigKerRhoCheckB}
\end{figure*}

The results of the mean density verifications are illustrated in Fig. \ref{FigRhoCheckCygAB}, and the averaging and cross-term kernels associated with some of these test cases are given in Figs. \ref{FigKerRhoCheckA} and \ref{FigKerRhoCheckB} together with the errors of the inversions in Tables \ref{tab:CheckupErrorsA} and \ref{tab:CheckupErrorsB}.

The mean density inversion is robust and quite accurate. However, as noted in \citet{Reese2012} and \citet{Buldgen2015tau}, it overestimates its actual precision. The derivation of the uncertainties of inverted results in the SOLA technique is not always optimal. Thus, for mean density inversions, the actual precision of the method must consider some dispersion coming from model dependence and from the imperfect reproduction of the target function of the inversion. This emphasizes the need to carry out the inversion from multiple reference models using a wide variety of physical ingredients. 

\subsection{Core condition indicators}\label{Sec:CoreCheckups}

\begin{figure*}
\begin{flushleft}
	\begin{minipage}{\textwidth}
  	\includegraphics[trim=5 5 5 5, clip, width=0.47\linewidth]{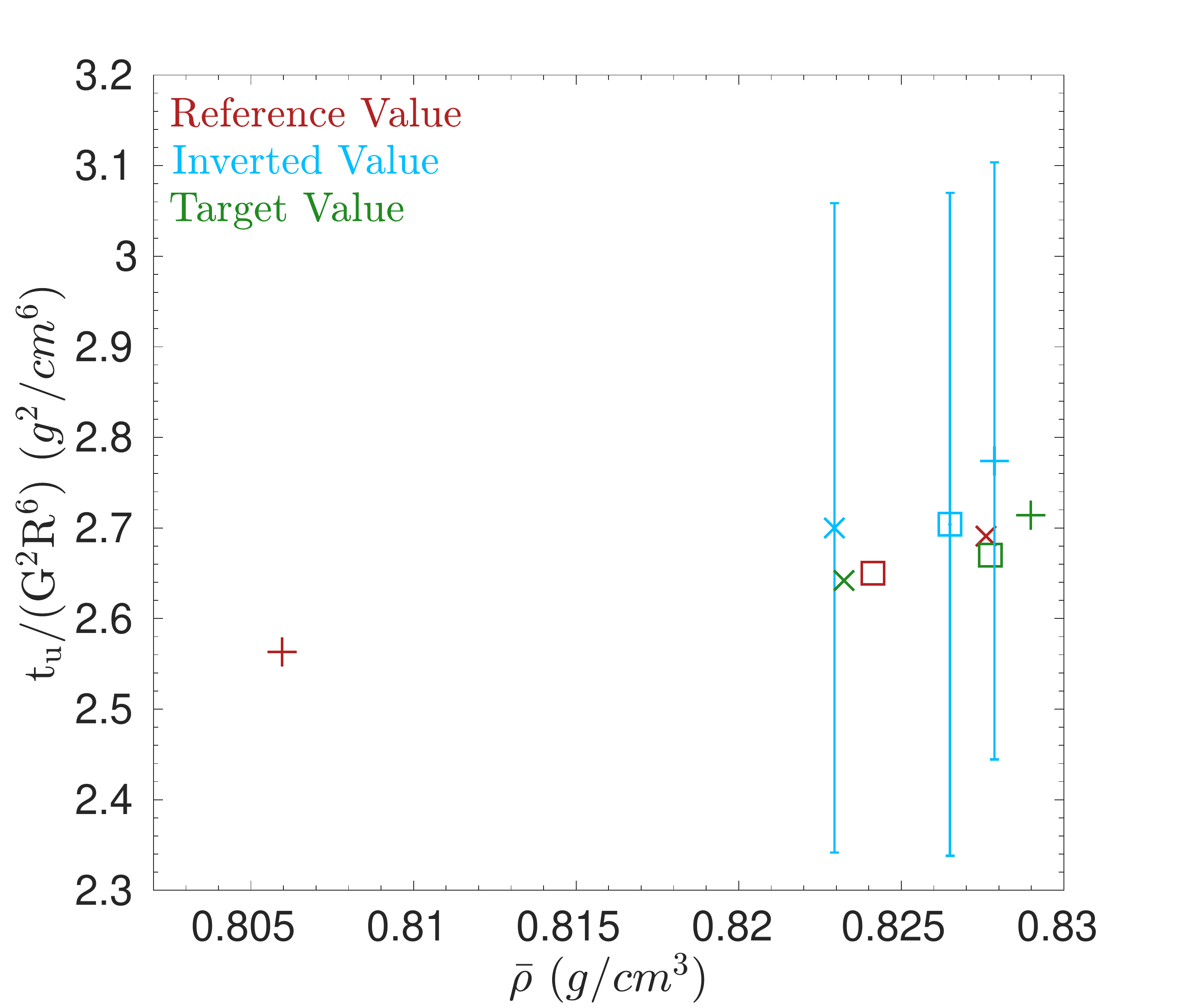}
 	\includegraphics[trim=5 5 5 5, clip, width=0.47\linewidth]{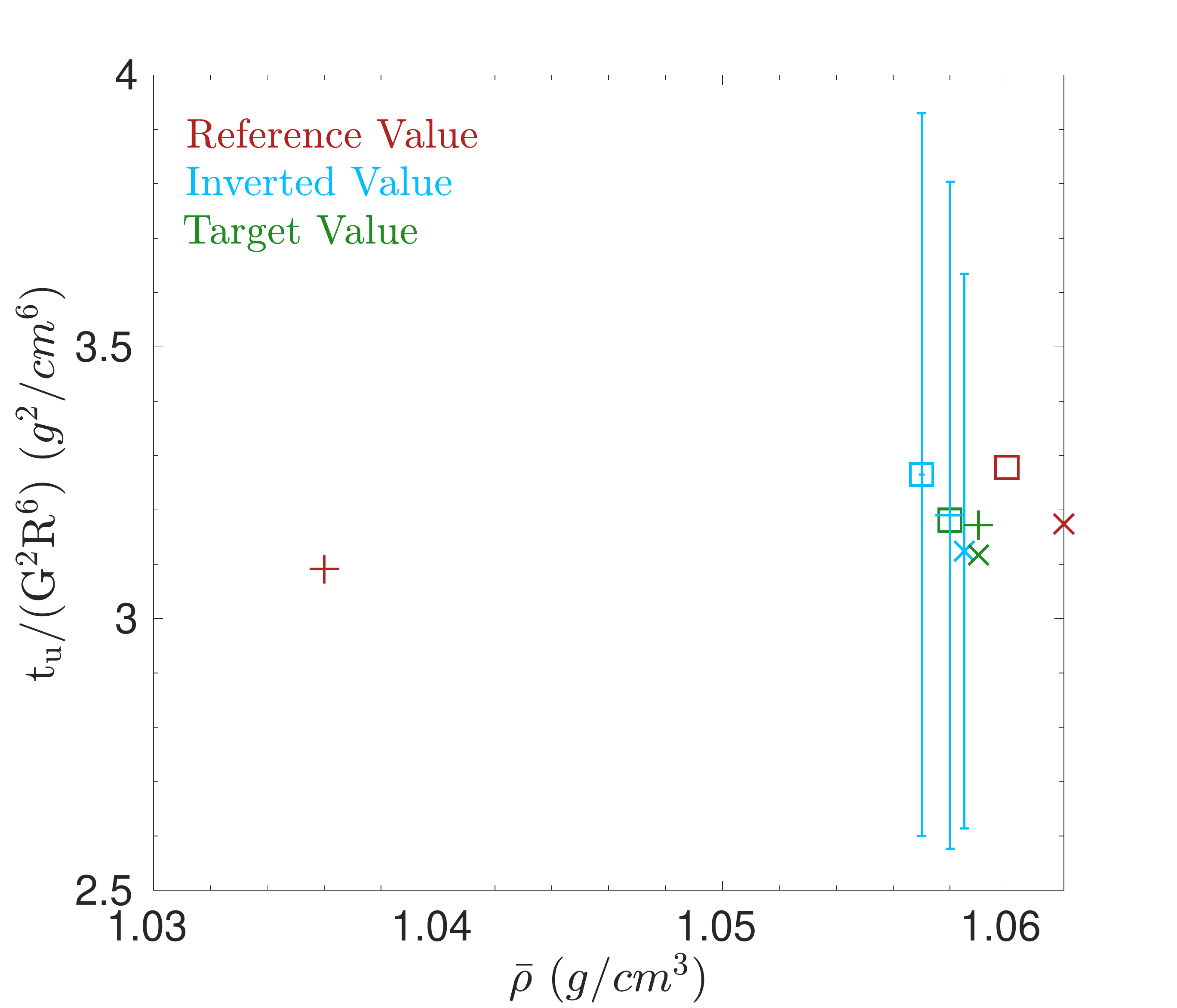}
 \end{minipage}	
 \end{flushleft}
	\caption{Inversion results for the core condition indicator, $t_{u}$, as a function of the mean density, using various stellar evolutionary models of 16CygA (left panel) and 16CygB (right panel) as artificial targets and the exact same dataset as the observed dataset in each case.}
		\label{FigTuCheckCygAB}
\end{figure*}

\begin{figure*}
\begin{center}
  	\includegraphics[trim=5 5 5 5, clip, width=0.80\linewidth]{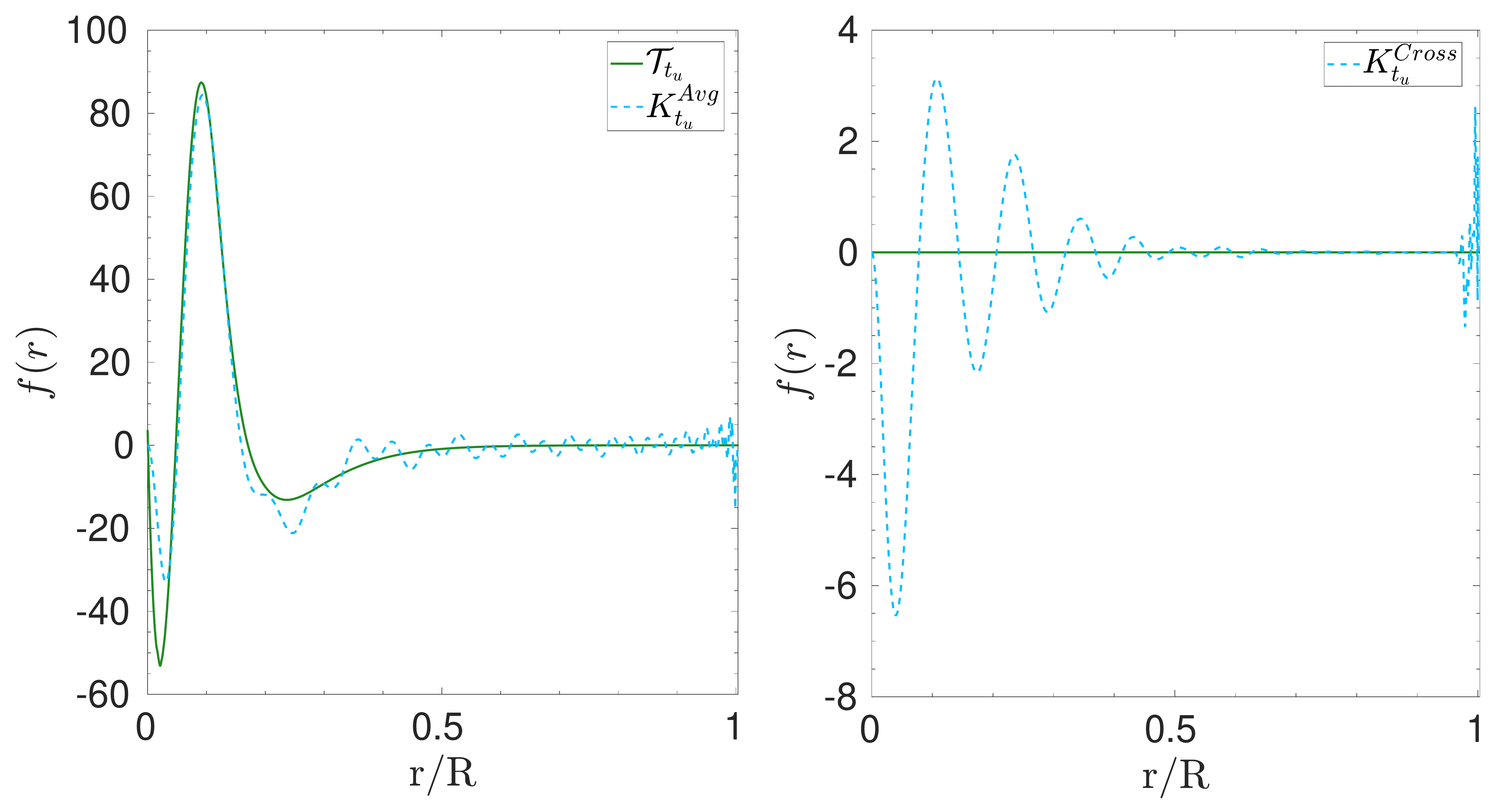}
 \end{center}
	\caption{Averaging kernel for the $t_{u}$ inversion of 16CygA (left, blue) and its associated target function (green). Cross-term kernel for the $t_{u}$ inversion of 16CygA (right, blue) and its associated target function (green).}
		\label{FigKerTuCheckA}
\end{figure*}

\begin{figure*}
\begin{center}
  	\includegraphics[trim=5 5 5 5, clip, width=0.80\linewidth]{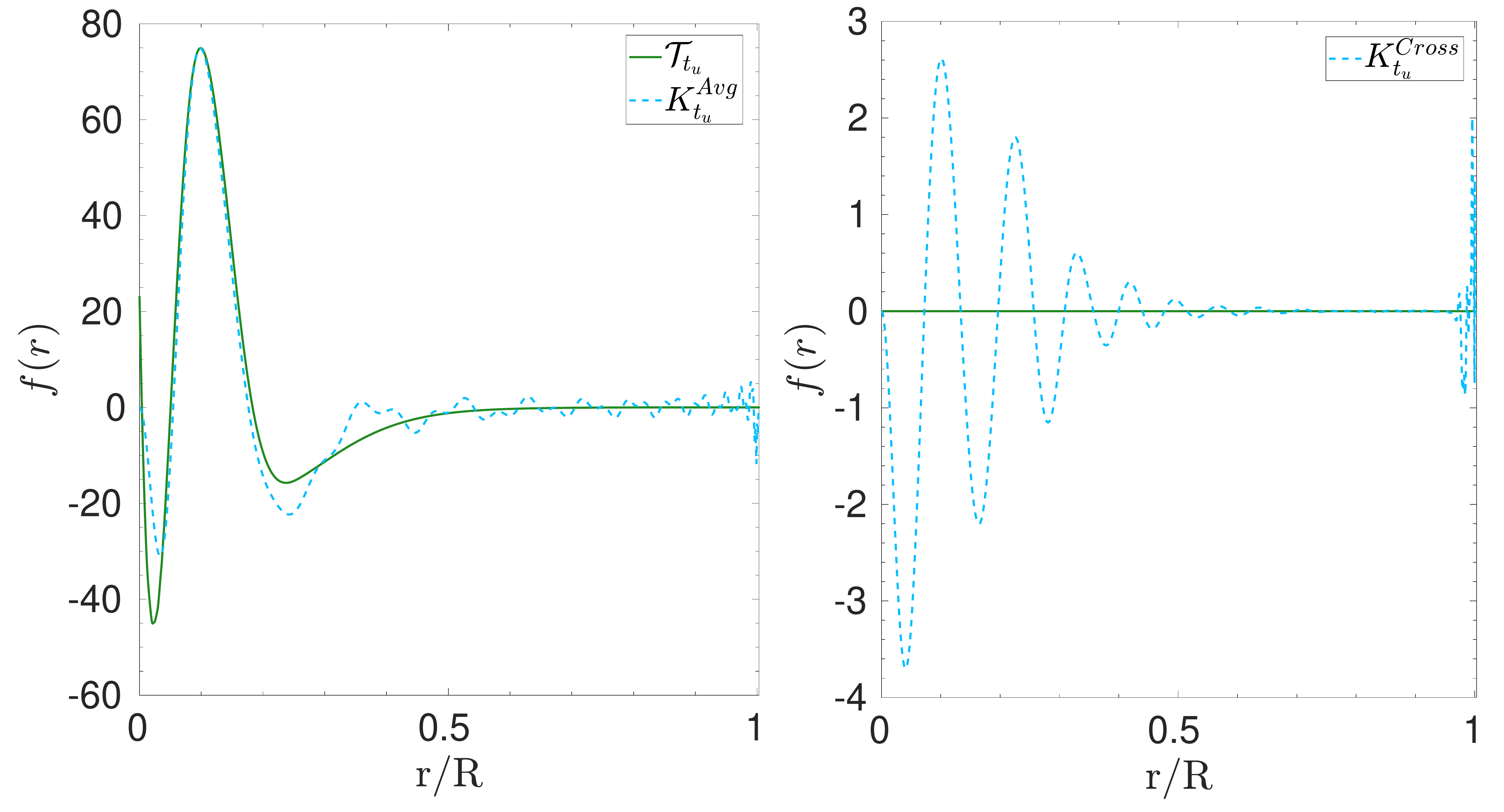}
 \end{center}
	\caption{Averaging kernel for the $t_{u}$ inversion of 16CygB (left, blue) and its associated target function (green). Cross-term kernel for the $t_{u}$ inversion of 16CygB (right, blue) and its associated target function (green).}
		\label{FigKerTuCheckB}
\end{figure*}

\begin{figure*}
\begin{flushleft}
	\begin{minipage}{\textwidth}
  	\includegraphics[trim=5 5 5 5, clip, width=0.47\linewidth]{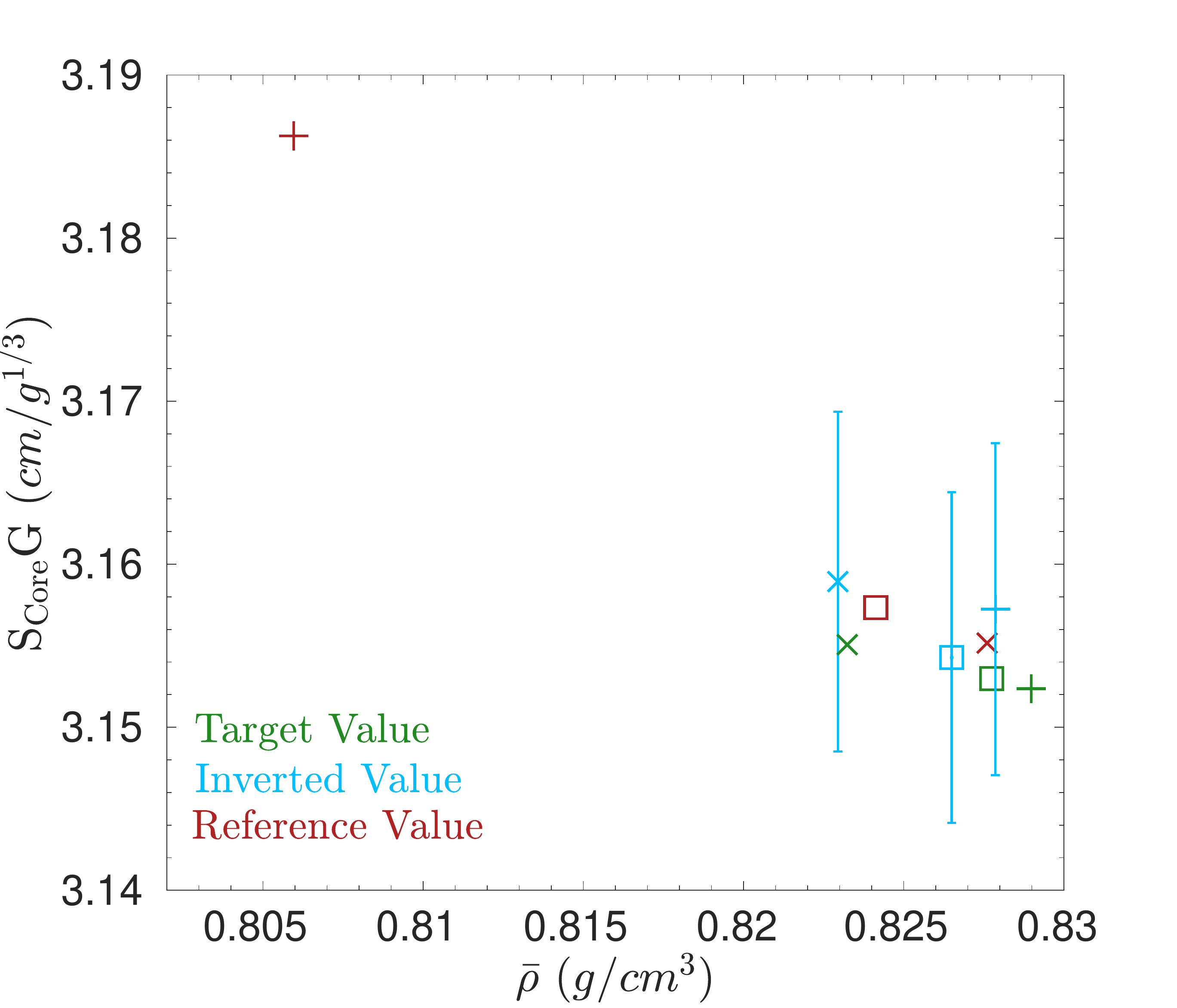}
 	\includegraphics[trim=5 5 5 5, clip, width=0.47\linewidth]{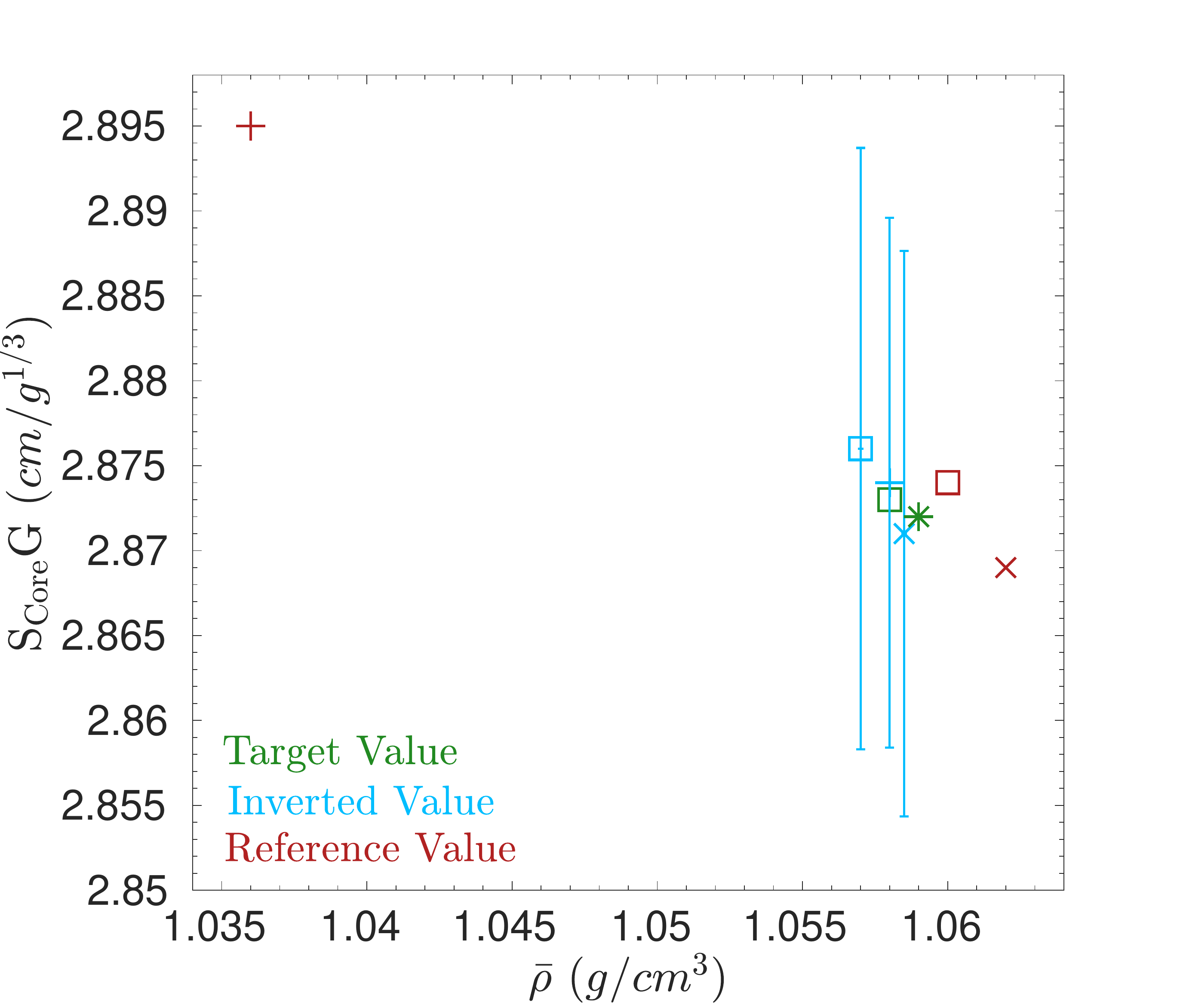}
 \end{minipage}	
 \end{flushleft}
	\caption{Inversion results for the core condition indicator, $S_{\rm{Core}}$, as a function of the mean density, using various stellar evolutionary models of 16CygA (left panel) and B (right panel) as artificial targets and the exact same dataset as the observed dataset for both stars.}
		\label{FigSCheckCygAB}
\end{figure*}

The results of the verifications are shown in Figs. \ref{FigTuCheckCygAB} and \ref{FigSCheckCygAB}. For both cases, the results are well within 1$\sigma$ of the actual target values for both $t_{u}$ and $S_{Core}$. The mean density values are taken from the mean density inversion carried out in Appendix \ref{Sec:RhoCheckups}. The inversion also remains stable because no large spurious corrections are observed. However, given the small variations found between the models, we can already foresee that the inversion will be at its resolution limit to provide constraints on the internal structure of 16CygA$\&$B. Illustrations of the averaging and cross-term kernels for specific test cases for the two stars are shown in Figs. \ref{FigKerTuCheckA} and \ref{FigKerTuCheckB}.

\begin{figure*}
\begin{center}
  	\includegraphics[trim=5 5 5 5, clip, width=0.80\linewidth]{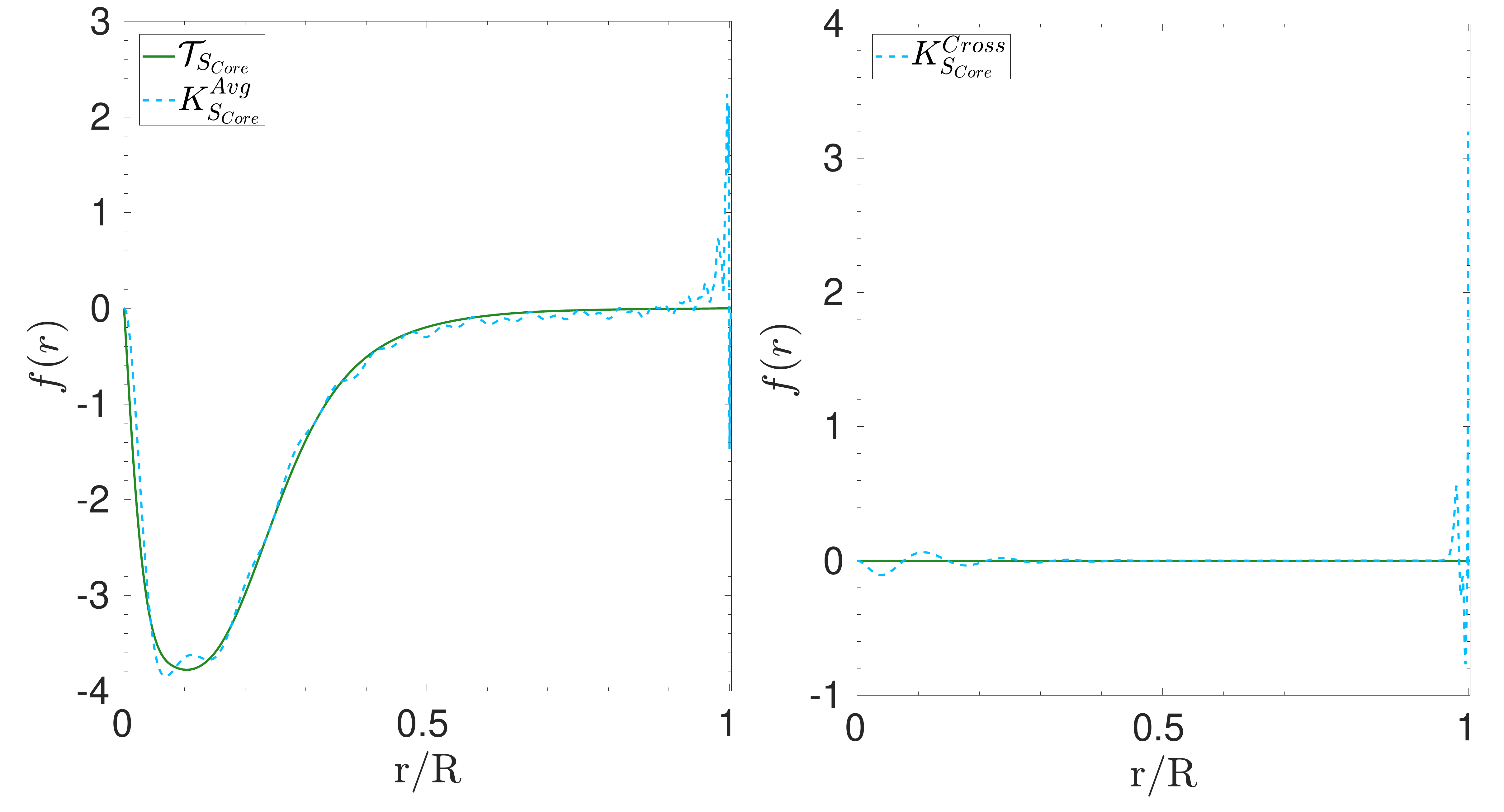}
 \end{center}
	\caption{Averaging kernel for the $S_{Core}$ inversion of 16CygA (left, blue) and its associated target function (green). Cross-term kernel for the $S_{Core}$ inversion of 16CygA (right, blue) and its associated target function (green).}
		\label{FigKerSCoreCheckA}
\end{figure*}

\begin{figure*}
\begin{center}
  	\includegraphics[trim=5 5 5 5, clip, width=0.80\linewidth]{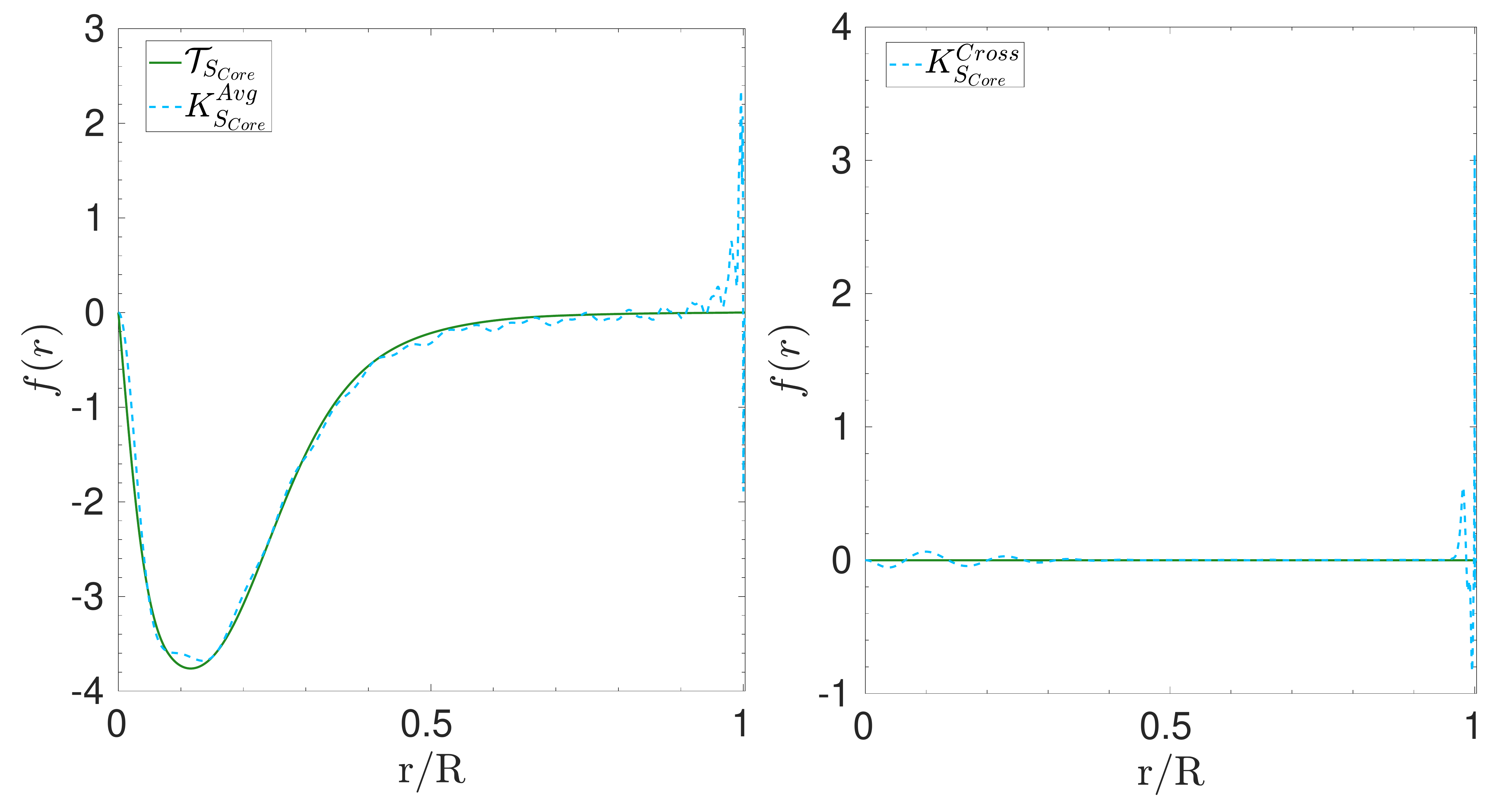}
 \end{center}
	\caption{Averaging kernel for the $S_{Core}$ inversion of 16CygB (left, blue) and its associated target function (green). Cross-term kernel for the $S_{Core}$ inversion of 16CygB (right, blue) and its associated target function (green).}
		\label{FigKerSCoreCheckB}
\end{figure*}

In the case of the $S_{Core}$ inversion, the results are stable and can also lead to meaningful corrections, as illustrated by the crosses on the left in the two panels of Fig. \ref{FigSCheckCygAB}. Thus, the inversion appears to be accurate and stable. The averaging and cross-term kernels for some of these inversions are shown in Figs. \ref{FigKerSCoreCheckA} and \ref{FigKerSCoreCheckB}.

\subsection{Envelope indicators}\label{Sec:EnvCheckups}

\begin{figure*}
\begin{flushleft}
	\begin{minipage}{0.95\textwidth}
  	\includegraphics[trim=5 5 5 5, clip, width=0.47\linewidth]{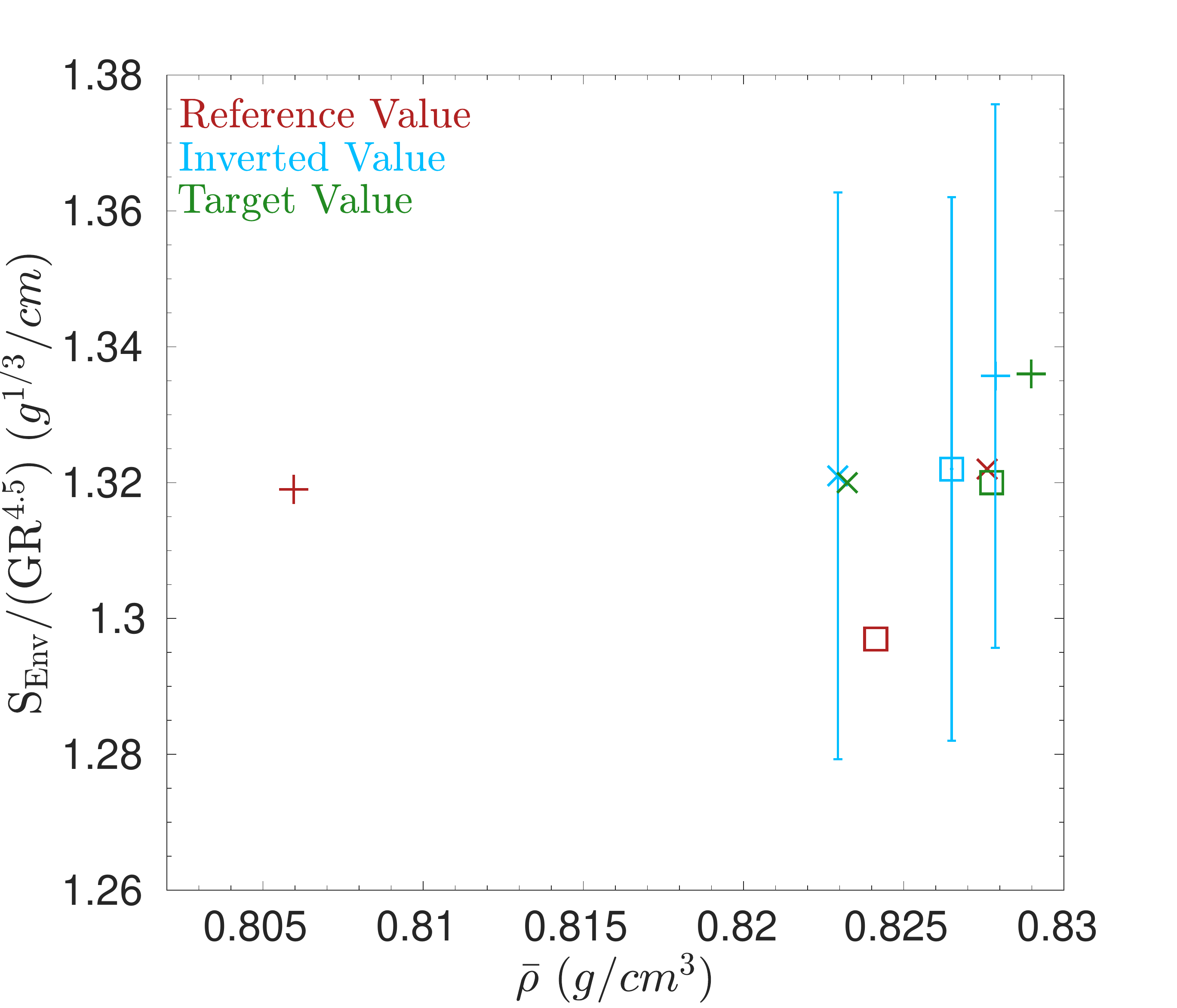}
 	\includegraphics[trim=5 5 5 5, clip, width=0.47\linewidth]{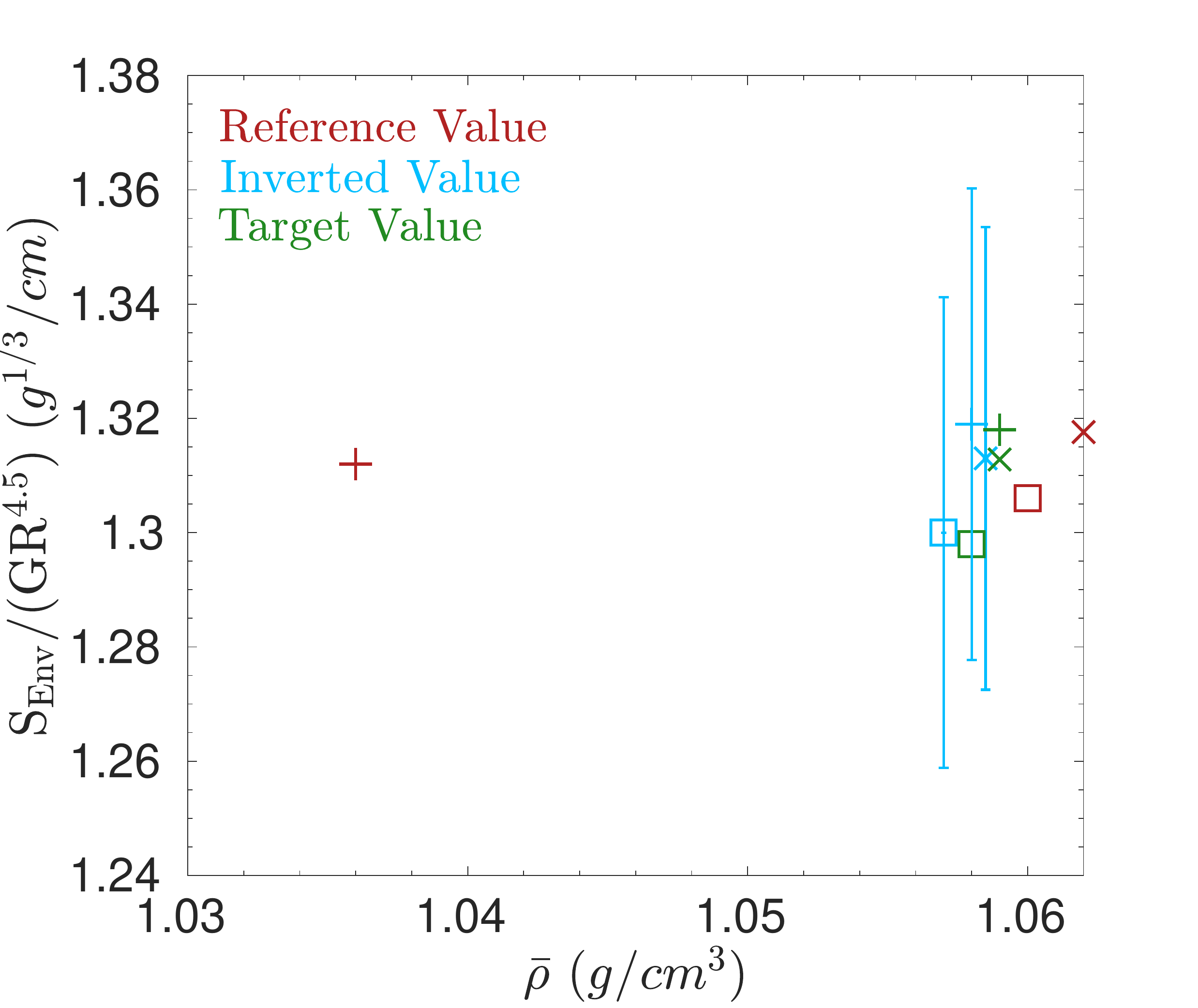}
 \end{minipage}	
 \end{flushleft}
	\caption{Inversion results for the envelope condition indicator, $S_{\rm{Env}}$, as a function of the mean density, using various stellar evolutionary models of 16CygA (left panel) and B (right panel) as artificial targets and the exact same dataset as the observed dataset in both cases.}
		\label{FigSEnvCheckAB}
\end{figure*}

\begin{figure*}
\begin{center}
  	\includegraphics[trim=5 5 5 5, clip, width=0.80\linewidth]{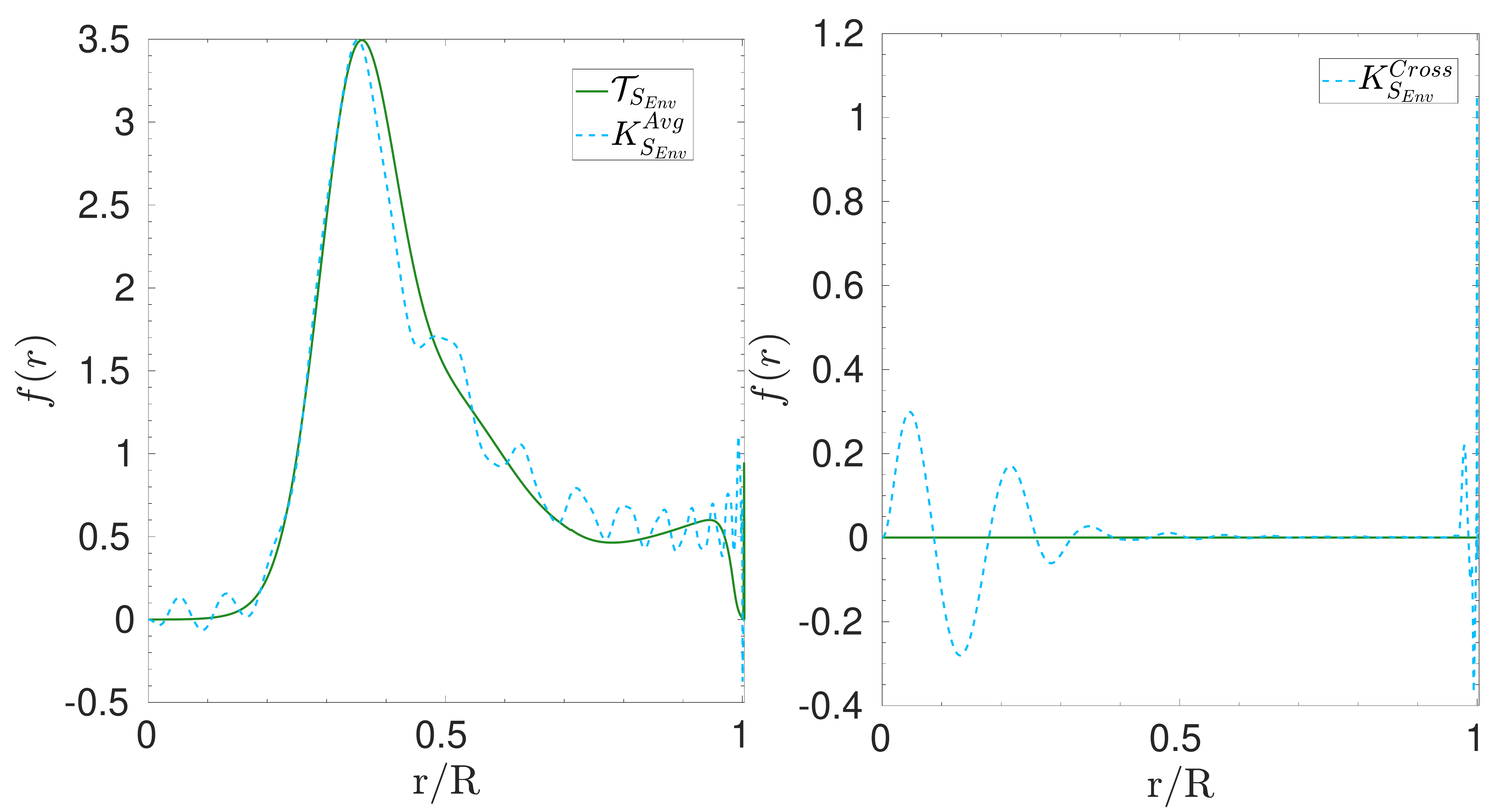}
 \end{center}
	\caption{Averaging kernel for the $S_{Env}$ inversion of 16CygA (left, blue) and its associated target function (green). Cross-term kernel for the $S_{Env}$ inversion of 16CygA (right, blue) and its associated target function (green).}
		\label{FigKerSEnvCheckA}
\end{figure*}

\begin{figure*}
\begin{center}
  	\includegraphics[trim=5 5 5 5, clip, width=0.80\linewidth]{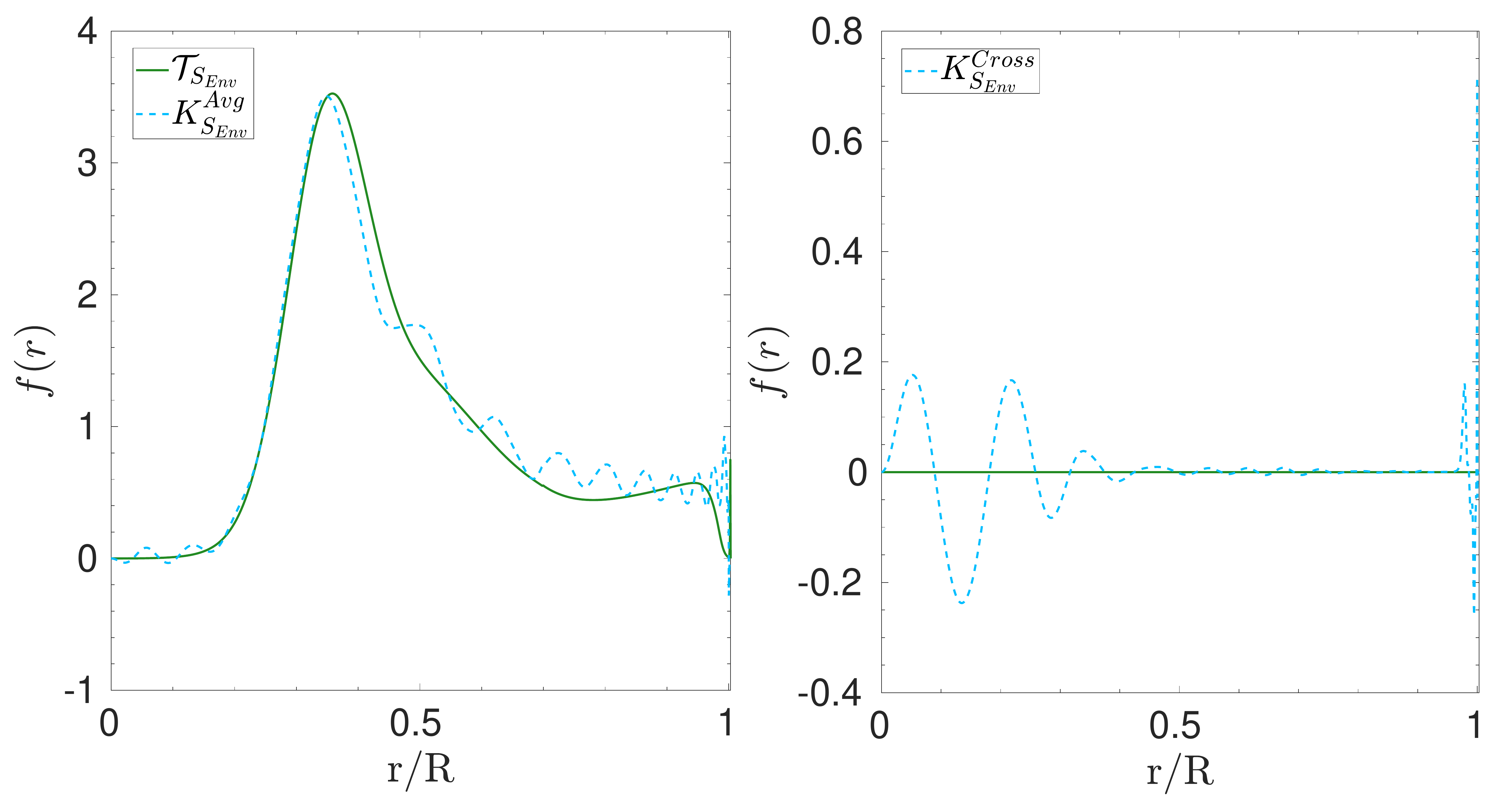}
 \end{center}
	\caption{Averaging kernel for the $S_{Env}$ inversion of 16CygB (left, blue) and its associated target function (green). Cross-term kernel for the $S_{Env}$ inversion of 16CygB (right, blue) and its associated target function (green).}
		\label{FigKerSEnvCheckB}
\end{figure*}

As shown in Fig. \ref{FigSEnvCheckAB}, the inversion remains stable and does not provide corrections between the reference model and the target. This may again indicate that it might not be able to provide any significant corrections for this indicator, but confirms the reliability of the models. This inversion is particularly difficult because the properties of the convective envelope are difficult to extract using only low-degree modes. Nevertheless, a relatively good agreement between the averaging kernels and the target function can be achieved, as illustrated for both stars for a specific test case in Figs. \ref{FigKerSEnvCheckA} and \ref{FigKerSEnvCheckB}.

\subsection{Localised inversions}\label{Sec:LocCheckups}

\begin{figure*}
\begin{flushleft}
  	\includegraphics[trim=5 5 5 5, clip, width=0.95\linewidth]{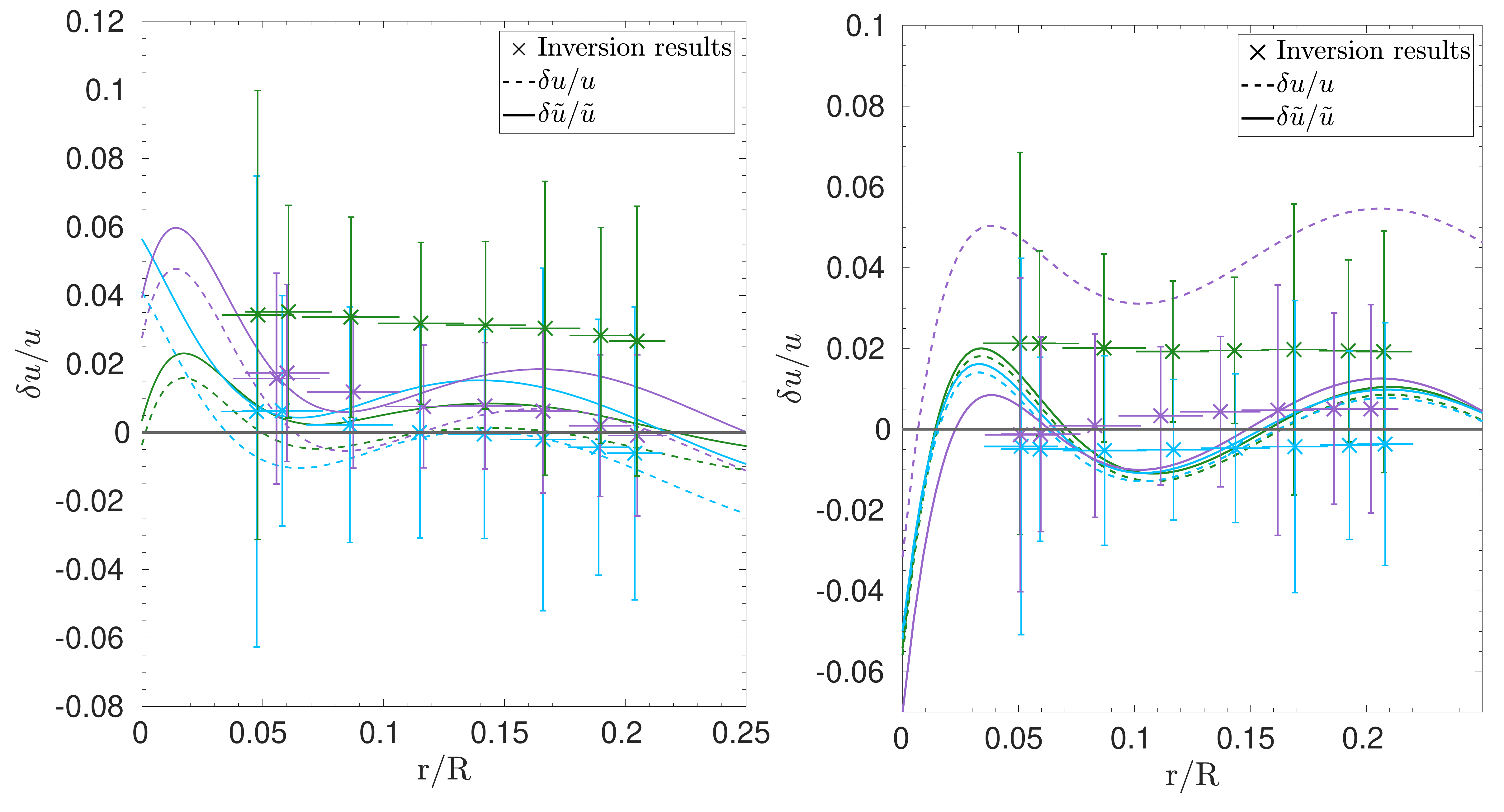}
 \end{flushleft}
	\caption{Inversion of relative differences of the squared isothermal sound speed as a function of the normalised radius using various evolutionary models as references and targets for both 16CygA (left panel) and 16CygB (right panel). The inverted differences are compared in terms of the dimensional, $u$, and adimensional squared adiabatic sound speed, $\tilde{u}$. The colours are associated with three different pairs of reference models and targets, all computed in Paper I. The continuous and dashed lines correspond to the actual differences between the target and reference models in adimensional and dimensional squared isothermal sound speed.}
		\label{FigStrucInvABCheckup}
\end{figure*}

We illustrate the results of the robustness tests of localised inversions in Fig. \ref{FigStrucInvABCheckup}, where the left panel shows the inversions of 16CygA models and the right panel shows those for 16CygB models. The crosses are the corrections obtained using the SOLA technique, and the dashed and plain lines are the actual dimensional and adimensional differences between the two models in squared isothermal sound speed, defined following
\begin{align}
\frac{u^{\rm{Obs}}-u^{\rm{Ref}}}{u^{\rm{Ref}}} = \frac{(M_{\rm{Obs}}/R_{\rm{Obs}})\tilde{u}^{\rm{Obs}}-(M_{\rm{Ref}}/R_{\rm{Ref}})\tilde{u}^{\rm{Ref}}}{(M_{\rm{Ref}}/R_{\rm{Ref}})\tilde{u}^{\rm{Ref}}},
\end{align} 
with $M_{\rm{Obs}}$ and $R_{\rm{Obs}}$ the mass and radius of the observed target and $M_{\rm{Ref}}$ and $R_{\rm{Ref}}$, those of the reference model. 

In both panels, green symbols are related to a target model with a mean density difference of about $2\%$ with respect to the reference model. In these conditions neither the dimensional nor adimensional squared isothermal sound speed can be estimated reliably.

The blue and magenta symbols show results for models with mean density values well within $1\%$ of each other. In the left panel, the results are very similar for the adimensional and the dimensional isothermal sound speed within $1\sigma$. The actual differences between the models remain within $1\%$, except for the very deep core, that is not probed by pure p-modes. A similar situation is found in helioseismology, where the sound speed inversions usually stop at about $0.05\rm{R}_{\odot}$. 

In the right panel, the magenta results are related to specific models with the same mean density, but very different $M/R$. In these conditions, no meaningful information is obtained on the dimensional isothermal sound speed. The inversion results also almost disagree with the actual differences of $\tilde{u}$ around $0.12$R. The blue symbols are related to models with almost exactly the same mass and radius, thus $u$ and $\tilde{u}$ are essentially the same. In this case, the inversion results agree excellently with the actual differences.

From our tests, we conclude that inversions of $\tilde{u}$ are only meaningful if the mean density is determined well below $1\%$. Similarly, the dimensional sound speed can only be determined in the presence of reliable mass and radius estimates (beyond the reaches of classical seismic modelling when model-independent inversions results are the goal). In these conditions, spectroscopic binary systems or targets with independent radius determination using interferometry are prime candidates. 

However, these test cases also indicate a similar behaviour for most of the indicators, namely that the SOLA inversions will mostly validate the quality of the models built by \citet{Farnir2020}. 

\begin{figure*}
\begin{center}
  	\includegraphics[trim=5 5 5 5, clip, width=0.80\linewidth]{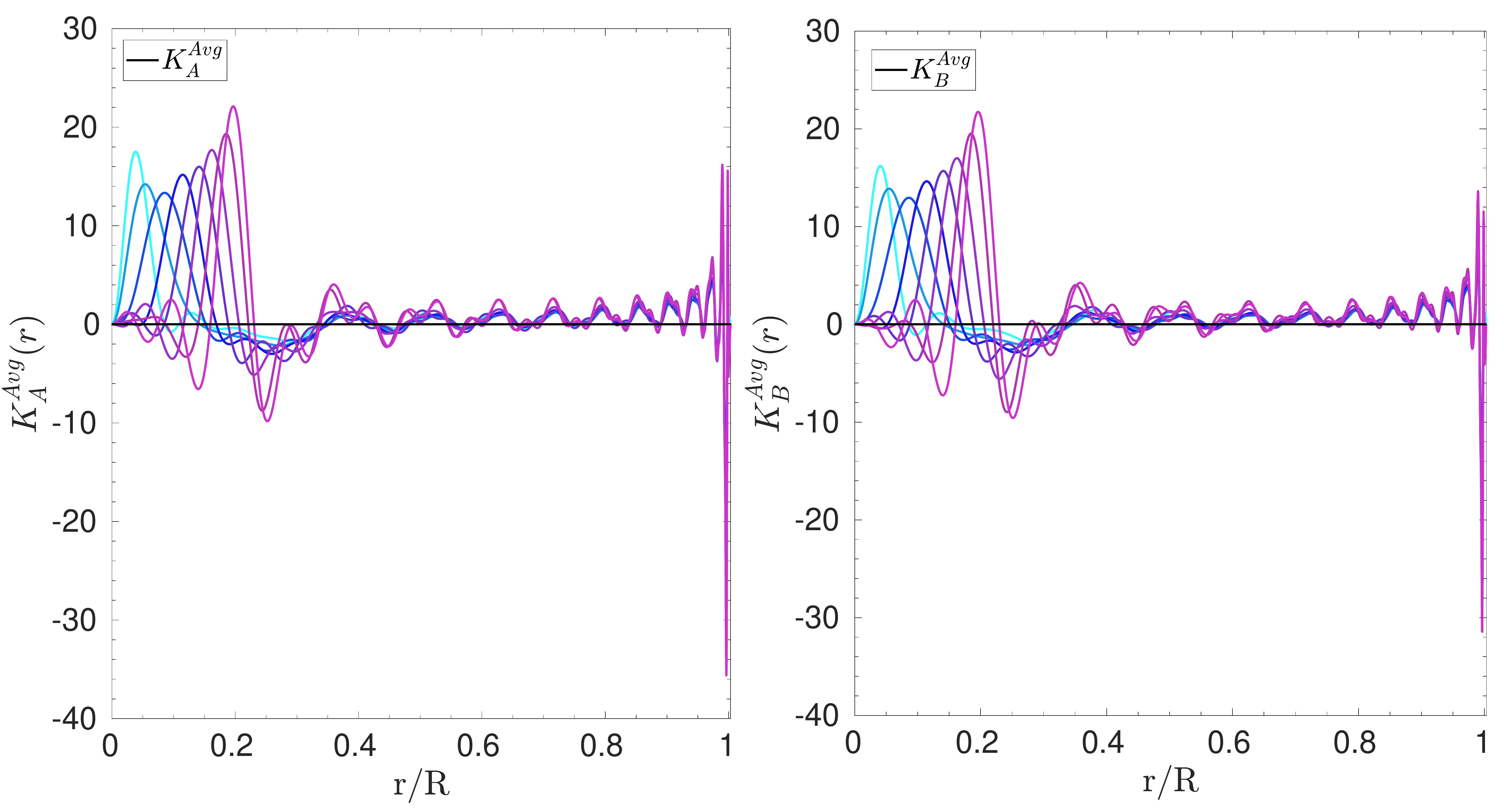}
 \end{center}
	\caption{Averaging kernel for the localised inversion of adimensional squared isothermal sound speed of 16CygA for various positions (left). Same for 16CygB (right).}
		\label{FigKerLocalGauss}
\end{figure*}

\section{Error contributions of the inversion robustness tests}\label{SecSupplTables}

We provide here some additional data regarding the verification inversions between the various models in Tables \ref{tab:CheckupErrorsA} and \ref{tab:CheckupErrorsB} including the errors as defined in Eqs. \ref{eq:AvgError}, \ref{eq:CrossError}, and \ref{eq:ResError}. 

\begin{table*}[t]
\caption{Results of the verfication inversions for the 16CygA models for the various indicators $A^{j}$, $j$ denoting the test number.}
\label{tab:CheckupErrorsA}
  \centering
\begin{tabular}{r | c | c | c }
\hline \hline
 & $\varepsilon_{\mathrm{Avg}}$& $\varepsilon_{\mathrm{Cross}}$&$\varepsilon_{\mathrm{Res}}$\\ \hline
 $\bar{\rho}^{1}$& $-7.909\times 10^{-4}$& $1.400\times 10^{-4}$& $2.904\times 10^{-4}$\\
 $\bar{\rho}^{2}$ &  $-1.497 \times 10^{-3}$& $-9.219 \times 10^{-5}$ & $8.256 \times 10^{-5}$\\
 $\bar{\rho}^{3}$ & $-1.256 \times 10^{-3}$& $-6.993 \times 10^{-5}$& $-2.027 \times 10^{-5}$ \\  
 $t^{1}_{u}$ & $3.395 \times 10^{-2}$& $3.124\times 10^{-3}$& $-2.875\times 10^{-2}$\\
  $t^{2}_{u}$ &  $-2.250 \times 10^{-3}$& $-6.282\times 10^{-4}$& $2.485\times 10^{-2}$\\
   $t^{3}_{u}$ & $1.312 \times 10^{-3}$& $1.070\times 10^{-3}$& $ 6.274\times 10^{-3}$\\  
  $S^{1}_{Core}$ & $1.110 \times 10^{-3}$& $2.824\times 10^{-5}$& $8.551\times 10^{-5}$\\
  $S^{2}_{Core}$ & $3.663 \times 10^{-4}$& $-1.105\times 10^{-5}$& $5.533\times 10^{-5}$\\
   $S^{3}_{Core}$ & $1.454 \times 10^{-3}$& $2.498\times 10^{-5}$& $ 6.274\times 10^{-3}$\\  
     $S^{1}_{Env}$ & $2.130 \times 10^{-3}$& $-1.294\times 10^{-4}$& $-1.076\times 10^{-3}$\\
  $S^{2}_{Env}$ & $-2.327 \times 10^{-3}$& $2.145\times 10^{-5}$& $3.703\times 10^{-3}$\\
  $S^{3}_{Env}$ & $1.985 \times 10^{-3}$& $-2.291\times 10^{-5}$& $-4.251\times 10^{-4}$\\
\hline
\end{tabular}
\end{table*}

\begin{table*}[t]
\caption{Results of the verification inversions for the 16CygB models for the various indicators $A^{j}$, $j$ denoting the test number.}
\label{tab:CheckupErrorsB}
  \centering
\begin{tabular}{r | c | c | c }
\hline \hline
 &$\varepsilon_{\mathrm{Avg}}$& $\varepsilon_{\mathrm{Cross}}$&$\varepsilon_{\mathrm{Res}}$\\ \hline
 $\bar{\rho}^{1}$& $-4.742\times 10^{-4}$& $-5.886\times 10^{-5}$& $1.238\times 10^{-5}$\\
 $\bar{\rho}^{2}$ & $-5.564\times 10^{-4}$& $8.324 \times 10^{-4}$ & $5.954 \times 10^{-7}$\\
 $\bar{\rho}^{3}$ & $-4.741 \times 10^{-4}$& $-6.366 \times 10^{-4}$& $-3.193 \times 10^{-5}$ \\
 
 $t^{1}_{u}$ &$3.750 \times 10^{-4}$& $-1.248\times 10^{-4}$& $-1.542\times 10^{-3}$\\
  $t^{2}_{u}$ & $3.067 \times 10^{-2}$& $6.149\times 10^{-4}$& $-3.870\times 10^{-2}$\\
  $t^{3}_{u}$ & $2.194 \times 10^{-4}$& $2.604\times 10^{-3}$& $3.595\times 10^{-3}$\\
  
   $S^{1}_{Core}$ & $6.603 \times 10^{-4}$& $2.780\times 10^{-5}$& $6.607\times 10^{-5}$\\
  $S^{2}_{Core}$ & $2.232\times 10^{-4}$& $-7.470\times 10^{-5}$& $-2.537\times 10^{-4}$\\
  $S^{3}_{Core}$ & $6.362 \times 10^{-4}$& $1.041\times 10^{-4}$& $1.776\times 10^{-4}$\\
  
     $S^{1}_{Env}$ & $3.166 \times 10^{-4}$& $-3.740\times 10^{-5}$& $2.640\times 10^{-4}$\\
  $S^{2}_{Env}$ & $1.191 \times 10^{-3}$& $3.484\times 10^{-5}$& $-6.779\times 10^{-4}$\\
  $S^{3}_{Env}$ & $2.353\times 10^{-4}$& $-1.181\times 10^{-4}$& $9.694\times 10^{-5}$\\
\hline
\end{tabular}
\end{table*}

\end{document}